\documentclass[journal]{IEEEtran}
\IEEEoverridecommandlockouts
\usepackage{cite}
\usepackage{amsthm,amsmath,amssymb,amsfonts}
\usepackage{algorithm}
\usepackage{algorithmic}
\usepackage{graphicx}
\usepackage{textcomp}
\usepackage{xcolor}
\usepackage{subcaption}
\usepackage{multirow}
\usepackage{mathtools}
\usepackage{xcolor}

\usepackage[font=footnotesize]{caption}
\usepackage[protrusion=true]{microtype}
\usepackage{enumitem} % For bullet indent

\newtheorem{theorem}{Theorem}
\newtheorem{lemma}{Lemma}
\newtheorem{remark}{Remark}

\renewenvironment{IEEEbiography}[1]
  {\IEEEbiographynophoto{#1}}
  {\endIEEEbiographynophoto}

\def\BibTeX{{\rm B\kern-.05em{\sc i\kern-.025em b}\kern-.08em
    T\kern-.1667em\lower.7ex\hbox{E}\kern-.125emX}}

\begin{document}

\title{
Minimum \textcolor{black}{Overhead} Beamforming and Resource Allocation in D2D Edge Networks
%Joint Optimization of MIMO Signal Design and Resource Allocation in D2D Edge Computing
%\\
}

%\author{\IEEEauthorblockN{Junghoon Kim\IEEEauthorrefmark{1},
%Christopher G. Brinton\IEEEauthorrefmark{1}, Taejoon Kim\IEEEauthorrefmark{2}, Morteza Hashemi\IEEEauthorrefmark{2} and David J. Love\IEEEauthorrefmark{1} }
%\IEEEauthorblockA{\IEEEauthorrefmark{1}Electrical and Computer Engineering, Purdue University, IN, USA
%\\
%\IEEEauthorrefmark{2}Electrical Engineering and Computer Science, University of Kansas, KS, USA \\
%%Wherever\\
%Email: \IEEEauthorrefmark{1}{\{kim3220, cgb, djlove\}}@purdue.edu,
%\IEEEauthorrefmark{2}{ \{taejoonkim, mhashemi\}}@ku.edu
%}}

\author{Junghoon~Kim,~\IEEEmembership{Student Member,~IEEE,}
        Taejoon~Kim,~\IEEEmembership{Senior Member,~IEEE,}
        Morteza~Hashemi,~\IEEEmembership{Member,~IEEE,}
        Christopher G. Brinton,~\IEEEmembership{Senior Member,~IEEE,}
        and David J. Love,~\IEEEmembership{Fellow,~IEEE}% <-this % stops a space
%\thanks{
%\author{Junghoon~Kim,
%        Taejoon~Kim,
%        Morteza~Hashemi,
%        Christopher G. Brinton,
%        and David J. Love%\IEEEmembership{Fellow,~IEEE}% <-this % stops a space
\thanks{
This work was supported in part by the National Science Foundation (NSF) under grants CNS1642982, CCF1816013, and CNS 1955561, and National Spectrum Consortium (NSC) under grant W15QKN-15-9-1004.
Also, this work was supported in part by the Office of Naval Research under grant N00014-21-1-2472.
This work was presented in part at the 2020 IEEE International Conference on Computer Communications (INFOCOM) \cite{kim2020joint}.}
\thanks{Junghoon Kim, Christopher G. Brinton, and David J. Love are with the Department
of Electrical and Computer Engineering, Purdue University, West Lafayette, IN, 47907 USA
(e-mail: kim3220@purdue.edu; cgb@purdue.edu; djlove@purdue.edu).}% <-this % stops a space
\thanks{Taejoon~Kim and Morteza~Hashemi are with the Department of Electrical Engineering and Computer Science, University of Kansas, KS, 66045 USA (email: taejoonkim@ku.edu; mhashemi@ku.edu).}% <-this % stops a space
}

\maketitle

\begin{abstract}
Device-to-device (D2D) communications is expected to be a critical enabler of distributed computing in edge networks at scale. A key challenge in providing this capability is the requirement for judicious management of the heterogeneous communication and computation resources that exist at the edge to meet processing needs. In this paper, we develop an optimization methodology that considers the network topology jointly with device and network resource allocation to minimize total D2D overhead, which we quantify in terms of time and energy required for task processing. Variables in our model include task assignment, CPU allocation, subchannel selection, and beamforming design for multiple-input multiple-output (MIMO) wireless devices. We propose two methods to solve the resulting non-convex mixed integer program: semi-exhaustive search optimization, which represents a ``best-effort'' at obtaining the optimal solution, and efficient alternate optimization, which is more computationally efficient. As a component of these two methods, we develop a novel coordinated beamforming algorithm which we show obtains the optimal beamformer for a common receiver characteristic. Through numerical experiments, we find that our methodology yields substantial improvements in network overhead compared with local computation and partially optimized methods, which validates our joint optimization approach. Further, we find that the efficient alternate optimization scales well with the number of nodes, and thus can be a practical solution for D2D computing in large networks.
\end{abstract}

\begin{IEEEkeywords}
% Wireless device-to-device (D2D) edge computing,  minimum communication overhead beamforming (MCOB), central processing unit (CPU) allocation, subchannel allocation.
Wireless edge networks, device-to-device (D2D) communications, multiple-input-multiple-output (MIMO), beamforming, network optimization.
\end{IEEEkeywords}

% \vspace{-2mm}
\section{Introduction}
\label{sec:intro}

\IEEEPARstart{T}{he} number of wireless devices is now over $8.6$ billion, and with the advent of new 5G-and-beyond technologies, this is expected to grow to $12.3$ billion by 2022 \cite{chiang2016fog}. 
Many of these devices will be data-processing-capable nodes in the hands of users that facilitate rapidly growing data-intensive applications running at the network edge, e.g., social networking, video streaming, and distributed data analytics. 
Given the bursty nature of user demands, when certain devices are occupied 
with processing for computationally-intensive applications, e.g., face recognition, 3D modeling/gaming, and augmented/virtual reality (AR/VR), it may be desirable for them to offload their data to devices with underutilized resources~\cite{yao2019edgeflow, yang2018communication, 8319323}.
Traditionally, cloud computing architectures, such as Amazon Web Services and Microsoft Azure, have been adopted for such data intensive applications, but the exponential rise in data generation at the edge is making centralized architectures infeasible for providing latency-sensitive quality of service at scale~\cite{chiang2016fog}.

\begin{figure}
\centering
\begin{subfigure}{.48\columnwidth}
  \centering
  \includegraphics[width=\linewidth]{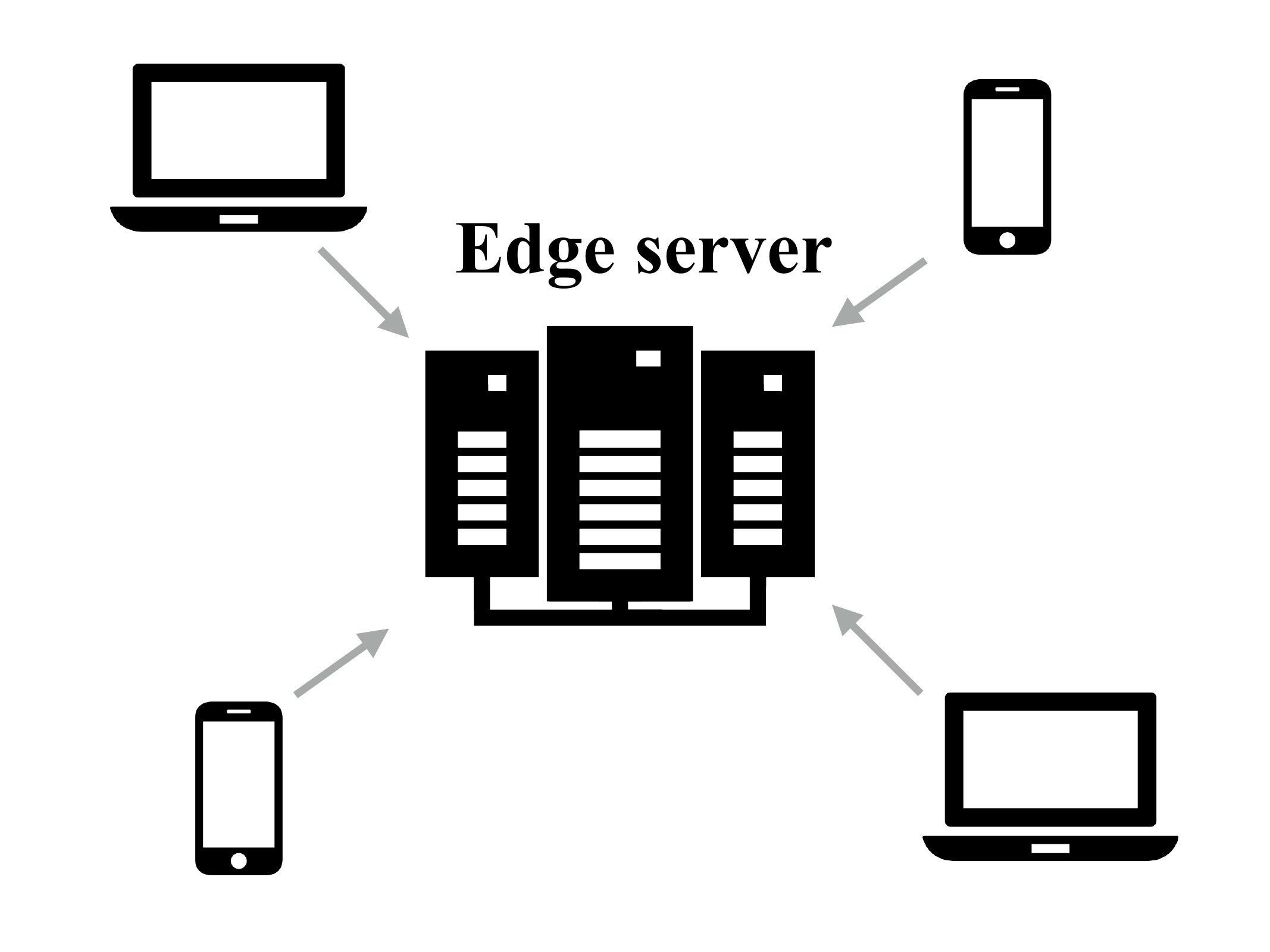}
  \vspace{-8mm}
  \caption{MEC}
  \label{fig:topo:MEC}
\end{subfigure}
\begin{subfigure}{.48\columnwidth}
  \centering
  \includegraphics[width=\linewidth]{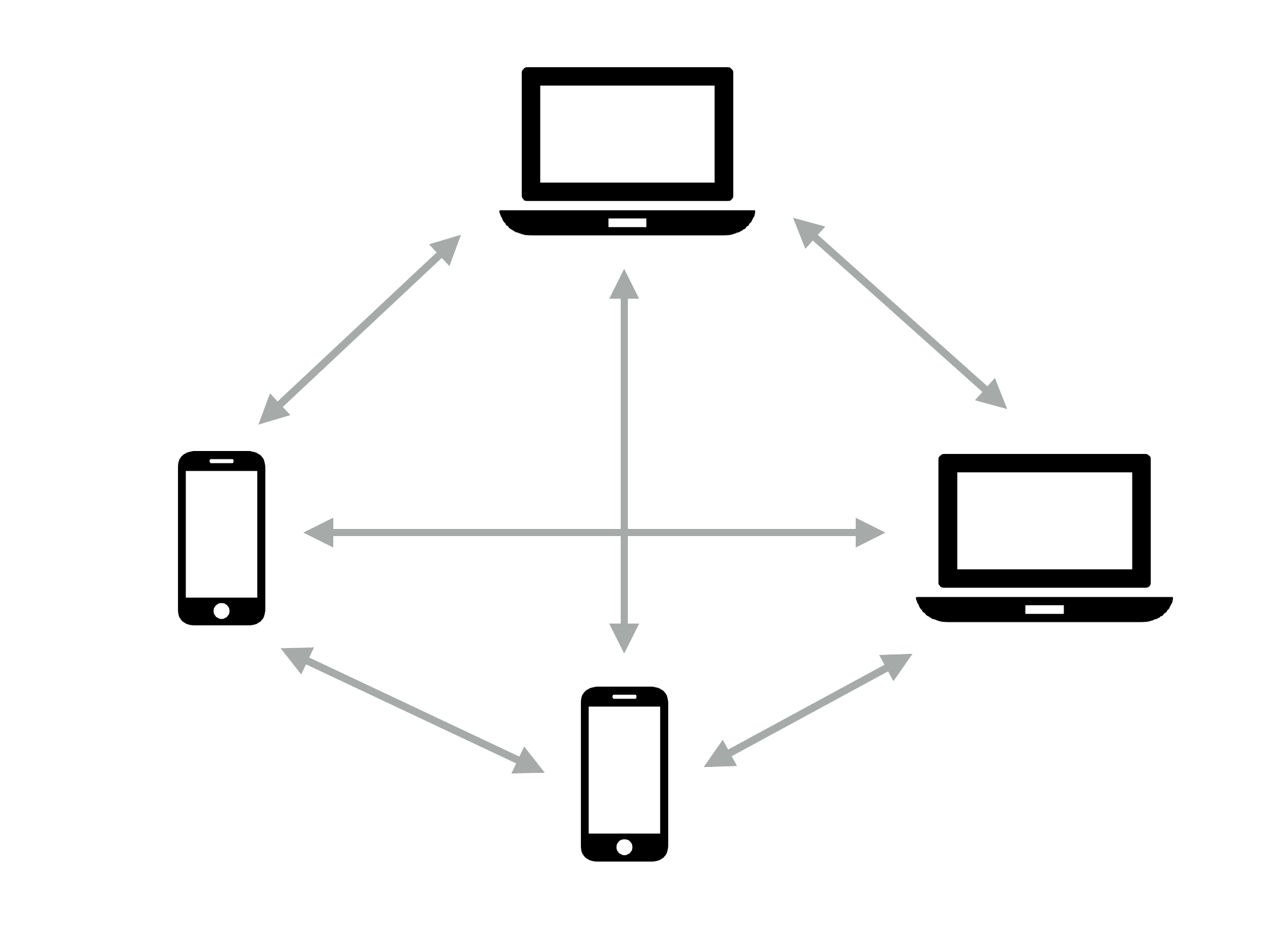}
  \vspace{-8mm}
  \caption{D2D}
  \label{fig:topo:D2D}
\end{subfigure}
\vspace{-1mm}
\caption{High-level comparison between the topologies of (a) mobile edge computing (MEC) systems and (b) device-to-device (D2D) networks. MEC topology is typically fixed and predetermined, while D2D topology is not and can support offloading between devices.}
\label{fig:topo}
\vspace{-4mm}
%\vspace{-0.1in}
\end{figure}

As a current trend in wireless networks is reducing cell sizes \cite{Sultan18}, many 5G networks will be dense 
with small cell coverage areas and networks composed of several smaller subsets \cite{Cisco19}. 
% with short distances, forming several smaller subnets 
Networks of small subnets combined with improved computational and storage capabilities of edge devices
are enabling mobile edge computing (MEC) architectures.
At a high level, MEC leverages radio access networks (RANs) 
to increase the amount of computing power located close to the end user, which enables the end user to offload computations (e.g., using a central processing entity)
% to boost computing power in close proximity to end-users, thus enabling the users to offload their computations to an edge server (central processing entity)
as shown in Fig. \ref{fig:topo}(\subref{fig:topo:MEC})~\cite{chen2015efficient,le2017efficient,naderializadeh2019energy,chen2018task,zhang2017energy,nguyen2019computation,sardellitti2015joint,barbarossa2017enabling,liu2019joint,wang2017joint}. 
In an MEC architecture, the edge servers 
have high-performance computing units which can process large amounts of computationally intensive tasks efficiently. 
This concept has been extended to ``helper" edge server architectures as well, where devices with idle computation resources become (small) edge servers~\cite{cao2018joint, diao2019joint, chen2017exploiting, xing2019joint,he2019d2d,kai2019energy}.
%to exploit the computation resources of idle devices
%\cite{xing2019joint,cao2018joint}. 

The current trend in distributed computing, though, is a migration
to architectures that are more decentralized than MEC.
%even more decentralized architectures.
This is due to the fact that all edge nodes can take part in data offloading at different times because of the advances in 5G communication technologies in conjunction with improved computational capabilities of individual devices.
For this reason, device-to-device (D2D) network architectures (in Fig. \ref{fig:topo}(\subref{fig:topo:D2D})) that were previously studied in 4G LTE standards 
now hold the promise of
%have the potential of 
providing distributed computing at scale \cite{hassan2019edge}.

Unlike the MEC system in Fig. \ref{fig:topo}(\subref{fig:topo:MEC}), distributed computing in the D2D network of Fig. \ref{fig:topo}(\subref{fig:topo:D2D}) will
have more complicated topology management needs 
that must be considered together with the management of device resources.
%given the additional coordination requirements.
%
From a computation perspective, the edge nodes that receive offloaded tasks must have a suitable strategy for allocating its central processing unit (CPU) and/or storage resources to the tasks.
From a communication perspective, wireless transmissions between edge nodes will inevitably incur inter-channel interference due to concurrent data offloading. 
In contemporary wireless networks, devices are often equipped with multiple antennas to support multiple input multiple output (MIMO) communications, and the ability to communicate over different subchannels. Theoretically, having both multiple transmit and multiple receive antennas (i.e., a MIMO channel) provides an additional spatial dimension for communication and yields a degree-of-freedom gain \cite{Tse05}, which can be leveraged to mitigate such inter-channel interference. 
This motivates interference management techniques that consider the joint effects of 
subchannel allocations, transmission powers, MIMO beamforming, and other device transmit resources.

The focus of this paper is on addressing these challenges. Specifically, we develop methodologies
that jointly optimize computation and communication resources together with topology configuration in D2D wireless edge computing systems. These methodologies will aim to minimize the \textit{overhead} incurred from communication and computation, measured in terms of time delay/latency and energy consumption incurred from processing tasks in the D2D network.
% networks to adapt to 
% minimize overhead\rev{\footnote{\rev{Note that our use of the term “overhead” refers to the time delay/latency and energy consumption incurred from processing tasks in the network.}}} in edge computing systems.

% \rev{From a communication perspective, wireless transmissions between edge nodes 
% will inevitably incur inter-channel interference due to concurrent data offloading. 
% Therefore, interference management is required using techniques that optimize 
% subchannel allocation, transmission powers, beamforming using multiple antennas for multiple-input multiple-output (MIMO) systems, and other device transmit resources.
% Theoretically, having both multiple transmit and multiple receive antennas (i.e., a MIMO channel) provides an additional spatial dimension for communication and yields a degree-of-freedom gain \cite{Tse05},
% which can be leveraged to mitigate the unavoidable inter-channel interference in a D2D network.}
% The focus of this paper is on addressing these challenges. We develop methodologies
% that jointly optimize computation and communication resources together with topology configuration in D2D networks to adapt to 
% minimize overhead\rev{\footnote{\rev{Note that our use of the term “overhead” refers to the time delay/latency and energy consumption incurred from processing tasks in the network.}}} in edge computing systems.

\vspace{-2mm}
\subsection{Related Work and Differentiation}
\label{ssec:related}
We discuss related works 
on task offloading, resource management, and edge computing. We divide our analysis into two main categories: MEC and D2D.
%of edge computing focusing on task offloading and resource management, and address the differentiations of our work compared to previous ones.
\subsubsection{MEC systems}
Researchers have developed methods for resource management and offloading decision-making to maximize MEC system performance.
%In MEC systems, both of the offloading decision and resources management are important where the resources contain CPU, subchannels, transmission power, and multuple-input multiple-output (MIMO) antenna array gain.
Offloading decisions were thoroughly studied
%The offloading decision was solely conducted 
in \cite{chen2015efficient}, where management of device resources is assumed to be fixed.
% without consideration of resource management.
On the other hand, under the assumption that  offloading decisions are given, 
studies have considered optimal allocations of
CPU and subchannel resources \cite{le2017efficient}, and have also considered these
%resources management can be conducted e.g., the CPU and subchannel allocation are considered in \cite{le2017efficient}, and also 
together with beamformer design for MIMO systems~\cite{sardellitti2015joint, barbarossa2017enabling}.
%In a large network with limited subchannels, beamforming design is essential to mitigate inevitable inter-channel interferences for robust data transfer and optimization.
Recently, offloading decisions have been considered together with 
management of resources in MEC systems
such as CPU~\cite{zhang2017energy,nguyen2019computation,chen2018task,naderializadeh2019energy}, subchannels~\cite{zhang2017energy,wang2017joint}, transmit powers~\cite{zhang2017energy,nguyen2019computation,liu2019joint}, and beamformer design \cite{nguyen2019computation}.
% naderializadeh2019energy,chen2018task
%Although many of these works have considered some computation and communication resources, they have 
%not yet addressed all of the important variables in a unified optimization problem.

Though we focus on D2D in this paper, as mentioned previously, newer MEC architectures allow idle devices in close proximity to be dedicated computing nodes.
Therefore, optimization in MEC systems can be viewed as a special case of  D2D networks, where offloading is restricted to specific devices unidirectionally.
On the contrary, D2D networks allow multi-directional task offloading between devices. This requires additional optimization components to capture the multi-directional task offloading, CPU allocation across possibly multiple tasks at each device, and MIMO combiner design at receive devices, which are not considered in MEC systems.

% that we study in this paper (see Fig. \ref{fig:topo}).

\subsubsection{D2D networks}
Several prior works have focused on
optimizing communication quality in D2D systems,
where the objectives have been to maximize sum-rate \cite{wang2013joint,6858049,wei2013device,han2012subchannel, zhao2017joint, feng2013device}, spectral efficiency \cite{lin2015interplay}, or signal-to-noise ratio (SINR) \cite{tang2013cooperative}, 
%or to minimize total transmission power \cite{wen2013energy}
% han2012subchannel, zhao2017joint, feng2013device
with consideration of device and channel resources such as subchannels~\cite{wang2013joint,6858049,han2012subchannel, zhao2017joint, feng2013device}, transmit powers~\cite{6858049, zhao2017joint, feng2013device}, 
%wen2013energy
and beamformer design for MIMO systems~\cite{lin2015interplay,tang2013cooperative}.
For beamforming specifically, a coordinated beamforming strategy, such as the weighted minimum mean square error (WMMSE) approach, can be used to maximize 
standard communication measures (e.g., sum-rate)
%  general utility functions 
 \cite{Qingjiangshi}.
However, conventional beamforming strategies are not designed to consider time delay and energy consumption in a joint overhead metric for D2D.
% in an edge computing scenario. Furthermore, 
% beamforming alone is not sufficient to fully explore the potential of D2D edge computing that requires joint optimization on other system parameters, such as subchannels, CPU, and topology configuration.
Furthermore, in D2D edge computing, the total network overhead is also impacted by other system parameters, such as subchannels, CPU allocation, and topology configuration.

Works on D2D in edge computing have primarily focused on D2D-enabled (or D2D-assisted) MEC systems where several helper nodes are available as dedicated nodes for computing together with the  edge server.
In this respect, within a fixed topology, \cite{cao2018joint} investigated energy minimization based on CPU and transmission power allocation, and \cite{diao2019joint} studied joint time and energy minimization based on CPU, subchannel, and transmission power allocation.
%With given topology, the total energy is minimized with CPU and transmission power allocation in \cite{cao2018joint} and the total time and energy are both minimized with CPU, subchannel, and transmission power allocation in \cite{diao2019joint}. 
On the other hand, for a given set of system resources, the strategy of topology reconfiguration was discussed to minimize total energy in \cite{chen2017exploiting}.
Some recent works have addressed topology configuration together with the allocation of specific device resources such as CPU \cite{xing2019joint,he2019d2d,kai2019energy} and power \cite{xing2019joint,he2019d2d}.
Advanced offloading strategies for topology configuration have also been introduced in vehicular fog computing \cite{zhou2019reliable,zhou2019energy}.
In \cite{zhou2019reliable}, a task offloading method leveraging pricing-based matching was proposed to minimize the total network delay under utility constraints. 
In \cite{zhou2019energy}, an ADMM-based approach was proposed for energy minimization via partial offloading decisions subject to computing capability and latency constraints.
% Nevertheless, physical-layer communication resources such as beamformers, combiners, and subchannels
% are not considered in \cite{zhou2019reliable,zhou2019energy}.
Overall, we are not aware of any work that has addressed computation, communication, and topology configuration together in a unified optimization model for D2D edge computing, which is the focus of our paper.

\vspace{-2mm}
\subsection{Summary of Contributions}

Compared to the related works discussed in Section \ref{ssec:related}, the contributions of this paper are as follows:
\begin{itemize}[leftmargin=3.8mm]
\item We formulate a unified optimization model for D2D edge computing networks that minimizes  total network overhead, defined as the weighted sum of time and
energy consumption required to process a given task.
In doing so, we consider different dimensions of MIMO wireless for combating interference caused by concurrent data offloading, where each device is equipped with multiple antennas.
Our model includes
a framework for joint topology configuration, CPU allocation, subchannel allocation, and beamformer design for MIMO systems
(Sections \ref{sec:model} and \ref{sec:opt}).

\item We propose two methods for minimizing the total network overhead in our model,
which we refer to as semi-exhaustive search optimization and efficient alternate optimization.
We compare these two methods in terms of optimality guarantees and computational complexity in solving our non-convex problem. While the semi-exhaustive search optimization can be viewed as a ``best effort'' to obtaining the optimal solution, its complexity becomes problematic as the network size grows, which motivates the efficient alternate optimization
(Section \ref{sec:algorithm}).

\item In developing these
methods, we study the decomposition of the optimization into several subproblems: topology design, CPU allocation, and beamformer design. 
In doing so,
% we solve for beamformer design problem for fixed resource allocation as a sub-problem of overall optimization.
we develop a novel beamforming algorithm, minimum communication overhead beamforming (MCOB), that can be seen as a generalization of the WMMSE technique to utility functions incorporating both time and energy overhead.
% The beamforming problem is not classified as a conventional utility function, leading to the need for a new method to solve it.}
% The objective of beamforming is to minimize the network overhead, which is not classified as a conventional utility function, leading to the need for a new method.}
%we decompose the overall optimization into two sub-problems: resource allocation for fixed beamforming, and beamformer design for fixed resource allocation. 
% We derive minimum communication overhead beamforming (MCOB), 
% a coordinated beamforming algorithm which 
We prove that MCOB obtains the optimal beamformer for an MMSE receiver (Section \ref{sec:algorithm}).

%In developing these algorithms, we decompose the overall optimization into two sub-problems: resource allocation for fixed beamforming, and beamformer design for fixed resource allocation. In doing so, we derive minimum communication overhead beamforming (MCOB), a coordinated beamforming algorithm which we show obtains the optimal beamformer when the receiver  follows a well known characteristic (Section \ref{sec:algorithm}). 

%We develop CPU allocation algorithm, beamforming design algorithm, and greedy algorithm as inner algorithms
%in the framework for minimizing the total network overhead. 
%We propose minimum communication overhead beamforming (MCOB) to design a form of coordinated beamforming aimed at minimizing communication overhead measured in time and energy consumption.

\item 
We conduct several numerical experiments to evaluate the performance of our network overhead optimization methodology.
%Extensive simulations are presented to evaluate the integrated framework and our proposed algorithms. 
Our results show, for example, that our efficient alternate optimization algorithm can reduce the total overhead in D2D networks by 20\%-30\% compared to computation without offloading (Section \ref{sec:eval}).
\end{itemize}

\vspace{-1mm}
\section{Wireless Device-to-device \\ (D2D) Network Model}
\label{sec:model}

In this section, we develop our models for computational tasks, wireless signals, and the allocation of network resources in D2D systems.

%In what follows, 
%We will examine  task model, signal model, and task and resource allocation in D2D framework.

\vspace{-2mm}
\subsection{Task Model}
\label{ssec:task}

%Fig. \ref{fig:model:link} demonstrates wireless D2D network model of $K$ nodes, where the set of nodes is denoted by $\mathcal{K} = \{1, 2, ..., K \}$.

%\subsubsection{Task}
%We consider each node $k \in \mathcal{K}$ has task $k$, where $\mathcal{K} = \{1, 2, ..., K \}$ is the set of nodes in D2D network with total $K$ nodes.

We let $\mathcal{K} = \{1, 2, ..., K \}$ be the set of nodes in the D2D network, with a total of $K$ nodes.
Each node $k \in \mathcal{K}$ has a \textit{task} to be completed, consisting of computational work involved in data processing, where the objective of the data processing is to 
perform a transformation from input to output data.
In this paper, we adopt an indivisible task model~\cite{mao2017survey}, where a task is processed as a whole.\footnote{The computation task model widely adopted in edge computing includes two categories: a divisible task model and an indivisible task model. The former supports fine-grained task partitions 
% for complex tasks 
composed of multiple parallel segments, and the latter supports highly integrated or relatively simple tasks that cannot be partitioned for execution.
% such as highly integrated or relatively simple tasks. 
In this paper, we focus on the latter.}
A task is considered to be \textit{completed} when the
input data is successfully processed to the desired output.
In general, task completion requires computational resources including CPU, RAM, and storage.
In this paper, 
similar to previous works \cite{zhang2017energy,nguyen2019computation,chen2018task,xing2019joint,he2019d2d,kai2019energy},
we focus on CPU as the computation resource.
In case of mobile devices, 
many of today's tasks require computation-intensive processing with high CPU requirements, such as 3D-gaming and location-based augmented/virtual reality (AR/VR) \cite{yao2019edgeflow, yang2018communication, 8319323}.

To quantify the complexity of the task for node $k$ (which we will refer to as task $k$ for brevity),
we introduce the {\it{task size}} $I_k$ (in bits), which is the length of the bit stream of input data consisting of task $k$.
In other words, the bit stream of input data is represented as $\{0,1\}^{I_k}$. 
Then, the {\it{task} workload} is denoted as $\mu_k I_k$ (in cycles), where $\mu_k$ (in cycles/bit) is the processing density, meaning how many CPU cycles are required to process a bit of data.
That is, $\mu_k I_k$ represents total number of CPU cycles required to complete task $k$.
The processing density $\mu_k$ depends on the application;
for example, in the case of the audio signal detection in \cite{johnson2006modified}, since 500 cycles are required for processing 1 bit of data, $\mu_k $ is 500.

% {
%\begin{remark}
%    The computation task model in edge computing can be divided into two types: the task model for partial offloading and the one for binary offloading \cite{mao2017survey}. The former category supports fine-grained task partitions while the latter category supports the tasks that cannot be partitioned for execution. In practice, partial offloading is favorable for complex tasks composed of multiple parallel segments, while binary offloading is easier to implement and suitable for highly integrated or relatively simple tasks. In this paper, we adopt the latter category
%%    ; each node has a non-dividable task that should be processed as a whole via either local computing at the node or offloading to other nodes that perform computing instead.
%\end{remark}
%}

%%%%%%%%%%%%%%%%%%%%%%%%%%%%%%%%%%%%
%%%%%%%%%%%%%%%%%%%%%%%%%%%%%%%%%%%%
%%%%%%%%%%%%%%%%%%%%%%%%%%%%%%%%%%%%
 \vspace{-2mm}
\subsection{Signal Model}
\label{ssec:comm}

%%%%%%%%%
\begin{figure}[t]
    \includegraphics[width=.7\linewidth]{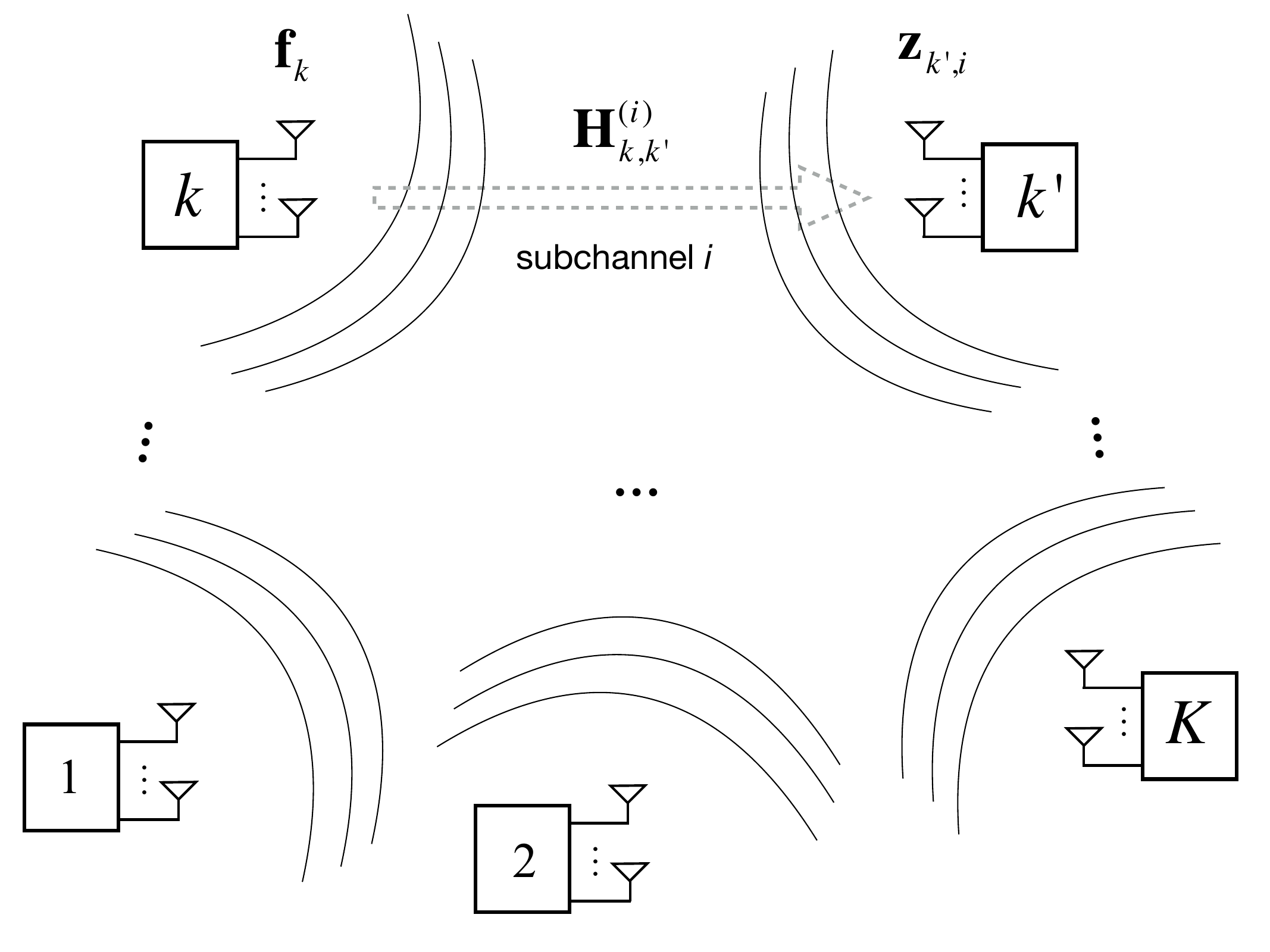}
    \centering
    \vspace{-2.5mm}
    \caption{Wireless device-to-device (D2D) network model 
    among $K$ nodes. Node $k$ transmits with a beamformer $\mathbf{f}_k$ to receive node $k'$ through subchannel $i$ characterized as $\mathbf{H}^{(i)}_{k,k'}$, which is decoded with a receive combiner $\mathbf{z}_{k',i}$. 
%    of a group of $K$ nodes. 
    }
\label{fig:model:link}
\vspace{-4mm}
\end{figure}
%%%%%%%%%%%%%

Fig. \ref{fig:model:link} demonstrates our wireless D2D network model among a set of $K$ nodes.
We assume that the nodes can transmit using multiple antennas on $S$ subchannels, where the set of subchannels is denoted $\mathcal{S} = \{1, 2, ..., S \}$.
Each node $k' \in \mathcal{K}$ receives a signal ${{\bf{y}}_{k',i}}$ through subchannel $i \in \mathcal{S}$ 
%in our model 
as
\vspace{-2mm}
\begin{align}
    {{\bf{y}}_{k',i}} = \sum\limits_{k =1}^K b_{k,i} {\bf{H}}^{(i)}_{k,k'} {{\bf{f}}_{k}}{s_{k}} + {\bf{n}}_{k',i} \in \mathbb{C}^{N_{k'}},
    \label{eq:signal}
    \vspace{-2mm}
\end{align}
% \vspace{-1mm}
where
${N_{k'}}$ is the number of antennas of node $k'$.
The scalar $s_{k} \in \mathbb C$ denotes 
the transmit signal sent by node $k$ with unit power ${\mathbb E}[|s_{k}|^2] = 1$, 
where $s_{k}$ can be understood as a single channel use of a Gaussian codeword vector that is encoded with ${I_k}$ bits per channel use.
%which is one of the encoded symbols (e.g., phase-shift keying) from a bit stream, $\{0,1 \}^{I_k}$.
%
\textcolor{black}{The vector ${\bf{f}}_{k} \in \mathbb{C}^{N_k}$ is the {\it{transmit beamformer}} of node $k$ with 
transmission power constraint $P_k$, i.e., $||{\bf{f}}_{k}||_2^2 \leq P_k$.}
Also, the matrix ${\bf{H}}^{(i)}_{k,k'} \in \mathbb{C}^{N_{k'} \times N_k }$ denotes a {\it multiple-input multiple-output (MIMO) channel} from transmit node $k$ to receive node $k'$ through subchannel $i$.
The noise vector ${\bf{n}}_{k',i} \in \mathbb{C}^{N_{k'}}$ is assumed to be complex additive Gaussian noise with zero mean and identity covariance matrix scaled by the noise power $\sigma^2$, i.e., ${\bf{n}}_{k',i} \sim {\mathcal C}{\mathcal N} ({\bf{0}},\sigma^2{\bf{I}})$.
%
%The ${P_{k}}$ is the transmit power of node $k$. 
The \textit{subchannel allocation variable} $b_{k,i} \in \{0, 1\}$ denotes whether transmit node $k$ uses subchannel $i$ for transmission.
We assume that the transmit node $k$ uses only one subchannel for transmission; if $b_{k,i}=1$, then $b_{k,j}=0$ $\forall j \ne i$.
%, i.e., if $b_{k,i}=1$, then $b_{k,j}=0$ for $\forall j \ne i$.

At receive node $k'$ on subchannel $i$, we consider a linear receive combiner ${\bf{z}}_{k',i} \in \mathbb{C}^{N_{k'}}$ so that the estimated value ${\hat y_{k',i}}$ is given by
\vspace{-.5mm}
\begin{equation}
    {\hat y_{k',i}} = {\bf{z}}_{k',i}^H {{\bf{y}}_{k',i}},
    \vspace{-.5mm}
\end{equation}
where the superscript $H$ denotes the conjugate transpose.

\vspace{-3mm}
\subsection{Task and Resource Allocation}
\label{ssec:TaskAssignment}

% {The computation task model in edge computing can be divided into two types: the task model for partial offloading and the one for binary offloading \cite{mao2017survey}. 
%The former category supports fine-grained task partitions while the latter category supports the tasks that cannot be partitioned for execution.
%Practically, the non-dividable task corresponds to highly integrated or relatively simple tasks, such as those in speech recognition and natural language translation \cite{mao2017survey}.
%In this paper, we adopt the latter category, i.e., 
%%For simplicity, we assume that
%each node has a non-dividable task that should be processed as a whole via either local computing at the node or offloading to other node.}

The assignment of tasks to either offloading or local processing determines the D2D network topology.
Constraints on how subchannels and processing resources are allocated must be specified based on these assignments.
%\textcolor{black}{This means the configuration of how $K$ nodes communicate with each other is established.}
%The corresponding resource allocation should be determined given the task assignment.

%\vspace{+1mm}
\subsubsection{Task assignment}
Each task $k$ can be either processed locally at node $k$ or offloaded to another node $k'$ for processing. 
We define $a_{k,k'} \in \{0, 1\}$ as the {\it{task assignment variable}} of whether 
task $k$ is assigned to
node $k'$ for $k,k' \in \mathcal{K}$.
If $a_{k,k} = 1$, then we have {\it{local processing}} of task $k$ at node $k$.
On the other hand, if $a_{k,k'} = 1$ for some $k' \ne k$, then we have {\it{offloaded processing}} where task $k$ is offloaded from $k$ to $k'$ and processed at node $k'$.
The set of task assignments is denoted by
\begin{equation}
    {\mathcal{A}} = \{ (k,k'): a_{k,k'} = 1 \;\;
    \forall k,k' \in {\mathcal{K}} \}.
\end{equation}
%
%
%Due to the assumption that 
Since each task should be processed as a whole, task $k$ should be assigned to only one node,
i.e.,
% which implies the constraint that
\vspace{-1mm}
\begin{equation}
    \sum\limits_{k'=1}^K a_{k,k'} = 1 \;\; \forall k.
    \vspace{-1mm}
\end{equation}
% Note that task assignment should be jointly determined with communication and computation resources. 
% For efficient offloading decisions, the channel matrices between nodes should be an important consideration, also with subchannels and available computing capabilities of nodes.
For efficient offloading decisions, it is important to consider the channel matrices between nodes, as well as the subchannels and available computing capabilities of nodes. {\footnote{  {In this paper,
we assume a standard channel state information (CSI) acquisition framework \cite{Tse05} in which the receiver 
can measure the channel matrix given in \eqref{eq:signal} through training signals sent by the transmitter.
Each entry of the channel matrix captures the effect of large/small scale fading.}}}
% The effects of the distance between nodes and the communication range are captured in the large scale fading as a form of the path loss exponent.
%For efficient offloading decision in D2D networks, the channel matrices between nodes should be an important consideration, also with available computing capabilities of nodes.
%
% {
%\begin{remark}
%    In this paper, we considered the input-output model of the wireless channel \cite{Tse05}, where the receiver measures the impulse response as the transmitter sends an impulse signal. Then, in MIMO system, the receiver can measure the channel matrix defined in \eqref{eq:signal}, where each component value of channel matrix captures the effect of large/small scale fading factor for each pair of transmitter and receiver elements. For efficient offloading decision in D2D networks, the channel matrices between nodes should be an important consideration, also with available computing capabilities of nodes. In this paper, task allocation is jointly determined with communication resources given the channel matrices and computation resource.
%\end{remark}
%}

\subsubsection{Subchannel allocation}
%Consider that all the $S$ subchannels have equal and non-overlapped bandwidth, $W$.
The task assignment specifies the configuration of how the $K$ nodes communicate with each other.
%We note that {\it{offloaded processing}} entails  transmission between nodes.
%In other words, $(k,k') \in {\mathcal{A}}_{\rm{offload}}$ denotes the transmission pair.
Therefore, the subchannel allocation variable $b_{k,i}$ is related to task assignment variable $a_{k,k'}$ as
\begin{align}
    \label{eq:suballoc}
    \sum\limits_{i = 1}^S {{b_{k,i}}}  = \left\{ \begin{array}{l}
    1\;\;{\rm{for}}\;\;k\;\;{\rm{with}}\;\;{a_{k,k}} = 0\\
    0\;\;{\rm{for}}\;\;k\;\;{\rm{with}}\;\;{a_{k,k}} = 1.
\end{array} \right.
\end{align}
$a_{k,k} = 0$ implies node $k$ is a {\it{transmit node}}, because task $k$ is not locally processed at node $k$, implying transmission to another node. In this case, transmit node $k$ uses one of the subchannels for transmission, i.e., $\sum\nolimits^S_{i=1} b_{k,i} =1$.
On the other hand, if node $k$ is not a transmit node, then $a_{k,k} = 1$ and there is no subchannel allocation for node $k$, i.e., $\sum\nolimits^S_{i=1} b_{k,i} =0$.

Each of the $S$ subchannels is assumed to have equal and non-overlapping bandwidth of width $W$. Consider, however, the case that node $k'$ receives multiple tasks from multiple transmit nodes. 
If same subchannel $i$ is used by these transmitters, the receive node must jointly decode the data of tasks, which leads to degraded decoding performance.
Therefore, in this paper, we follow prior work and  assume that the transmit nodes  that transmit to the same receive node use different subchannels \cite{wang2017joint}. In other words, for each receive node $k'$, 
we restrict the number of transmitters on subchannel $i$ according to
%there are no multiple subchannel allocation on subchannel $i$, i.e.,
\begin{equation}
    \sum\limits_{k \in \mathcal{A}_{k'}} b_{k,i} \le 1 \;\; \forall k', i,
\end{equation}
where $\mathcal{A}_{k'}$ denotes the set of transmit nodes that transmit to the receive node $k'$ given by
\begin{equation}
    \mathcal{A}_{k'} = \{ k: a_{k,k'} =1 \;\; \forall k \in {\mathcal{K}} \;\; {\rm{and}} \;\; k \ne k' \}.
    \label{eq:set:Ak'}
\end{equation}

%\vspace{+1mm}
\subsubsection{Computational resource allocation}
\label{sssec:compresource}

Consider that node $k'$ has multiple tasks to complete (its own and/or those offloaded to it).
Its \textit{computational resource} (CPU) ${F}_{k'}$ will be shared across these multiple tasks, where ${F}_{k'}$ (in cycles/sec or Hz) denotes the available CPU of node $k'$.
We define the amount of CPU resource of node $k'$ allocated to task $k$ as $F_{k,k'}$, which is subject to the constraints
\vspace{-.5mm}
\begin{gather}
    \sum\limits^K_{k=1} F_{k,k'} \le {F}_{k'} \;\; \forall k',
    \label{eq:delsum}
    \\
    F_{k,k'} = 0  \;\; 
%    \forall k,k' \;\; {\rm{with}} \;\;
    {\rm{if}} \;
    a_{k,k'} = 0,
%    \label{eq:del0}
%    \\
\quad
    F_{k,k'} \ge 0 \;\; \forall k,k'.
    \label{eq:del}
    % \vspace{-.5mm}
\end{gather}
In \eqref{eq:delsum}, 
the total CPU resource allocated cannot exceed the available CPU resource for each node $k'$.
In \eqref{eq:del}, $a_{k,k'} = 0$ implies that  task $k$ has not been assigned to node $k'$, 
so no CPU resources will be allocated to task $k$.
%In \eqref{eq:del}, 
 {In addition,} the allocated CPU $F_{k,k'}$ is restricted to a positive real value.

 \vspace{-2mm}
\section{D2D Network Optimization Model}
\label{sec:opt}

In this section, we formulate the optimization problem for minimizing D2D network task completion overhead. 
We define the total network overhead as a cost function to be minimized, consisting of both computation and communication overhead. 
%Also, we state a few standard assumptions for this paper.
%Then, we will formulate the overall optimization problem.

 \vspace{-2mm}
\subsection{Computation Overhead}
\label{ssec:compd}

We first define the computation overhead associated with node $k$ offloading to node $k'$. Based on the models from Section \ref{sec:model}, we can compute the \textit{computation time} $T_{\rm{comp}}(k,k')$ (in seconds) of task $k$ computed at node $k'$ according to
\vspace{-1mm}
\begin{equation}
    \label{eq:Tcomp}
    T_{\rm{comp}}(k,k') 
    = \frac{\mu_k I_k}{F_{k,k'}}.
    \vspace{-1mm}
\end{equation}
%We note that, when $k=k'$, $T_{\rm{comp}}(k,k)$ denotes local computation time of task $k$ computed at local node $k$. 
%
The \textit{computation energy consumption} $E_{\rm{comp}}(k,k')$ (in Joules) can be computed as
\vspace{-1mm}
\begin{equation}
    \label{eq:Ecomp}
    E_{\rm{comp}}(k,k') 
    = \kappa_{k'}  F_{k,k'}^2  {\mu_k I_k},
    \vspace{-1mm}
\end{equation}
where $\kappa_{k'}$ is the energy coefficient (in Joules $\cdot$ seconds$^2$/cycles$^3$) of node $k'$ that depends on the processor chip architecture \cite{wen2012energy}.
Here, $\kappa_{k'} F_{k,k'}^2$ denotes the energy consumption per cycle (in units of Joules/cycle).

We define the \textit{computation overhead} ${Y}_{\rm{comp}}(k,k')$ as the weighted sum of time and energy consumption, given by
%We  represent the computation demand  as
\vspace{-1mm}
\begin{align}
    {Y}_{\rm{comp}}(k,k') & = (1-\beta_k) T_{\rm{comp}}(k,k') + \beta_k E_{\rm{comp}}(k,k')
    \nonumber
    \\
    & = \big( (1-\beta_k)\frac{1}{F_{k,k'}} + \beta_k \kappa_{k'}  F_{k,k'}^2  \big)
%    \kappa {\delta^2_{k,k'}} F^2_{k'} ) 
    \mu_k I_k,
    \label{eq:Ccomp}
     \vspace{-1mm}
\end{align}
where $ \beta_k \in [0,1]$ is a demand overhead factor.
From \eqref{eq:Tcomp} and \eqref{eq:Ecomp}, note that the time consumption $T_{\rm{comp}}(k,k')$ and energy consumption $E_{\rm{comp}}(k,k')$ have tradeoff relationship with respect to computation resources: as more computation resources $F_{k,k'}$ are used, computation time $T_{\rm{comp}}(k,k')$ decreases while computation energy $E_{\rm{comp}}(k,k')$ increases. 
The overhead factor
$\beta_k$ trades off the importance of these two factors, and should be determined by the requirement of task $k$.
%For example, node $k$ with stringent requirement on task completion time can have a lower $\beta_k$ in order to place more importance on shortening the time at the expense of more energy consumption. 
Note that ${Y}_{\rm{comp}}(k,k)$ gives the local computation overhead in the case that task $k$ is locally processed at node $k$.

\vspace{-2mm}
\subsection{Communication Overhead}
\label{ssec:commd}

We now define the communication overhead associated with transmission of a task from node $k$ to $k'$. When $k \neq k'$,
%For two different nodes, $k$ and $k'$ with $k \ne k'$, 
we can write the \textit{signal to interference plus noise ratio} (SINR) from node $k$ to node $k'$ on subchannel $i$ as
%\begin{align}
%    \label{eq:SINR}
%    {\rm{SINR}}^{(i)}_{k,k'} = \left\{ \begin{array}{l}
%\frac{{{P_k}{{\left| {{\bf{z}}_{k',i}^H{\bf{H}}_{k,k'}^{(i)}{{\bf{f}}_k}} \right|}^2}}} {{\sum\limits_{l \ne k}^K {{b_{l,i}}{P_l}{{\left| {{\bf{z}}_{k',i}^H{\bf{H}}_{l,k'}^{(i)}{{\bf{f}}_l}} \right|}^2}}  + \left\| {{{\bf{z}}_{k',i}}} \right\|_2^2}} \;\; {\rm{for}} \;\; {b_{k,i}} = 1\\
%\hspace{2cm} 0 \hspace{2.1cm} \;\; {\rm{for}} \;\; {b_{k,i}} = 0,
%\end{array} \right.
%\end{align}
\vspace{-1mm}
\begin{align}
    \label{eq:SINR}
    {\rm{SINR}}^{(i)}_{k,k'} = 
\frac{ b_{k,i} {{{\left| {{\bf{z}}_{k',i}^H{\bf{H}}_{k,k'}^{(i)}{{\bf{f}}_k}} \right|}^2}}} {{\sum\limits_{\ell \ne k}^K {{b_{\ell,i}}{{\left| {{\bf{z}}_{k',i}^H{\bf{H}}_{\ell,k'}^{(i)}{{\bf{f}}_\ell}} \right|}^2}}  +  {\sigma^2} \left\| {{{\bf{z}}_{k',i}}} \right\|_2^2}} ,
\vspace{-2mm}
\end{align}
where 
all other transmit nodes $\ell \ne k$ using  subchannel $i$ are  interferences to the data stream of node $k$ 
%when it uses
on subchannel $i$.

%\begin{equation}
%    \label{eq:SINR}
%     {{\rm{SINR}}_{k,k'}} = \frac{{{P_{k,i_k}}{{\left| {{\bf{z}}_{k',i_k}^H{{\bf{H}}^{(i_k)}_{k,k'}}{{\bf{f}}_{k,i_k}}} \right|}^2}}}{{\sum\limits_{l \ne k}^{K} { {P_{l,i}}\left| {{\bf{z}}_{k',i}^H{{\bf{H}}^{(i)}_{l,k'}}{{\bf{f}}_{l,i}}} \right|^2}  + \left\| {{\bf{z}}_{k',i}} \right\|_2^2}}
%\end{equation}

%\cite{bjornson2013optimal}
%from Shannon's capacity formula \cite{cover2012}

Assuming perfect channel state information (CSI), we can write the \textit{maximum achievable data rate} $R^{(i)}_{k,k'}$ (in bits/second) from node $k$ to node $k'$ on subchannel $i$ as
\vspace{-2mm}
\begin{align}
    \label{eq:ratei}
    {R}^{(i)}_{k,k'} =
    W{\log _2} \big( 1 +
    {{\rm{SINR}}^{(i)}_{k,k'}}
    \big),
    \vspace{-2mm}
\end{align}
where 
$W$ is the bandwidth of each frequency subchannel.
Then, the total maximum achievable data rate from node $k$ to node $k'$ over all subchannels is
\vspace{-2mm}
\begin{gather}
    \label{eq:ratesum}
      {R}_{k,k'} = \sum\limits_{i=1}^S {R}^{(i)}_{k,k'}.
      \vspace{-1.5mm}
\end{gather}
When node $k$ is a transmitter, by \eqref{eq:suballoc}, 
only one subchannel is active. 
In other words,
%For transmit node $k$, only one subchannel is active in \eqref{eq:suballoc}. 
when $b_{k,i}=1$, $b_{k,j}=0$ for $j \ne i$, leading to ${R}^{(j)}_{k,k'}=0$.
Letting $i(k)$
%= \underset{i}{\arg } \{i: b_{k,i}=1\} 
be the active subchannel for node $k$, 
i.e., satisfying $b_{k,i(k)}=1$,
% which satisfies $b_{k,i}=1$, 
%i.e., $i_k = \underset{i}{\arg} \{b_{k,i}=1\}$,
the achievable rate is
%$i(k) = \underset{i}{\arg \max} b_{k,i}$
%When $b_{k,i}=1$, the achievable rate on subchannel $i$ is
%That is, for $i$ with $b_{k,i} =1$,
%$R_{k,k'}$ can be represented as
%
%\begin{equation}
%    {R_{k,k'}} = 
%W{\log _2} \left(1 + \frac{{{{\left| {{\bf{z}}_{k',i}^H{\bf{H}}_{k,k'}^{(i)}{{\bf{f}}_k}} \right|}^2}}}{{\sum\limits_{\ell \ne k}^K {{b_{\ell,i}}{{\left| {{\bf{z}}_{k',i}^H{\bf{H}}_{\ell,k'}^{(i)}{{\bf{f}}_\ell}} \right|}^2}}  + \left\| {{{\bf{z}}_{k',i}}} \right\|_2^2}} \right).
%\label{eq:rate}
%\end{equation}
\vspace{-1mm}
% \begin{align}
%     & {R_{k,k'}} = 
%     \label{eq:rate}
%     \\
%     &  W{\log _2} \negmedspace  \left( \negmedspace  1 + \frac{{{{\left| {{\bf{z}}_{k',i(k)}^H{\bf{H}}_{k,k'}^{(i(k))}{{\bf{f}}_k}} \right|}^2}}}{{\sum\limits_{\ell \ne k}^K {{b_{\ell,i(k)}}{{\left| {{\bf{z}}_{k',i(k)}^H{\bf{H}}_{\ell,k'}^{(i(k))}{{\bf{f}}_\ell}} \right|}^2}}  +   {\sigma^2} \left\| {{{\bf{z}}_{k',i(k)}}} \right\|_2^2}} \negmedspace  \right).
%     \vspace{-1mm}
%     \nonumber
% \end{align}
\begin{equation}
    % \hspace{-.2mm} %.05 
    \resizebox{.99\hsize}{!}{$ % 925
    {{R}_{k,k'}}  \negmedspace =  \negmedspace
    W{\log _2} \negmedspace  \left( \negmedspace  1 + \frac{{{{\left| {{\bf{z}}_{k'\negmedspace ,i(k)}^H{\bf{H}}_{k,k'}^{(i(k))}{{\bf{f}}_k}} \right|}^2}}}{{\sum\limits_{\ell \ne k}^K {{b_{\ell,i(k)}}{{\left| {{\bf{z}}_{k'\negmedspace ,i(k)}^H{\bf{H}}_{\ell,k'}^{(i(k))}{{\bf{f}}_\ell}} \right|}^2}}  +   {\sigma^2} \left\| {{{\bf{z}}_{k'\negmedspace ,i(k)}}} \right\|_2^2}} \negmedspace  \right)\negmedspace.$} 
    % \hspace{-6mm}
    % \vspace{-1mm}
    \label{eq:rate}
\end{equation}
% \begin{equation}
% \hspace{+1mm}
%  \resizebox{0.93\hsize}{!}{$
%     {\bf h}_{\rm eff}({\bf c}[t],t) = {\bf h}^{\rm UB}[t]
%     + {\bf H}^{\rm IB}[t] 
%     \sum_{\ell =1}^{L[t]} {\boldsymbol \Phi}( {\bf c}[t] , \theta_\ell[t] ) {\bf h}^{\rm UI}_\ell(\theta_\ell[t], t),$}
%     \label{eq:effchan} \hspace{-6mm}
%     % \vspace{-.5mm}
% \end{equation}

%
Given the data rate, we can compute the \textit{communication time} $T_{\rm{comm}}(k,k')$ (in seconds) 
%from offloading node $k$'s task to $k'$ as
for offloading task $k$ to node $k'$ as
%of task $k$ offloaded from node $k$ to $k'$ as 
\vspace{-1mm}
\begin{equation}
    \label{eq:Tcomm}
    T_{\rm{comm}}(k,k') = \frac{I_k}{R_{k,k'}}.
    \vspace{-1mm}
\end{equation}
\textcolor{black}{
The \textit{communication energy consumption} for node $k$ corresponding to the link from $k$ to $k'$ is 
\vspace{-1mm}
\begin{equation}
    \label{eq:Ecomm}    
    E_{\rm{comm}}(k,k') = (||{\bf{f}}_{k}||_2^2+ P_{\rm c} )\frac{I_k}{R_{k,k'}},
\vspace{-1mm}
\end{equation}
}where $P_{\rm c}$ is the circuit power including power dissipations in the transmit filter, mixer, and digital-to-analog converter, which are independent of the actual transmit power $||{\bf{f}}_{k}||_2^2$.

With these expressions for $T_{\rm{comm}}(k,k')$ and $E_{\rm{comm}}(k,k')$, the \textit{communication overhead} ${Y}_{\rm{comm}}(k,k')$ is defined with respect to the overhead factor $\beta_k$
as
%To quantify network demand with respect to communication, we define communication demand $U_{\rm{comm}}(k,k')$ containing both time and energy as
\begin{align}
    {Y}_{\rm{comm}}(k,k') &= (1-\beta_k) T_{\rm{comm}}(k,k') + \beta_k E_{\rm{comm}}(k,k')
    \nonumber
    \\
    & =  (1-\beta_k + \beta_k ||{\bf{f}}_{k}||_2^2 + \beta_k P_{\rm c}) \frac{I_k}{R_{k,k'}}.
    \label{eq:Ccomm}
\end{align}
%where the overhead factor $\beta_k$ determines the weight between time and energy for the communication overhead in the same way as the computation overhead.
We allocate the same $\beta_k$ to the computation and communication overhead definitions (in \eqref{eq:Ccomp} and \eqref{eq:Ccomm}) because  $\beta_k$ is the weighting factor between time delay and energy consumption for task $k$, which intuitively should be the same for both types of overhead.\footnote{If different overhead factors are preferred due to different resource restrictions
on communication and computation,
our methodology can be easily extended by defining $\beta_k^{\rm comp}$ in \eqref{eq:Ccomp} and $\beta_k^{\rm comm}$ in  \eqref{eq:Ccomm}.}
% Our methodology can be easily extended to 
% the case of different overhead factors by defining $\beta_k^{\rm comp}$ in \eqref{eq:Ccomp} and $\beta_k^{\rm comm}$ in  \eqref{eq:Ccomm}.}}}
There is a tradeoff between
$T_{\rm{comm}}(k,k')$ and $E_{\rm{comm}}(k,k')$ with respect to the transmit power $||{\bf{f}}_{k}||_2^2$: as more power $||{\bf{f}}_{k}||_2^2$ is applied, 
$T_{\rm{comm}}(k,k')$ decreases due to the increasing data rate $R_{k,k'}$ in \eqref{eq:rate}, while $E_{\rm{comm}}(k,k')$ increases because $||{\bf{f}}_{k}||_2^2/R_{k,k'}$ increases.

%time minimization both in computation and communication.
%, i.e., both in computation and communication.

%the task completion time in both computation and communication.
%saving more energy in both computation and communication.

\vspace{-1mm}
\subsection{Total Network Overhead}
Recall that there are two possibilities for task $k$: (i) local processing, i.e., $a_{k,k} = 1$, and (ii) offloaded processing, i.e., $a_{k,k'} = 1$ for some $k' \neq k$.
%Local processing means that the node $k$ processes its own task $k$, i.e., $a_{k,k} = 1$.
%However, offloaded processing means that the node $k$ transfers the task $k$ to another node $k'$ which then processes it, i.e., $a_{k,k'} = 1$, for some $k' \ne k$.
Local processing only incurs computation overhead ${Y}_{\rm{comp}}(k,k)$ while offloaded processing incurs both communication and computation overhead, ${Y}_{\rm{comm}}(k,k') + {Y}_{\rm{comp}}(k,k')$.
%%
%That is, the network demand for completion of task $k$ in each of these cases is
%\begin{equation}
%    {D}_{\rm{comp}}(k,k), \qquad {D}_{\rm{comm}}(k,k') + {D}_{\rm{comp}}(k,k').
%    \label{eq:dp}
%\end{equation}
%
%
With this, for a given D2D network topology configuration, we can write the {\it{total network overhead}} to complete all tasks in the network as
 \vspace{-2mm}
\begin{multline}
{Y}_{\rm{total}} = 
\sum_{k=1}^{K} \Bigg( 
a_{k,k} {Y}_{\rm{comp}}(k,k) +
\\
\sum_{k' \neq k}^K a_{k,k'}  \Big( {Y}_{\rm{comm}}(k,k') + {Y}_{\rm{comp}}(k,k') \Big)\Bigg).
\label{eq:obj:Utotal}
 \vspace{-2mm}
\end{multline}
%The objective of this paper is to minimize the total network demand ${D}_{\rm{total}}$.

 \vspace{-2mm}
\subsection{Optimization Formulation}
\label{ssec:opt}

We now formulate the problem of jointly optimizing 
the D2D network parameters to achieve the minimum total network overhead $Y_{\rm total}$.
The degrees of freedom available are the 
task assignments $\{a_{k,k'}\}$, computational resource allocations $\{F_{k,k'}\}$, subchannel allocations 
$\{b_{k,i}\}$, and 
beamformer design variables involving 
transmit beamformers $\{{\bf{f}}_{k}\}$ and receive combiners $\{{\bf{z}}_{k',i}\}$.
% to achieve minimum total network overhead $Y_{\rm{total}}$ in \eqref{eq:obj:Utotal}.
%Note that beamforming design refers to the design of the transmit beamformers and receive combiners.
%
% which will lead to minimum total network demand ${D}_{\rm{total}}$, given in \eqref{eq:obj:Utotal}. 
The optimization problem is given by:
\begin{align}
& \mathop {\text{minimize} }\limits_{
\{{\bf{f}}_{k}\}, \{{\bf{z}}_{k',i}\}, \{F_{k,k'}\}, \{a_{k,k'}\}, \{b_{k,i}\}}
Y_{\rm{total}} \; \mbox{in  \eqref{eq:obj:Utotal}}
\label{eq:obj:optCtotal} \\
& \hspace{-0.2cm} \text{subject to} \;\;
\sum\nolimits_{k' =1}^K a_{k,k'} = 1 \; \forall k, \;\; a_{k,k'} \in \{0,1\} \; \forall k,k',
\label{eq:con:a1}
%\sum\nolimits_{k' =1}^K a_{k,k'} = 1 \;\; \forall k, 
%\label{eq:con:a1} \\
%& & & a_{k,k'} \in \{0,1\} \;\; \forall k,k',
%\label{eq:con:a2} 
%\nonumber 
\\
&  
\sum\nolimits^S_{i=1} b_{k,i} = \left\{ {\begin{array}{*{20}{l}}
1 \quad \forall k \;\;
{\rm{with}} \;\; a_{k,k} = 0
\\
0 \quad \forall k \;\;
{\rm{with}} \;\; a_{k,k} = 1,
\end{array}} \right.
\label{eq:con:b1}
\\
& \sum\nolimits_{k \in \mathcal{A}_{k'}} b_{k,i} \le 1 \;\; \forall k',i,
%\label{eq:con:b2}
%\\
%& 
\quad
b_{k,i} \in \{0,1\} \;\; \forall k,i,
\label{eq:con:b3} 
%\nonumber 
\\
& R_{k,k'} \; \mbox{defined in (\ref{eq:rate})},
%\label{eq:con:R} 
%\\
%& 
\quad
||{\bf{f}}_{k}||_2^2 \leq P_k \;\; \forall k,
\label{eq:con:f} \\
&  
\sum\nolimits^K_{k =1} F_{k,k'} \le {F}_{k'} \;\; \forall k',
%\nonumber 
\label{eq:con:F1} 
\\
& F_{k,k'} = 0  \;\; 
%    \forall k,k' \;\; {\rm{with}} \;\;
    {\rm{if}} \;
    a_{k,k'} = 0,
%\label{eq:con:F2} 
%\\
\quad
F_{k,k'} \ge 0 \;\; \forall k,k'
\label{eq:con:F3} 
\end{align}

Constraints \eqref{eq:con:a1}-\eqref{eq:con:b3} and \eqref{eq:con:F1}-\eqref{eq:con:F3} account for task assignment, subchannel allocation, and CPU allocation requirements, which were described in Section \ref{ssec:TaskAssignment}. 
\eqref{eq:con:f} captures the constraint for the transmission power budget $P_k$ of an individual node. Note that there is no constraint on $\{{\bf{z}}_{k',i}\}$ such as a maximum magnitude restriction because the data rate $R_{k,k'}$ is not affected by the magnitude of ${\bf{z}}_{k',i}$.
%We can scale to unit-norm $|| {\bf{z}}_{k',i} ||_2 = 1$ after optimization.
%where the constraint $||{\bf{z}}_{k',i} ||_2 = 1$ can be removed because the data rate $R_{k,k'}$ is not affected by the magnitude of the receive combiner ${\bf{z}}_{k',i}$. Once ${\bf{z}}_{k',i}$ is determined, we just have to scale to ${\bf{z}}_{k',i}/||{\bf{z}}_{k',i}||_2$ to meet the constraint eventually.

%, where in \eqref{eq:con:P1} we impose an transmission power budget $P$ on individual node.

Assuming all nodes have $N$ antennas, meaning that $N_k=N$ for all $k$,
the optimization is a mixed integer program (MIP) with $K(N+NS+K)$ non-integer variables from  
$\{{\bf{f}}_{k}\}$, $\{{\bf{z}}_{k',i}\}$, $\{F_{k,k'}\}$,
and $K(K + S)$ integer variables from $\{a_{k,k'}\}$, $\{b_{k,i}\}$. 
\textcolor{black}{
The function $Y_{\rm comm}(k,k')$ is non-convex with respect to $\{{\bf{f}}_{k}\}$ and $\{{\bf{z}}_{k',i}\}$, which makes the problem a non-convex MIP.}
%Also, since $O_{\rm comm}(k,k')$ is non-convex with respect to $\{{\bf{f}}_{k}\}$ or $\{{\bf{z}}_{k',i}\}$, the problem is a non-convex MIP.
Existing solvers for non-convex MIPs do not scale well with the number of variables \cite{burer2012non}, and even in a relatively small D2D setting with $K=20$ nodes, $S=5$ subchannels, and $N=10$ antennas, our problem has already more than 2000 variables.
In Section \ref{sec:algorithm}, we turn to addressing the challenge of solving this optimization at scale.
%One of the challenges is developing an effective algorithm to solve this problem. 
%To this end, we propose two algorithms which are addressed in Section \ref{sec:algorithm}.

\vspace{-2mm}
\subsection{D2D Network Optimization Assumptions}
\label{ssec:assumption}
A few assumptions made on the D2D model in this section are noteworthy. 
First, although the network states will be dynamic over time, 
%we assume quasi-static scenario where $K$ nodes remains unchanged and channels are stable during offloading,
we assume a quasi-static scenario with $K$  active nodes and fixed channels during one codeword block, similar to previous works
\cite{le2017efficient,sardellitti2015joint, barbarossa2017enabling,zhang2017energy,nguyen2019computation,cao2018joint,diao2019joint,xing2019joint,he2019d2d}. 
The algorithms we develop for solving the optimization \eqref{eq:obj:optCtotal}-\eqref{eq:con:F3} in Section \ref{sec:algorithm} could then be applied to each quasi-static scenario as the number of nodes and channel conditions change, or at some suitable time interval. Related to this, we focus our optimization on the tasks generated at nodes in a single time period. A straightforward way to extend this approach to a dynamic task generation scenario is through queuing~\cite{ross2014introduction}, where new tasks are queued at nodes and processed by the optimization in a series of frames. We provide an experiment on this in Appendix D-C.

%  {\footnote{ {In this paper, we focus on a single-time optimization  given the tasks at nodes. 
% We could straightforwardly extend our approach to a dynamic task scenario. When a new task is generated at a node, it can be queued and processed by the optimization as a series of frames. We provide an additional experiment and the corresponding discussion in Appendix D-C.}}}

Second,  {we consider operator-assisted D2D networks \cite{tehrani2014device,pu2016d2d,mumtaz2014smart}, where a network operator  functions as a centralized controller for resource management.
The network operator, e.g., a base station (BS), executes our algorithms based on knowledge of the required network information, such as task size, CSI, and availability of subchannels and CPUs. %
This aligns with the current framework of cellular systems found in practice, where mobile devices typically have their radio access and resource allocation controlled through information sharing with the BS \cite{LTEstandard}.
Our methodology can thus be seen as enabling distributed task processing in the data plane via centralized optimization for variable design in the control plane.
The tasks generated by the nodes are processed in a distributed way, i.e., through offloading between devices or local processing, while the network operator will solve the optimization for orchestrating communication and computation resources.}
 {
\begin{remark} 
    The BS, as a network operator, can acquire the CSI of devices through  standard protocols of existing cellular network architectures.
    First, for CSI estimation at each device, concurrent channel estimation from multiple transmit devices can be conducted via well-established frameworks in the Long-Term Evolution (LTE)  \cite{LTEstandard} and 5G new radio (5G-NR)  \cite{lien20175g} standards, using UE-specific reference signals. 
    Since each device's wireless channel is temporally correlated in practice,
    CSI can be estimated with only a few parameters via  channel tracking techniques \cite{larew2019adaptive,booth2019multi,choi2014downlink}.
    % ozdemir2007channel
    Second, 
    for CSI acquisition at the BS, the BS can obtain measures such as channel quality information (CQI), precoding matrix indicator (PMI), and rank indicator (RI) from the devices, without receiving entire channel matrices, through  standard CSI reporting frameworks \cite{LTEstandard}.
    Since the number of CSI bits received at the BS  would be very small compared to the sizes of modern computing tasks,
    % For example, in the case of facial recognition,
    % the image (task) size is normally larger than 128 Kbytes = 1 Mbits = $10^6$ bits \cite{chen2015efficient,lyu2016multiuser}. This implies that
    the communication overhead for CSI acquisition can be safely ignored in the optimization.
    Although the CSI acquired at the BS will be  imperfect, 
    we will demonstrate in Section \ref{sec:eval} that 
    the effect of imperfect CSI (e.g., refer to \cite{love2008overview, love2004value, love2003grassmannian}) has only a small effect on our  proposed network 
    overhead optimization.
    % which will be demonstrated in Fig. \ref{fig:sim:CSI} of Section \ref{sec:eval}.
\end{remark}}

Third, we do not consider the process of transferring the computation result of an offloaded task  back to the source node,  {similar to previous works, e.g., \cite{chen2015efficient,lyu2016multiuser}.
%\cite{chen2015efficient,lyu2016multiuser,huang2012dynamic,xian2007adaptive,liu2019joint}. 
This assumption is reasonable since for many applications (e.g., facial recognition), the size of the output data (e.g., the recognized individuals) is in general much smaller than the size of the input data (e.g., the original full images).
% For many applications, e.g., face recognition, the size of the output data, in general, is much smaller than the size of computation input data.
%\cite{chen2015efficient,huang2012dynamic
%}
Therefore,}
we consider that the output data is negligible in size compared with the task so that it can be transferred through the network with minimal load.

\vspace{-2mm}
\section{Optimization Algorithms}
\label{sec:algorithm}

In this section, we develop two methods for solving the minimum overhead optimization problem \eqref{eq:obj:optCtotal}-\eqref{eq:con:F3}. The first method, semi-exhaustive search, provides a best-effort attempt to obtain the optimal solution, but has exponential complexity. The second method, efficient alternate optimization, reduces the complexity to polynomial time, for which we use semi-exhaustive search as an optimality benchmark.
As a component of these two methods, we will derive a new algorithm for MIMO beamforming (Section \ref{sssec:MIMO}) which minimizes communication overhead in D2D wireless edge networks.

%%%%%%%%%%%%%%%%%%%%%%%%%%%%%%%%%%%%%%%%%%%%%%%%%%%%%%%%%%%%%%%%
%%%%%%%%%%%%%%%%%%%%%%%%%%%%%%%%%%%%%%%%%%%%%%%%%%%%%%%%%%%%%%%%
\vspace{-2mm}
\subsection{Semi-Exhaustive Search Optimization}
\label{ssec:nro}

Given the task assignments $\{a_{k,k'}\}$ and subchannel allocations $\{b_{k,i}\}$ variables are binary, an intuitive approach to solving the optimization is to exhaustively search through all of their possibilities, so long as the search space is not prohibitively large. Then, for each possibility, we can solve for the non-integer variables $\{{\bf{f}}_{k}\}$, $\{{\bf{z}}_{k',i}\}$, and $\{F_{k,k'}\}$.
%In order to solve the optimization problem, an intuitive approach is to search exhaustively on
%all possible solution spaces.
%Motivated by exhaustive search,
%we propose the semi-exhaustive search optimization, which 
%has the exhaustive search method for task assignment and subchannel allocation. The overall algorithm is described in Algorithm \ref{al:Exh}.
%The key idea is that for all possible combinations of binary variables $\{a_{k,k'}\}$ and $\{b_{k,i}\}$, the non-integer variables $\{{\bf{f}}_{k}\}$, $\{{\bf{z}}_{k',i}\}$, and $\{F_{k,k'}\}$ are optimized.
We refer to this method as \textit{semi-exhaustive search}. The overall procedure is described in Algorithm 1: each choice of $\{a_{k,k'}\}$ and $\{b_{k,i}\}$ satisfying  constraints \eqref{eq:con:a1}-\eqref{eq:con:b3} is considered.
For given task assignments $\{a_{k,k'}\}$, we solve the CPU allocation problem for the processing resources $\{F_{k,k'}\}$, which is a convex problem.
In addition, for fixed task assignments $\{a_{k,k'}\}$ and subchannel allocations $\{b_{k,i}\}$, we solve the problem with respect to the beamformers $\{{\bf{f}}_{k}\}$ and combiners $\{{\bf{z}}_{k',i}\}$, which is a beamformer design problem.
We will develop solutions to these two problems in the following.
%

%\vspace{+1mm}
\subsubsection{CPU allocation}
With task assignments $\{a_{k,k'}\}$ determined, the optimization problem \eqref{eq:obj:optCtotal}-\eqref{eq:con:F3} with respect to CPU allocations $\{F_{k,k'}\}$ can be reduced to 
\vspace{-1mm}
\begin{align}
    & \mathop {\text{minimize} }\limits_{\{F_{k,k'}\}} & & 
\sum\limits_{k' = 1}^K {\sum\limits_{k = 1}^K {{a_{k,k'}}}   {Y_{{\rm{comp}}}}(k,k')} 
\label{eq:obj:CPU} \\
& \text{subject to}
& &  \mbox{Constraints} \; \eqref{eq:con:F1}-\eqref{eq:con:F3}
\label{eq:obj:CPU:con} 
%\\
%& \text{variables} 
%& &  \{F_{k,k'}\}. \nonumber
\vspace{-1mm}
\end{align}
%\mathop {\text{minimize} }\limits_{
The problem can be decomposed into $K$ independent subproblems: each node can allocate its own CPU regardless of the others.
For each node $k' \in \mathcal{K}$, the optimization problem is given as
\vspace{-2mm}
\begin{align}
    & \mathop {\text{minimize}}_{F_{k,k'} \; \forall k} & & 
\sum\limits_{k = 1}^K {{a_{k,k'}}} ((1 - {\beta _k})\frac{1}{{{F_{k,k'}}}} + {\beta _k}\kappa F_{k,k'}^2){\mu _k}I_k 
\label{eq:opt:CPU}
\\
& \text{subject to}
& &  \sum\limits^K_{k =1} F_{k,k'} \le {F}_{k'}, \;\; F_{k,k'} \ge 0 \;\; \forall k,
\\
& & & F_{k,k'} = 0  \;\; 
%    \forall k,k' \;\; {\rm{with}} \;\;
    {\rm{if}} \;
    a_{k,k'} = 0
\label{eq:opt:CPU:const}
%\\
%& \text{variables} 
%& &  F_{k,k'} \;\; \forall k. \nonumber
\vspace{-2mm}
\end{align}
Note that ${Y_{{\rm{comp}}}}(k,k')$ is convex with respect to $\{F_{k,k'}\}$ (since all parameters in ${Y_{{\rm{comp}}}}(k,k')$ are positive) and the constraints \eqref{eq:con:F1}-\eqref{eq:con:F3} are also convex. 
Therefore, optimization \eqref{eq:obj:CPU}-\eqref{eq:obj:CPU:con}  is convex.
The decomposed subproblem \eqref{eq:opt:CPU}-\eqref{eq:opt:CPU:const} for each $k'$ is also a convex problem that can be solved accordingly.
%where the problem is a standard form of geometric program (GP) \cite{boyd2007gp} and can be easily solved with geometric programming. 

%%%%%%%%% Algorithm
 \begin{algorithm}[t]
 \caption{Semi-exhaustive search optimization}
 \label{al:Exh}
 \begin{algorithmic}[1]
 \footnotesize
%  \small
  \STATE \textbf{Initialization.} 
  Set $\mathcal{G}^\star =  \emptyset$ and $Y_{{\rm{total}}}^\star = {\Upsilon}$ (e.g., ${\Upsilon} = 10^5$).
  \REPEAT
  %%%%%%%%%%%%%%%%%%%%%%%% 
    \STATE{ Generate new $\{ {a_{k,k'\!}}\}$ and $\{ {b_{k,i}}\}$ which satisfy  \eqref{eq:con:a1}-\eqref{eq:con:b3}. }    
    \STATE {
    {\it CPU allocation}: Solve for $\{ {F_{k,k'\!}}\} $ with $\{ {a_{k,k'\!}}\}$ from \eqref{eq:opt:CPU}-\eqref{eq:opt:CPU:const}
    }
    \STATE {\it \textcolor{black}{Beamformer design}}: Solve for $\{ {{\bf{f}}_k}\}$ and $\{ {{\bf{z}}_{k',i}}\} $ with $\{ {a_{k,k'}}\}$ and $\{ {b_{k,i}}\}$ from Algorithm \ref{al:MIMO}.
    \STATE Calculate ${Y_{{\rm{total}}}}$ in \eqref{eq:obj:Utotal} with the solution set ${\cal G} = \{ \{ {a_{k,k'}}\} ,{\rm{ \{ }}{b_{k,i}}\} ,{\rm{ \{ }}{{\bf{f}}_k}\}, {\rm{ \{ }}{{\bf{z}}_{k',i}}\} ,{\rm{ \{ }}{F_{k,k'}}\} \} $.
    \IF{${Y_{{\rm{total}}}} <Y_{{\rm{total}}}^\star $}
    \STATE Update $Y_{{\rm{total}}}^\star = {Y_{{\rm{total}}}}$ and $\mathcal{G}^\star \leftarrow \mathcal{G}$.
    \ENDIF
  \UNTIL { There is no possible case of $\{ {a_{k,k'}}\}$ and $\{ {b_{k,i}}\}$
  }
  \RETURN $ \{ {a_{k,k'}}\} ,{\rm{ \{ }}{b_{k,i}}\} ,{\rm{ \{ }}{{\bf{f}}_k}\} , {\rm{ \{ }}{{\bf{z}}_{k',i}}\} ,{\rm{ \{ }}{F_{k,k'}}\} $ in ${{\cal G}^\star}$
 \end{algorithmic}
 \end{algorithm}
%%%%%%%%%

%\vspace{+1mm}
\subsubsection{\textcolor{black}{Beamformer design}}
\label{sssec:MIMO}
With task assignments $\{a_{k,k'}\}$ and subchannel allocations $\{b_{k,i}\}$ determined, the optimization problem \eqref{eq:obj:optCtotal}-\eqref{eq:con:F3}  with respect to the beamformer design variables 
%$\{{\bf{f}}_{k}\}_{k \in \mathcal{K}}$ and $\{{\bf{z}}_{k',i}\}_{k' \in \mathcal{K}, i \in \mathcal{S}}$,
${\bf{f}}_{k}$ and ${\bf{z}}_{k',i}$ $\forall {k,k' \in \mathcal{K}}, i \in \mathcal{S}$,
can be reduced to
\vspace{-2mm}
\begin{align}
    & \mathop {\text{minimize}}_{
    {\bf{f}}_{k}, \; {\bf{z}}_{k',i} \;\; \forall {k,k' \in \mathcal{K}}, \; i \in \mathcal{S}
    } & & 
\sum\limits_{k = 1}^K {\sum\limits_{k' \ne k}^K {{a_{k,k'}}{Y_{{\rm{comm}}}}(k,k')} } 
\label{eq:bf:obj}
\\
& \hspace{8mm} \text{subject to}
& &  
\mbox{Constraint} \; \eqref{eq:con:f}
\label{eq:bf:con}
%\textcolor{black}{||{\bf{f}}_{k}||_2^2 \leq P_k \;\; \forall k \in \mathcal{K}} 
%\\
%& \text{variables} 
%& &  {\bf{f}}_{k}, \; {\bf{z}}_{k',i} \;\; \forall {k,k' \in \mathcal{K}}, \; i \in \mathcal{S}. \nonumber
\vspace{-3mm}
\end{align}

%This problem is a beamformer design problem with specific objective function.
We refer to this as the \textit{minimum communication overhead beamforming} (MCOB) problem.
Conventionally, objective functions in beamforming resource allocation problems take the form of
sum rate or sum harmonic rate utility functions \cite{hong2014signal}.
In our D2D setting, the objective instead becomes the weighted sum of time and energy consumption for transmission.
%Therefore, we cannot follow the same methods used in the conventional utility maximization problems, and we should come up with a different solution.
%Note that this problem is essential to enhance the benefit of data offloading in D2D network with efficient exploitation of communication resources.

We are interested in determining the variables ${\bf{f}}_{k}$ and ${\bf{z}}_{k',i}$ related to active data streams, i.e., for $k$, $k'$, and $i$ with $a_{k,k'}=1$ and $b_{k,i}=1$.
%As a first step, we define the tuple $(k,k',i)$ of the data stream, meaning the transmit node $k$ offloads to receive node $k'$ through subchannel $i$.
Denote the set of all transmit nodes as $\mathcal{K}_{\rm Tx} = \bigcup\nolimits_{k' \in \mathcal{K}} {\mathcal{A}_{k'}}  \subset \mathcal{K}$ from \eqref{eq:set:Ak'}.
Since each node $k \in \mathcal{K}_{Tx}$ offloads to one $k'$ on one subchannel $i$, we index this datastream as the tuple $(k, k', i)$.\footnote{Once $\{a_{k,k'}\}$ and $\{b_{k,i}\}$ are determined, the tuple $(k,k',i)$ is specified by $k$ and can be written as $(k,k'(k),i(k))$. For convenience, we are 
omitting the dependency of $k'$ and $i$ on $k$.}
%Using the tuple\footnote{Once $\{a_{k,k'}\}$ and $\{b_{k,i}\}$ are determined, the tuple $(k,k',i)$ is dependent on $k$ and can be written as $(k,k'_k,i_k)$. For convenience, we are abbreviating $k'_k$ as $k'$ and $i_k$ as $i$, with the understanding that they are dependent on $k$.} $(k,k',i)$, 
Our problem can be then rewritten as 
\vspace{-1mm}
\begin{align}
    \label{eq:opt:MIMO1}
    & \hspace{-2mm} \mathop {\text{minimize}}_{
    {\bf{f}}_{k}, {\bf{z}}_{k',i} \; \forall k \in \mathcal{K}_{\rm Tx}
    }
\sum\limits_{k \in {{\cal K}_{{\rm{Tx}}}}}^{} {(1 - {\beta _k} + {\beta _k}||{\bf{f}}_{k}||_2^2    + {\beta _k}{P_{\rm{c}}}) \frac{{I_k}}{{{R_{k,k'}}}}} 
\\
& \text{subject to} \quad
\label{eq:opt:MIMO1:const}
\textcolor{black}{||{\bf{f}}_{k}||_2^2  \leq P_k \;\; \forall k \in \mathcal{K}_{\rm Tx}} 
%\\
%& \text{variables} 
%& &  {\bf{f}}_{k}, \; {\bf{z}}_{k',i} \;\; \forall k \in \mathcal{K}_{\rm Tx}. \nonumber
\vspace{-1.5mm}
\end{align}
%where the constraint $||{\bf{z}}_{k',i} ||_2 = 1$ can be removed because the data rate $R_{k,k'}$ is not affected by the magnitude of the receive combiner ${\bf{z}}_{k',i}$. Once ${\bf{z}}_{k',i}$ is determined, we just have to scale to ${\bf{z}}_{k',i}/||{\bf{z}}_{k',i}||_2$ to meet the constraint eventually.

%The data rate $R_{k,k'}$  is given by
%\begin{align}
%    \label{eq:MIMOrate}
%    {R_{k,k'}} = 
%W{\log _2}(1 + \frac{{{{\left| {{\bf{z}}_{k',i}^H{\bf{H}}_{k,k'}^{(i)}{{\bf{f}}_k}} \right|}^2}}}{{\sum\limits_{l \ne k}^K {{b_{l,i}}{{\left| {{\bf{z}}_{k',i}^H{\bf{H}}_{l,k'}^{(i)}{{\bf{f}}_l}} \right|}^2}}  + \left\| {{{\bf{z}}_{k',i}}} \right\|_2^2}})
%\end{align}

 {This communication overhead minimization problem} is non-convex and hard to solve
due to the logarithm term in the data rate  ${R_{k,k'}}$ in \eqref{eq:rate}.
%Once the beamformers {f_k} are fixed
However, if the beamformers
$\{{{\bf{f}}_k}\}$ are fixed,\footnote{In this case, the notation $\{ {{\bf{f}}_k}\}$ is short for $\{ {{\bf{f}}_k}\}_{k \in \mathcal{K}_{\rm Tx}}$ denoting all variables ${{\bf{f}}_k}$ with $k \in \mathcal{K}_{\rm Tx}$. Throughout the paper, the context will make the distinction clear. The same simplification is applied for $\{{\bf{z}}_{{k'},{i}}\}$, $\{w_k\}$, $\{\lambda_k\}$, and $\{\gamma_k\}$.} minimizing \eqref{eq:opt:MIMO1} leads to the well known minimum mean square error (MMSE) receiver.
%, which will be given in \eqref{eq:optsol:z}.
If we restrict ourselves to using MMSE receiver,  we can transform the data rate into a quadratic form with the following lemma.
%, similar to \cite{Qingjiangshi,he2013coordinated}.

\begin{lemma}
With an MMSE-designed receiver, the data rate in \eqref{eq:rate} can be represented in quadratic form as
\vspace{-2mm}
\begin{equation}
    {R_{k,k'}} = \mathop {\max }\limits_{{{\bf{z}}_{k',i}},{w_k}} 
    {u_k}(\{ {{\bf{f}}_k}\} ,{{\bf{z}}_{k',i}},{w_k}),
    \label{eq:Rh}
    \vspace{-3.5mm}
\end{equation}
where
\vspace{-2.5mm}
\begin{gather}
    \hspace{-1.5mm}  {u_k}(\{ {{\bf{f}}_k}\} ,{{\bf{z}}_{k'\!,i}},{w_k}) \!= \!  - w_k^{ \!-\! 1} e_k^{{\rm{mse}}}(  \{ {{\bf{f}}_k}\} ,{{\bf{z}}_{k',i}}  ) \! - \negmedspace \log {w_k} \negmedspace + \negmedspace   {\sigma^2}\!,
    \label{eq:h}
    \vspace{-1mm}
\end{gather}
$w_k \in \mathbb{R^+}$ is an auxiliary variable, and the term $e_k^{{\rm{mse}}}$ is the MSE of receive node $k'$ given by
\vspace{-1mm}
\begin{gather}
    e_k^{{\rm{mse}}}(  \{ {{\bf{f}}_k}\} ,{{\bf{z}}_{k',i}}  ) = {(1 - {\bf{z}}_{{k'},{i}}^H{\bf{H}}_{k,{k'}}^{({i})}{{\bf{f}}_k})^H}(1 - {\bf{z}}_{{k'},{i}}^H{\bf{H}}_{k,{k'}}^{({i})}{{\bf{f}}_k}) 
    \nonumber
    \\
    + {\bf{z}}_{{k'},{i}}^H(\sum\limits_{\ell \ne k}^K {{b_{\ell,{i}}}{\bf{H}}_{\ell,{k'}}^{({i})}{{\bf{f}}_\ell}{\bf{f}}_\ell^H{\bf{H}}{{_{\ell,{k'}}^{({i})H}}} +  {\sigma^2} {\bf{I}}} ){\bf{z}}_{{k'},{i}}^{}.
    \vspace{-1mm}
\end{gather}
\end{lemma}
%
%Since the proof is immediate from \cite{Qingjiangshi}, we refer readers to past work.
\vspace{-1mm}
The proof follows from Theorem 1 in \cite{Qingjiangshi}. Since
%It is easily known that 
$u_k$ in \eqref{eq:h} is concave with respect to each of the variables
$\{ {{\bf{f}}_k}\}$, ${{\bf{z}}_{k',i}}$ and $w_k$, the optimal solution to \eqref{eq:Rh} is
%${\bf{z}}_{{k'},{i}}$ and $w_k$ in \eqref{eq:Rh} are
\begin{gather}
    {\bf{z}}_{{k'},{i}}^\star = 
    {\bf{J}}_k^{ - 1}{\bf{H}}_{k,{k'}}^{({i})}{{\bf{f}}_k},
    \label{eq:optsol:z}
    \\
    w_k^\star = e_k^{{\rm{mse}}}(\{ {{\bf{f}}_k}\} ,{\bf{z}}_{{k'},{i}}^\star),
    \label{eq:optsol:w}
    \vspace{-1mm}
\end{gather}
where ${{\bf{J}}_k} = \sum\limits_{\ell = 1}^K {{b_{\ell,{i}}}{\bf{H}}_{\ell,{k'}}^{({i})}{{\bf{f}}_\ell}{\bf{f}}_\ell^H{\bf{H}}_{\ell,{k'}}^{({i})} +  {\sigma^2} {\bf{I}}} $.
Note that ${\bf{z}}_{{k'},{i}}^\star$ is the MMSE receiver solution.

%With the obtained optimal solutions, the right hand side of \eqref{eq:Rh} is therefore $- \log (e_k^{{\rm{mse}}}(\{ {{\bf{f}}_k}\} ,{\bf{z}}_{{k'},{i}}^\star))$.

Using the formulation in Lemma 1,
the optimization problem \eqref{eq:opt:MIMO1}-\eqref{eq:opt:MIMO1:const} can be written as
\vspace{-1mm}
\begin{align}
    & \mathop {\text{minimize}}_{\{{\bf{f}}_{k}\}, \; \{{\bf{z}}_{k',i}\}, \; \{w_k\}} & & 
    \label{eq:opt:MIMO2}
    \sum\limits_{k \in {K_{{\rm{Tx}}}}}^{} {I_k\frac{{{g_k}({{\bf{f}}_k})}}{{{u_k}(\{ {{\bf{f}}_k}\} ,{{\bf{z}}_{k',i}},{w_k})}}} 
    \\
    \label{eq:opt:MIMO2:const}
    & \hspace{6mm} \text{subject to} & &  ||{\bf{f}}_{k}||_2^2   \le P_k \;\; \forall k \in \mathcal{K}_{\rm Tx}
%    \\
%    & \text{variables} & &  
%    \{{\bf{f}}_{k}\}, \; \{{\bf{z}}_{k',i}\}, \; \{w_k\}, \nonumber
\end{align}
% \vspace{-1mm}
where 
% \vspace{-1mm}
\begin{equation}
    \label{eq:f}
    {g_k}({{\bf{f}}_k}) = 1 - {\beta _k} + {\beta _k} ||{\bf{f}}_{k}||_2^2  + {\beta _k}{P_{\rm{c}}}.
\end{equation}

For a given $\{{{\bf{f}}_k}\}$, the optimal solutions of ${\bf{z}}_{k',i}$ and $w_k $ for \eqref{eq:opt:MIMO2}-\eqref{eq:opt:MIMO2:const} are given by \eqref{eq:optsol:z} and \eqref{eq:optsol:w}. Moreover, for given ${\bf{z}}_{k',i}$ and $w_k $, the function $g_k$ is convex and $u_k$ is concave with respect to $\{{{\bf{f}}_k}\}$. 
Optimization \eqref{eq:opt:MIMO2}-\eqref{eq:opt:MIMO2:const} with respect to $\{{{\bf{f}}_k}\}$ is thus a convex-concave multiple-ratio fractional programming problem \cite{shen2018fractional}, which is not convex.
Motivated by \cite{he2013coordinated}, we will exploit the fractional programming approach to solve it.

Specifically, we have the following theorem, which introduces an equivalent problem that is convex
%Theorem \ref{thm1} makes the optimization problem more tractable because the problem \eqref{eq:opt:MIMO4}-\eqref{eq:opt:MIMO4:const} is a convex problem 
with respect to each individual set of variables $\{{\bf{f}}_{k}\}$, $\{{\bf{z}}_{k',i}\}$, and $\{w_k\} $ when two other sets of variables $\{\lambda_k\}$ and $\{ \gamma_k \}$ are introduced.

\begin{theorem}
\label{thm1}
Consider the optimization problem 
\vspace{-2mm}
\begin{align}
    & \hspace{-3mm} \mathop{\text{minimize}}_{\{{\bf{f}}_{k}\},  \{{\bf{z}}_{k',i}\}, \{w_k\}} 
    \label{eq:opt:MIMO4}
    \sum\limits_{k \in {\mathcal{K}_{{\rm{Tx}}}}}^{} \! {{\lambda _k}({g_k}({{\bf{f}}_k}) - {\gamma _k}{u_k}(\{ {{\bf{f}}_k}\} ,{{\bf{z}}_{k',i}},{w_k}))} 
    \\
    \label{eq:opt:MIMO4:const}
    & \hspace{3mm} \text{subject to} \quad ||{\bf{f}}_{k}||_2^2  \le P_k \;\; \forall k \in \mathcal{K}_{\rm Tx}
%    \\
%    & \text{variables} & & 
%    \{{\bf{f}}_{k}\}, \; \{{\bf{z}}_{k',i}\}, \; \{w_k\}, \nonumber 
\end{align}
and the system equations
%\begin{align}
%    {\lambda _k} & = \frac{{I_k}}{{{u_k}(\{ {{\bf{f}}_k}\} ,{{\bf{z}}_{k',i}},{w_k})}},
%    \label{eq:lambda}
%    \\
%    {\gamma _k} & = \frac{{{g_k}({{\bf{f}}_k})}}{{{u_k}(\{ {{\bf{f}}_k}\} ,{{\bf{z}}_{k',i}},{w_k})}}.
%    \label{eq:gamma}
%\end{align}
\vspace{-1.5mm}
\begin{equation}
    {\lambda _k} = \frac{{I_k}}{{{u_k}(\{ {{\bf{f}}_k}\} ,{{\bf{z}}_{k',i}},{w_k})}},
    \quad
    {\gamma _k} = \frac{{{g_k}({{\bf{f}}_k})}}{{{u_k}(\{ {{\bf{f}}_k}\} ,{{\bf{z}}_{k',i}},{w_k})}}.
    \label{eq:gamma}
\end{equation}
If $\{ { \tilde{ \bf{f}}_k}\}$, $\{ {\tilde {\bf{ z}}_{k',i}}\}$, and $\{ {\tilde w_k}\} $ are solutions of the problem \eqref{eq:opt:MIMO4}-\eqref{eq:opt:MIMO4:const} and also simultaneously satisfy the system equations in \eqref{eq:gamma}, then 
%$\{ {\tilde {\bf{f}}_k}\}$, $\{ {\tilde{\bf{ z}}_{k',i}}\}$, and $\{ {\tilde w_k}\}$
they are  optimal solutions to  {\eqref{eq:opt:MIMO2}-\eqref{eq:opt:MIMO2:const}}.
\end{theorem}
The proof of Theorem \ref{thm1} is relegated to Appendix A.
Optimization \eqref{eq:opt:MIMO2}-\eqref{eq:opt:MIMO2:const} is equivalent to \eqref{eq:opt:MIMO4}-\eqref{eq:gamma} in the sense 
that they have the same globally optimal solutions.
%However, solving for all variables jointly to obtain global solutions makes the problem non-tractable.
Using the fact that optimization \eqref{eq:opt:MIMO4}-\eqref{eq:opt:MIMO4:const} is  convex with respect to each set of variables $\{{\bf{f}}_{k}\}$, $\{{\bf{z}}_{k',i}\}$, and $\{w_k\} $,
we will solve for each set, iteratively, which will yield solutions with $\{\lambda_k\}$ and $\{ \gamma_k \}$ being fixed.
Specifically, we propose an iterative algorithm to solve \eqref{eq:opt:MIMO4}-\eqref{eq:opt:MIMO4:const} and satisfy the system equations \eqref{eq:gamma} simultaneously:
%Note that $\{\lambda_k\}$ and $\{ \gamma_k \}$ are the functions of $\{{\bf{f}}_{k}\}$, $\{{\bf{z}}_{k',i}\}$, and $\{w_k\} $.
%That is, 
given $\{{\lambda _k}\}$ and $\{{\gamma _k}\}$, we solve for $\{ { { \bf{f}}_k}\}$, $\{ { {\bf{ z}}_{k',i}}\}$, and $\{ {w_k}\} $, and then 
% (which is convex problem with respect to each set of variables), and then 
update $\{{\lambda _k}\}$ and $\{{\gamma _k}\}$ from the updated variables $\{ { { \bf{f}}_k}\}$, $\{ { {\bf{ z}}_{k',i}}\}$, and $\{ {w_k}\} $.

%%%%%%%%% Algorithm
 \begin{algorithm}[t]
 \caption{\textcolor{black}{Minimum communication overhead beamforming (MCOB) algorithm}}
 \label{al:MIMO}
 \begin{algorithmic}[1]
%  \small
    \footnotesize
  \STATE \textbf{Initialization.} 
  Choose arbitrary $\{{\bf{f}}_k^{(0)}\}$ with  $||{{\bf{f}}_{k}^{(0)}}||_2^2 = P_k$ where ${\bf{f}}^{(0)}_k \in \mathbb{C}^{N_k}$.
  Update $\{{\bf{z}}_{{k'},{i}}^{(0)}\}$ and $\{w_k^{(0)}\}$ from \eqref{eq:optsol:z} and \eqref{eq:optsol:w}.
    Update the system equations $\{\lambda_k^{(0)}\}$ and $\{ \gamma_k^{(0)} \}$ from \eqref{eq:gamma} with $\{{\bf{f}}_k^{(0)}\}$, $\{{\bf{z}}_{{k'},{i}}^{(0)}\}$, and $\{w_k^{(0)}\}$.
    Set $\rho^{(0)} = 1$. Set the iteration number $j=1$.
  \REPEAT
  %%%%%%%%%%%%%%%%%%%%%%%% 
%    \STATE{ Update $\rho'$ = $\rho$. }    
    \STATE Solve for $\{{\bf{f}}_k^{(j)}\}$ from \eqref{eq:opt:MIMOeach}-\eqref{eq:opt:MIMOeach:const}.
    \STATE Update $\{{\bf{z}}_{{k'},{i}}^{(j)}\}$ and $\{w_k^{(j)}\}$ from \eqref{eq:optsol:z} and \eqref{eq:optsol:w}.
    \STATE Calculate the objective function $\rho^{(j)}$ in \eqref{eq:opt:MIMO4} with $\{{\bf{f}}_k^{(j)}\}$, $\{{\bf{z}}_{{k'},{i}}^{(j)}\}$, and $\{w_k^{(j)}\}$.
    \STATE {
    Update the system equations $\{\lambda_k^{(j)}\}$ and $\{ \gamma_k^{(j)} \}$ from \eqref{eq:gamma} with $\{{\bf{f}}_k^{(j)}\}$, $\{{\bf{z}}_{{k'},{i}}^{(j)}\}$, and $\{w_k^{(j)}\}$.
    }
    \STATE Calculate the system equation error $\zeta^{(j)}$ from \eqref{eq:syserr}.
    \STATE Set $j=j+1$.
  \UNTIL { $\left| {{\rho^{(j)}} - {\rho}^{(j-1)}} \right| \le \varepsilon $ and $\zeta^{(j)}  \le \varepsilon$
   (e.g., $\varepsilon=10^{-4}$)}
\STATE Obtain the solutions, $\{ {{\bf{f}}_k}\} = \{ {{\bf{f}}_k^{(j)}}\}$ and $\{ {{\bf{z}}_{k',i}}\} = \{ {{\bf{z}}_{k',i}^{(j)}}\}$
\RETURN $\{ {{\bf{f}}_k}\}$, $\{ {{\bf{z}}_{k',i}}\}$
 \end{algorithmic}
 \end{algorithm}
%%%%%%%%%

To solve \eqref{eq:opt:MIMO4}-\eqref{eq:opt:MIMO4:const} for fixed $\{\lambda_k\}$ and $\{ \gamma_k \}$,
we use the block coordinate descent (BCD) method~\cite{bisection}, where each set of the variables is solved fixing the other two.
In particular, with $\{{\bf{f}}_{k}\}$ and $\{w_k\} $ fixed, the optimal solution of each ${\bf{z}}_{k',i}$ is given in \eqref{eq:optsol:z}. With $\{{\bf{f}}_{k}\}$ and $\{{\bf{z}}_{k',i}\}$ fixed, the optimal solution of each $w_k$ is given in \eqref{eq:optsol:w}.
The remaining part is to solve for $\{{\bf{f}}_{k}\}$ with $\{{\bf{z}}_{k',i}\}$ and $\{w_k\} $ fixed,  {which is not-trivial.}
To solve for $\{{\bf{f}}_{k}\}$, we reorganize
the objective function in \eqref{eq:opt:MIMO4} 
% can be organized as follows
by replacing $u_k$ and $g_k$ with \eqref{eq:h} and \eqref{eq:f}:
\vspace{-1mm}
\begin{align}
   & \sum\limits_{k \in \mathcal{K}_{\rm Tx} } {{\lambda _k}{g_k}({{\bf{f}}_k})}  - \sum\limits_{k \in K_{\rm Tx}}^{} {{\lambda _k}{\gamma _k}{u_k}(\{ {{\bf{f}}_k}\} ,{{\bf{z}}_{k',i}},{w_k})}
   \nonumber
    \\
    & = \sum\limits_{k \in \mathcal{K}_{\rm Tx} } {\lambda _k} \bigg(1 - {\beta _k} + {\beta _k}P_{\rm c} - {\gamma _k}w_k^{ - 1} - {\gamma _k}w_k^{ - 1}{\bf{z}}_{k',i}^H{\bf{z}}_{k',i}
    \nonumber
    \\
    & \hspace{1.5cm} - {\gamma _k}\log {w_k} + {\gamma _k}\bigg) 
    + \sum\limits_{k \in \mathcal{K}_{\rm Tx} } {{\bf{f}}_k^H{{\boldsymbol \Sigma} _k}{\bf{f}}_k}
    \nonumber
    \\
    & + \sum\limits_{k \in \mathcal{K}_{\rm Tx} } {{\lambda _k}\bigg({\beta _k} ||{\bf{f}}_{k}||_2^2 - 2{\gamma _k}} w_k^{ - 1}{\mathop{\rm Re}\nolimits} [{\bf{z}}_{k',i}^H{\bf{H}}_{k,k'}^{({i})}{{\bf{f}}_k}] \bigg)
%    \nonumber
%    \\
%    & \hspace{0.5cm} + \sum\limits_{k \in \mathcal{K}_{\rm Tx} } {{\bf{f}}_k^H{{\boldsymbol \Sigma} _k}{\bf{f}}_k},
    \label{eq:opt:MIMO6}
\end{align}
where 
\vspace{-2mm}
\begin{equation}
    \label{eq:Sig}
    {{\boldsymbol \Sigma} _k}  \! = \!\! {\sum\limits_{\ell \in \mathcal{K}_{\rm Tx}}^{} \!\! {{\lambda _\ell}{\gamma _\ell}w_\ell^{ - 1}{b_{\ell,{i(\ell)}}}{\bf{H}}{{_{k,k{'(\ell)}}^{({i(\ell)})  H}}}{\bf{z}}_{k{'(\ell)},{i(\ell)}}^{}{\bf{z}}_{k{'(\ell)},{i(\ell)}}^H{\bf{H}}_{k,k{'(\ell)}}^{({i(\ell)})}} }.
    \vspace{-1mm}
\end{equation}
In \eqref{eq:Sig}, for the tuple $(\ell, k'(\ell),i(\ell))$,
$k'(\ell)$ denotes the receive node of the transmit node $\ell$ and $i(\ell)$ denotes the subchannel that $\ell$ uses. 
Since the first term in \eqref{eq:opt:MIMO6} is constant with respect to $\{{\bf{f}}_{k}\}$, we are only interested in the second and third terms. The optimization can be decoupled into $|\mathcal{K}_{\rm Tx}|$ independent subproblems, one for each ${\bf{f}}_{k}$, as
\vspace{-1mm}
\begin{align}
    & \mathop{\text{minimize}}_{{\bf{f}}_{k}} & & 
    {\lambda _k}{\beta _k} ||{\bf{f}}_{k}||_2^2  - 2{\lambda _k}{\gamma _k}w_k^{ - 1}{\mathop{\rm Re}\nolimits} [{\bf{z}}_{k',i}^H{\bf{H}}_{k,k'}^{({i})}{{\bf{f}}_k}] 
    \nonumber
    \\
    \label{eq:opt:MIMOeach}
    & & & + {\bf{f}}_k^H {\boldsymbol \Sigma}_k {{\bf{f}}_k}
    \\
    \label{eq:opt:MIMOeach:const}
    & \text{subject to} & & ||{\bf{f}}_{k}||_2^2  \le P_k
%    \\
%    & \text{variables} & &  {\bf{f}}_{k} \nonumber
\end{align}

A closed-form solution can be derived for \eqref{eq:opt:MIMOeach}-\eqref{eq:opt:MIMOeach:const} with the Karush–Kuhn–Tucker (KKT) conditions \cite{boyd2004convex}, since
the reduced problem \eqref{eq:opt:MIMOeach}-\eqref{eq:opt:MIMOeach:const} is a quadratically constrained quadratic program (QCQP). 
The detailed derivation of this procedure is given in Appendix C.
Through this process, we have transformed the original communication overhead minimization problem \eqref{eq:bf:obj}-\eqref{eq:bf:con} to a form in \eqref{eq:optsol:z}-\eqref{eq:optsol:w}, \eqref{eq:opt:MIMOeach}-\eqref{eq:opt:MIMOeach:const}
which allows each beamformer to be designed individually, under the BCD framework.
The solution ${\bf{f}}_k$ 
%is the beamformer for node $k$ which 
minimizes the communication overhead caused by concurrent task offloading, for a given topology configuration, subchannel allocation, and MMSE combiner at the receive nodes.

With $\{ {{\bf{f}}_k}\}$, $\{ {{\bf{  z}}_{k',i}}\}$, and $\{ {  w_k}\}$ in hand, we can then update $\{ {  \lambda_k}\}$ and $\{ {  \gamma_k}\}$ using \eqref{eq:gamma}.
The overall MCOB algorithm is demonstrated in Algorithm 
\ref{al:MIMO}, which determines $\{ {{\bf{f}}_k}\}$, $\{ {{\bf{  z}}_{k'\!,i}}\}$, $\{ {w_k}\}$, $\{ {\lambda_k}\}$, and $\{ { \gamma_k}\}$ that are the solutions to \eqref{eq:opt:MIMO4}-\eqref{eq:gamma}.
%where $\{ {{\bf{f}}_k}\}$, $\{ {{\bf{  z}}_{k',i}}\}$, and $\{ {  w_k}\}$ are the solutions for \eqref{eq:opt:MIMO4}-\eqref{eq:opt:MIMO4:const} and 
%$\{ {  \lambda_k}\}$ and $\{ {  \gamma_k}\}$ satisfy the system equations \eqref{eq:lambda}-\eqref{eq:gamma} simultaneously.
The algorithm runs until the objective function value $\rho$ in \eqref{eq:opt:MIMO4} 
%$\{ {{\bf{f}}_k}\}$, $\{ {{\bf{  z}}_{k',i}}\}$, and $\{ {  w_k}\}$
changes less than a threshold and the system equation error is also less than that. We define the system equation error as 
\vspace{-1mm}
\begin{equation}
    \hspace{-.mm} \zeta^{(j)}  = \sum\nolimits_{k \in \mathcal{K}_{\rm Tx}} {\big( {{{\big| {\lambda _k^{(j)}}-{\lambda _k^{(j-1)}} \big|}^2} + {{\big| {\gamma _k^{(j)}} - {\gamma _k^{(j-1)}} \big|}^2}} \big)}.
    \hspace{-.5mm}
    \label{eq:syserr}
    \vspace{-2mm}
\end{equation}
%where  low system equation error indicates the convergence of ${\lambda _k}$ and ${\gamma _k}$.

%%%%%%%%%%%%%%%%%%%%%%%%%%%%%%%%%%%%%%%%%%%%%%%%%%%%%%%%%%%%%%%%
%%%%%%%%%%%%%%%%%%%%%%%%%%%%%%%%%%%%%%%%%%%%%%%%%%%%%%%%%%%%%%%%
\vspace{-2mm}
\subsection{Efficient Alternate Optimization}
\label{ssec:greedy}

In this section, we propose a
%The semi-exhaustive search optimization (Algorithm \ref{al:Exh}) requires a lot of computations caused by a tremendous number of combinations of the binary variables $\{{a_{k,k'}}\}$ and $\{{b_{k,i}}\}$.
%Therefore, we propose the 
computationally efficient alternative to the semi-exhaustive search optimization (Algorithm \ref{al:Exh}) that avoids the brute force strategy of handling the binary variables $\{a_{k,k'}\}$ and $\{b_{k,i}\}$. This method, which we term \textit{efficient alternate optimization}, is demonstrated in Algorithm \ref{al:alt}.
The key idea is that we divide the optimization  \eqref{eq:obj:optCtotal}-\eqref{eq:con:F3}  into two subproblems and solve them alternately. The first problem is the \textcolor{black}{beamformer design} for the variables $\{ {{\bf{f}}_k}\}$ and $\{ {{\bf{z}}_{k',i}}\}$ given task assignments $\{ {a_{k,k'}}\}$ and subchannel allocations $\{ {b_{k,i}}\}$, which we already developed in Algorithm \ref{al:MIMO}.
The second problem is the resource allocation design for $\{ {a_{k,k'}}\}$, $\{ {b_{k,i}}\}$, and CPU allocation $\{{F_{k,k'}}\} $ with given beamformer design variables $\{ {{\bf{f}}_k}\}$ and $\{ {{\bf{z}}_{k',i}}\}$.
%
% In doing so, we are assuming the optimization \eqref{eq:obj:optCtotal}-\eqref{eq:con:F3} is separable along these dimensions, which is not guaranteed to be true, as we discuss in Section \ref{ssec:analysis}.
While each sub-algorithm is dedicated to each sub-problem, the overall composition via iterative alternating optimization 
is aiming to solve the overall problem  \eqref{eq:obj:optCtotal}-\eqref{eq:con:F3}.
The experiments in Section V demonstrate that this efficient alternate optimization achieves a substantial reduction in network overhead compared to local~processing.
% This will be demonstrated in our simulations in Section V.

%%%%%%%%% Algorithm
 \begin{algorithm}[t]
 \caption{Efficient alternate optimization}
 \label{al:alt}
 \begin{algorithmic}[1]
%  \small
 \footnotesize
  \STATE \textbf{Initialization.} 
  {Set $Y_{{\rm{total}}}^{\rm cur} = {\Upsilon}$ (e.g., ${\Upsilon} = 10^5$)}.
  {Generate arbitrary $\{ {a_{k,k'}}\}$ and $\{ {b_{k,i}}\}$, which satisfy \eqref{eq:con:a1}-\eqref{eq:con:b3}.}   
  \REPEAT
  %%%%%%%%%%%%%%%%%%%%%%%% 
    \STATE {Update $Y_{{\rm{total}}}^{\rm prev} = Y_{{\rm{total}}}^{\rm cur}$.}
    \STATE {\it \textcolor{black}{Beamformer design}}: Solve for $\{ {{\bf{f}}_k}\}$ and $\{ {{\bf{z}}_{k',i}}\} $ with $\{ {a_{k,k'}}\}$ and $\{ {b_{k,i}}\}$, using Algorithm \ref{al:MIMO}.
    \STATE
    {\it Greedy algorithm}: Solve for $\{ {a_{k,k'}}\}$, $\{ {b_{k,i}}\}$, and $\{ {F_{k,k'}}\} $ with $\{ {{\bf{f}}_k}\}$ and $\{ {{\bf{z}}_{k',i}}\} $, using Algorithm \ref{al:gre}.
    \STATE Calculate ${Y_{{\rm{total}}}^{\rm cur}}$ in \eqref{eq:obj:Utotal} with $ \{ {a_{k,k'}}\}$, ${\rm{ \{ }}{b_{k,i}}\}$, ${\rm{ \{ }}{{\bf{f}}_k}\}$, ${\rm{ \{ }}{{\bf{z}}_{k',i}}\}$, and ${\rm{ \{ }}{F_{k,k'}}\} $.    
  \UNTIL { 
 $|Y_{{\rm{total}}}^{\rm cur}-Y_{{\rm{total}}}^{\rm prev}| < \varepsilon$ (e.g., $\varepsilon = 10^{-4}$)   
  }
  \RETURN $ \{ {a_{k,k'}}\} ,{\rm{ \{ }}{b_{k,i}}\} ,{\rm{ \{ }}{{\bf{f}}_k}\}, {\rm{ \{ }}{{\bf{z}}_{k',i}}\} ,{\rm{ \{ }}{F_{k,k'}}\} $
 \end{algorithmic}
 \end{algorithm}
%%%%%%%%%

Algorithm \ref{al:gre} demonstrates our approach for the resource allocation problem. 
The key idea is that at each step, we determine the data stream tuple $(k,k',i)$ that 
provides the most reduction in overhead, and allocate these resources accordingly.
%\textcolor{red}{by offloading and allocating these resources accordingly.}
%and allocate these resources accordingly.
%We keep determining $(k,k',i)$ 
The process continues until there are no cases that any tuple will improve the optimization objective.
The maximizer for the current step is determined as
%Therefore, for every step of the greedy algorithm,
%we will choose $(k,k',i)$ that maximizes the offloading benefit ${\eta }_{k,k'}^{(i)}$.
%The maximizer $(\tilde k,\tilde k',\tilde i)$ is determined as
\vspace{-1mm}
\begin{equation}
    ( \tilde k,\tilde k',\tilde i )
%    = \mathop {\arg \max }\limits_{\scriptstyle k \in {{\cal K}_{{\rm{Tx}}}}, \; k' \in {{\cal K}_{{\rm{Rx}}}}, \; k \ne k', \; i \in {\cal I} } \eta_{k,k',i},
    = \mathop {\arg \max }\limits_{\scriptstyle k \in {K_{{\rm{Tx}}}},\;k' \in {K_{{\rm{Rx}}}},\;k \ne k',\;i \in I,\hfill\atop
    \scriptstyle k,k',i \; {\rm{satisfy}} \; \eqref{eq:con:a1}-\eqref{eq:con:b3}} \eta_{k,k',i},
    \label{eq:offbemax}
    \vspace{-1mm}
\end{equation}
where ${{\cal K}_{{\rm{Tx}}}}$ denotes the candidate set of transmit nodes, ${{\cal K}_{{\rm{Rx}}}}$ denotes the candidate set of receive nodes, and
$\eta_{k,k',i}$ is the \textit{offloading benefit} provided by tuple $(k,k',i)$. The offloading benefit is defined as
\vspace{-1mm}
\begin{equation}
    \eta_{k,k',i} = {Y^{\rm loc}} - {Y^{\rm off}},
    \label{eq:offbe}
    \vspace{-1mm}
\end{equation}
which quantifies the reduction in network overhead  by offloading from node $k$ to $k'$ on subchannel $i$ on top of the current resource allocations.
%with given ${\cal A}$ and ${\cal B}$.
${Y^{\rm loc}}$ denotes the total network overhead in case of no offloading from $k$ to $k'$, while  ${Y^{\rm off}}$ denotes the total network overhead in case of offloading.

Algorithm \ref{al:gre} begins with 
${{\cal K}_{{\rm{Tx}}}} = \mathcal{K}$, ${{\cal K}_{{\rm{Rx}}}} = \mathcal{K}$, meaning that all of the nodes are candidates for transmit and receive.
With ${\cal A}$ denoting the task assignment set ${\cal A} = \{ (k,k'): a_{k,k'}=1 \} $ and ${\cal B}$ denoting the subchannel allocation set ${\cal B} = \{ (k,i): b_{k,i}=1 \}$, initially $\mathcal{A} = \mathcal{B} = \emptyset$.
%With given ${\cal A}$ and ${\cal B}$, 
%the next data stream tuple $(k,k',i)$ is determined by calculating all of the offloading benefits ${\eta }_{k,k',i}$ for all $k,k',i$, and choosing the maximizer $(\tilde k,\tilde k',\tilde i) $.
%

%
%The remaining part is to calculate ${Y^{\rm loc}}$ and ${Y^{\rm off}}$ with given ${\cal A}$ and ${\cal B}$. 
%%
%We note that the CPU allocation variables $\{ {F_{k,k'}}\}$ can be optimized from \eqref{eq:opt:CPU}-\eqref{eq:opt:CPU:const} when the task assignment is determined.
%The ${Y^{\rm loc}}$ is given by
For a given $\mathcal{A}$ and $\mathcal{B}$, $Y^{\rm loc}$ is computed as
%\begin{align}
%    &
%    {Y^{{\rm{loc}}}} = 
%    \sum_{(k,k') \in \mathcal{A}^{loc}} Y^{\star}_{comp}(k,k')    
%%    \mathop {\min }\limits_{\{ {F_{k,k'}}\} } \left[ {\sum\limits_{(k,k') \in {{\cal A}^{{\rm{loc}}}}}^{} {{Y_{{\rm{comp}}}}(k,k')} } \right]
%    \nonumber \\
%    & \hspace{1.5cm} + \sum\limits_{(k,k') \in {{\cal A}^{{\rm{loc}}}},{\rm{ }}k \ne k',{\rm{ }}(k,i) \in {\cal B}}^{} {{Y_{{\rm{comm}}}}(k,k')},
%    \label{eq:Cloc}
%\end{align}
\vspace{-1mm}
\begin{align}
    {Y^{{\rm{loc}}}} = \!\!\!\!
    \sum_{(k,k') \in \mathcal{A}^{\rm loc}} 
    \!\!\!
    Y^{\star}_{\rm comp}(k,k')    
    + \!\!\!\! \sum\limits_{\scriptstyle (k,k') \in {{\cal A}^{{\rm{loc}}}},\hfil\atop 
    \scriptstyle k \ne k',{\rm{ }}(k,i) \in {\cal B}}  \!\!\!\! {{Y_{{\rm{comm}}}}(k,k')},
    \label{eq:Cloc}
    \vspace{-1mm}
\end{align}
where 
$Y^{\star}_{\rm comp}(k,k')$ is the value of $Y_{\rm comp}(k,k')$ obtained by the optimal solution to \eqref{eq:obj:CPU}-\eqref{eq:obj:CPU:con} for the allocation set $\mathcal{A}^{\rm loc}$,~and
\vspace{-4mm}
\begin{equation}
    {\cal A}^{\rm loc} = {\cal A} \cup \{ (k,k),(k',k')\}.
    \vspace{-3mm}
\end{equation}
${\cal A}^{\rm loc}$ denotes the new task assignment set when node $k$ and $k'$ process locally. 
%Therefore, the new task assignment $(k,k)$ and $(k',k')$ are added to the existing task assignment set $\mathcal{A}$.
In Algorithm \ref{al:gre}, $k'$ is added as a local processing node.
Otherwise, it might happen that at current step, task $k$ occupies all of the CPU of node $k'$ without consideration of allocating CPU to task $k'$. Then, $k'$ has no choice but to offload to other nodes at the next step.
%Since the greedy algorithm makes the locally optimal choice considering only the current stage and does not capture the previous state, it might happen that task $k$ occupies all of the CPU of  node $k'$ without consideration of allocating CPU to task $k'$. 
To overcome this, we consider the local processing of task $k'$ when task $k$ is being considered for offloading to node $k'$.
%, which is captured in calculating the offloading benefit.
%
%Note that the subchannel allocation set ${\mathcal{B}}$ remains the same because the offloading did not occur.

On the other hand, ${Y^{\rm off}}$ is given by
\vspace{-2mm}
\begin{equation}
%    &{Y^{{\rm{off}}}} = \mathop {\min }\limits_{\{ {F_{k,k'}}\} } \left[ {\sum\limits_{(k,k') \in {{\cal A}^{{\rm{off}}}}}^{} {{Y_{{\rm{comp}}}}(k,k')} } \right] \nonumber \\
    {Y^{{\rm{off}}}} = \!\!\!\!
    \sum_{(k,k') \in \mathcal{A}^{\rm off}} 
    \!\!\! Y^{\star}_{\rm comp}(k,k')   
    + \!\!\!\! \sum\limits_{ \scriptstyle (k,k') \in {{\cal A}^{{\rm{off}}}},\hfil\atop
    \scriptstyle k \ne k',{\rm{ }}(k,i) \in {{\cal B}^{{\rm{off}}}}}^{} \!\!\!\!{{Y_{{\rm{comm}}}}(k,k')},
    \label{eq:Coff}
    \vspace{-3mm}
\end{equation}
%In \eqref{eq:Cloc}, 
%In \eqref{eq:Coff}, 
where 
$Y^{\star}_{\rm comp}(k,k')$ is the optimal value for the allocation set $\mathcal{A}^{\rm off}$, and
\vspace{-2mm}
\begin{gather}
    {\cal A}^{\rm off} = {\cal A} \cup \{ (k,k'),(k',k')\}, \quad
    {\cal B}^{\rm off} = {\cal B} \cup \{ (k,i)\}.
    \vspace{-2mm}
\end{gather}
${\cal A}^{\rm off}$ denotes the new task assignment set when node $k$ offloads to $k'$.
%Therefore, the new task assignment $(k,k')$ and $(k',k')$ are added to the existing task assignment set $\mathcal{A}$.
${\cal B}^{\rm off}$ denotes the new subchannel allocation set when node $k$ uses subchannel $i$ for offloading. 
%Therefore, the new subchannel allocation $(k,i)$ is added to the existing subchannel allocation set $\mathcal{B}$.

%
%%%%%%%%% Algorithm
 \begin{algorithm}[t]
 \caption{Greedy algorithm for task assignment, subchannel allocation, and CPU allocation}
 \label{al:gre}
 \begin{algorithmic}[1]
%  \small
 \footnotesize
  \STATE \textbf{Initialization.} 
  Set ${{\cal K}_{{\rm{Tx}}}} = {\cal K}$, 
    ${{\cal K}_{{\rm{Rx}}}} = {\cal K} $, ${\cal A} = \emptyset$, and ${\cal B} = \emptyset$.
  \REPEAT
  %%%%%%%%%%%%%%%%%%%%%%%% 
    \STATE {
    $ ( \tilde k,\tilde k',\tilde i)
    = \mathop {\arg \max }\limits_{\scriptstyle k \in {K_{{\rm{Tx}}}},\;k' \in {K_{{\rm{Rx}}}},\;k \ne k',\;i \in I,\hfill\atop
    \scriptstyle k,k',i \; {\rm{satisfy}} \; \eqref{eq:con:a1}-\eqref{eq:con:b3}} \eta_{k,k',i}, $ where 
    \\
    $\eta_{k,k',i} = {Y^{\rm loc}} - {Y^{\rm off}}$. The ${Y^{\rm loc}}$ and ${Y^{\rm off}}$ are given in \eqref{eq:Cloc}\&\eqref{eq:Coff}.
%    ${D^{\rm loc}} = \mathop {\min }\limits_{\{ {F_{k,k'}}\} } {D_{{\rm{comp}}}}({\cal A}^{\rm loc},\{ {F_{k,k'}}\}) + {D_{{\rm{comm}}}}({\cal A}^{\rm loc},{\cal B}) $ 
%    \\
%    ${D^{\rm off}} = \mathop {\min }\limits_{\{ {F_{k,k'}}\} }      {D_{{\rm{comp}}}}({\cal A}^{\rm off},\{ {F_{k,k'}}\} )       + {D_{{\rm{comm}}}}({\cal A}^{\rm off},{\cal B}^{\rm off}) $
%    \\
%    where
%    ${\cal A}^{\rm loc} \leftarrow {\cal A} \cup \{ (k,k),(k',k')\} $, 
%    \\
%    ${\cal A}^{\rm off} \leftarrow {\cal A} \cup \{ (k,k'),(k',k')\} $, and ${\cal B}^{\rm off} = {\cal B} \cup \{ (k,i)\} $.
    }
    \IF{ $\eta_{\tilde k,\tilde k',\tilde i} \le 0$
    }
    \STATE Update ${\cal A} \leftarrow {\cal A} \cup \{ (k,k):k \in {{\cal K}_{{\rm{Tx}}}}\} $ and terminate the algorithm (set ${{\cal K}_{{\rm{Tx}}}} = \emptyset$).
    \ELSE
    \STATE
    Update
    ${{\cal K}_{{\rm{Tx}}}} \leftarrow {{\cal K}_{{\rm{Tx}}}} \setminus \{ \tilde k,\tilde k'\} $,\;
%    ${{\cal K}_{{\rm{Rx}}}} \leftarrow {\cal K}_{{\rm{Rx}}} - \{ \tilde k\} $,
    \\
    \hspace{0.9cm} ${\cal A} \leftarrow {\cal A} \cup \{ (\tilde k,\tilde k'),(\tilde k',\tilde k')\} $, and
    ${\cal B} \leftarrow {\cal B} \cup \{ (\tilde k,\tilde i)\} $.
    \ENDIF 
  \UNTIL { ${{\cal K}_{{\rm{Tx}}}} = \emptyset$
  }
\STATE Update $\{ {a_{k,k'}}\} $ with ${a_{k,k'}}  = 1$ for $(k,k') \in {\cal A}$ and ${a_{k,k'}}  = 0$ otherwise. Update $\{ {b_{k,i}}\} $ with ${b_{k,i}}  = 1$ for $(k,i) \in {\cal B}$ and ${b_{k,i}} = 0$ otherwise. Update $\{ {F_{k,k'}}\}$  
% as the solution to the optimization
from
\eqref{eq:obj:CPU}-\eqref{eq:obj:CPU:con}.
%$= \mathop {\arg \min }\limits_{\{ {F_{k,k'}}\} } {\sum\limits_{(k,k') \in {\cal A}}^{} {{{D}_{{\rm{comp}}}}(k,k')} } $.
\RETURN $\{ {a_{k,k'}}\} $, $\{ {b_{k,i}}\} $, $\{ {F_{k,k'}}\} $
 \end{algorithmic}
 \end{algorithm}
%%%%%%%%%

In each step of Algorithm \ref{al:gre}, as long as the best data stream $( \tilde k,\tilde k',\tilde i) $ from \eqref{eq:offbemax} has a positive offloading benefit $\eta_{\tilde k,\tilde k', \tilde i}$, then these resources are allocated.
%Based on the calculation of ${Y^{\rm loc}}$ and ${Y^{\rm off}}$, we determine the maximizer $( \tilde k,\tilde k',\tilde i) $ in \eqref{eq:offbemax}.
%If the maximum offloading benefit $\eta_{\tilde k,\tilde k', \tilde i}$ is less than 0, then we stop the algorithm because any data stream $(k,k',i)$ does not benefit by offloading.
%Otherwise, $( \tilde k, \tilde k',\tilde i) $ is determined, which 
This means 
task $\tilde k$ is offloaded to node $\tilde k'$ on subchannel $\tilde i$, and node $\tilde k'$ locally processes its own task $\tilde k'$.
As a result, we update ${\cal A} \! \leftarrow \! {\cal A} \cup \{ (\tilde k,\tilde k'),(\tilde k',\tilde k')\} $ and ${\cal B} \! \leftarrow \! {\cal B} \cup \{ (\tilde k,\tilde i)\} $.
Since nodes $\tilde k$ and $\tilde k'$ are no longer candidate transmit nodes, we update
% the task $\tilde k$ and $\tilde k'$ are already assigned, leading to 
${{\cal K}_{{\rm{Tx}}}} \! \leftarrow \! {{\cal K}_{{\rm{Tx}}}} \! \setminus \! \{ \tilde k,\tilde k'\} $.
Once there is no data stream with positive offloading benefit, the algorithm is terminated, and all remaining candidate transmit nodes are assigned to local processing.

\vspace{-2mm}
\subsection{Discussion of Optimality}
\label{ssec:analysis}

As mentioned previously, Algorithm \ref{al:Exh} (semi-exhaustive search) represents a best-effort approach for solving the optimization \eqref{eq:obj:optCtotal}-\eqref{eq:con:F3} with manageable complexity for small networks. We will explain this reasoning now. Then, in Section \ref{ssec:complexity}, we will compare the computational complexities between Algorithms \ref{al:Exh} and \ref{al:alt}.

%We first analyze our optimization problem.
%The optimization problem \eqref{eq:obj:optCtotal}-\eqref{eq:con:F3} is a non-convex mixed integer program (MIP). 
The optimal solution to \eqref{eq:obj:optCtotal}-\eqref{eq:con:F3} can be obtained (in theory)
by solving for the non-integer variables for all possible combinations of integer variables.
%As long as we can obtain the optimal solution for the non-integer variables with every given integer variables, we can get the optimal solution of overall problem.
%
%Since the objective function ${Y_{{\rm{total}}}}$ depends on $\{ {a_{k,k'}}\} $,  $\{ {b_{k,i}}\} $, ${{\{ {F_{k,k'}}\} }}$, ${{\{ {{\bf{f}}_k}\} }}$, and  ${\{ {{\bf{z}}_{k',i}}\} }$, it can be represented as the 
If we represent the objective function $Y_{\rm total}$ in its
functional form ${Y_{{\rm{total}}}} (\{a_{k,k'\!}\},\{b_{k,i}\},\{F_{k,k'\!}\},\{{\bf{f}}_k\}, \{ {\bf z}_{k'\!,i}\})$,
then by fixing the binary variables as $\{{\bar a_{k,k'\!}}\}$ and $\{{\bar b_{k,i}}\}$, we are left with 
% the problem
\begin{equation}
    \label{eq:opt:all}
    \mathop {\text{minimize} }\limits_{ \{F_{k,k'\!}\},\{{\bf{f}}_{k}\},\{ {\bf z}_{k'\!,i}\} }
    \!\!\!\!
    {Y_{{\rm{total}}}} (\{{\bar a}_{k,k'\!}\},\{{\bar b}_{k,i}\}, \{F_{k,k'\!}\},
    \{{\bf{f}}_{k}\}, \{ {\bf z}_{k'\!,i}\}),\!\!
\end{equation}
subject to the constraints.
Since the CPU allocation variable ${{\{ {F_{k,k'}}\} }}$ is not affected by the beamformer design variables $\{{\bf{f}}_{k}\}$ and $\{ {\bf z}_{k',i}\}$, and vice versa, this optimization can be divided into two independent problems given by
\begin{equation}
    \mathop {\text{minimize} }\limits_{\{{F_{k,k'}}\} } \; {Y_{{\rm{total}}}} (\{{\bar a}_{k,k'}\},\{{\bar b}_{k,i}\},\{F_{k,k'}\},
    \{{\bf{f}}_{k}\},\{ {\bf z}_{k',i}\}),
    \label{eq:opt:F}
\end{equation}
%and
\begin{equation}
    \hspace{-0.3mm} \mathop {\text{minimize} }\limits_{ \{{\bf{f}}_{k}\},\{ {\bf z}_{k',i}\} }  {Y_{{\rm{total}}}} (\{{\bar a}_{k,k'}\},\{{\bar b}_{k,i}\},\{F_{k,k'}\},
    \{{\bf{f}}_{k}\},\{ {\bf z}_{k',i}\}).
    \label{eq:opt:MIMO}
\end{equation}
%In summary, if we can obtain the optimal solution for \eqref{eq:opt:F} and \eqref{eq:opt:MIMO} for every given $\{{\bar a}_{k,k'}\}$ and $\{{\bar b}_{k,i}\}$, we can get a globally optimal solution of overall problem.
In summary, the optimization variables in \eqref{eq:opt:all} are separable, and the problem can be decomposed into \eqref{eq:opt:F} and \eqref{eq:opt:MIMO} for every combination of $\{{\bar a}_{k,k'}\}$ and $\{{\bar b}_{k,i}\}$.
%Once the optimal solution of \eqref{eq:opt:F} and \eqref{eq:opt:MIMO} is obtained for every given $\{{\bar a}_{k,k'}\}$ and $\{{\bar b}_{k,i}\}$, the global optimal solution will be obtained.
% leading to a globally optimal solution of overall problem.

Consider how the proposed semi-exhaustive search optimization addresses \eqref{eq:opt:F} and \eqref{eq:opt:MIMO}.
Problem \eqref{eq:opt:F} is convex: for this, we arrive at the convex problem \eqref{eq:opt:CPU}-\eqref{eq:opt:CPU:const} decomposed across nodes. Thus, we obtain the optimal solution ${\{ {F^\star_{k,k'}}\}}$ for this set of integer variables as
%For the problem \eqref{eq:opt:F}, it is a convex problem \eqref{eq:opt:CPU}-\eqref{eq:opt:CPU:const}, and, therefore the optimal solution ${\{ {F^\star_{k,k'}}\}}$ is obtained from
\begin{equation}
    \{ F^\star_{k,k'} \} = \mathop {\arg \min }\limits_{\{ {F_{k,k'}}\} } \; {Y_{{\rm{total}}}} (\{{\bar a}_{k,k'}\},\{{\bar b}_{k,i}\},\{F_{k,k'}\},
    \{{\bf{f}}_{k}\},\{ {\bf z}_{k'\!,i}\}).
\end{equation}
In contrast, problem \eqref{eq:opt:MIMO} is non-convex. To solve it, we developed
the MCOB algorithm  for optimizing the receive combiner $\{ {\bf z}_{k'\!,i}\}$ fixing the transmit beamformer $\{{\bf{f}}_{k}\}$ and vice versa (see Algorithm \ref{al:MIMO}).
The solution for $\{ {\bf z}_{k'\!,i}^\star\}$ for a fixed $\{ \bar {\bf{f}}_{k}\}$ based on an MMSE receiver is given in \eqref{eq:optsol:z}, such that
\vspace{-1mm}
\begin{equation}
    {\{ {\bf z}_{k',i}^\star\} } = \mathop {\arg \min }\limits_{\{ {\bf z}_{k',i}\} } \; {Y_{{\rm{total}}}} (\{{\bar a}_{k,k'}\},\{{\bar b}_{k,i}\},\{F_{k,k'}\},
    \{ \bar {\bf{f}}_{k}\},\{ {\bf z}_{k'\!,i}\}).
\end{equation}
The solution $\{{\bf{f}}_{k}^\star\}$ for a fixed $\{ \bar {\bf z}_{k'\!,i}\}$
is given in \eqref{eq:opt:MIMOeach}-\eqref{eq:opt:MIMOeach:const}, such that
\vspace{-1mm}
\begin{equation}
     {{\{ {\bf{f}}_{k}^\star\} }}=  \mathop {\arg \min }\limits_{ \{{\bf{f}}_{k}\},\{P_k\} }  {Y_{{\rm{total}}}} (\{{\bar a}_{k,k'}\},\{{\bar b}_{k,i}\},\{F_{k,k'}\},
     \{{\bf{f}}_{k}\},\{ \bar {\bf z}_{k'\!,i}\}).
    %  \vspace{-1mm}
\end{equation}

Although ${{\{ {\bf{f}}_{k}^\star\} }}$ and ${\{ {\bf z}_{k',i}^\star\}} $ are not guaranteed to be optimal solutions to the non-convex optimization in \eqref{eq:opt:MIMO}, 
they are practical solutions that have an efficient tradeoff between optimality and computational complexity.
%Since the problem \eqref{eq:opt:MIMO} is a non-convex problem, it is indeed computationally intensive to jointly optimize when the number of antennas is large.
Similar tradeoffs have been made in related works \cite{he2013coordinated,sun2014mimo,Qingjiangshi,zhang2005joint} for this reason.
However, $\{ {\bf z}_{k',i}^\star\}$ is an optimal solution for a given $\{ \bar {\bf{f}}_{k}\}$, and $\{ {\bf{f}}_{k}^\star\}$ is an optimal solution for a given $\{ \bar {\bf z}_{k',i}\}$, which is one of the main contributions of this paper.
%

%Therefore, the solution ${\{ {F^\star_{k,k'}}\}}$, ${{\{ {\bf{f}}_{k}^\star\} }}$, and ${\{ {\bf z}_{k',i}^\star\}} $ are the best-effort solutions to the problem \eqref{eq:opt:all} in the sense of the practicality.
%%
%In semi-exhaustive search optimization,
%all possible binary variables $\{{a_{k,k'}}\}$ and $\{{b_{k,i}}\}$ are considered, and the whole solutions $\{{a_{k,k'}}\}$, $\{{b_{k,i}}\}$, $\{F_{k,k'}\}$, $\{{\bf{f}}_{k}\}$, and $\{ {\bf z}_{k',i}\}$ that minimize ${Y_{{\rm{total}}}}$ are obtained.
%Therefore, the semi-exhaustive search optimization  will give the best-effort solution that we can realistically obtain.

\vspace{-1mm}
\subsection{Computational Complexity}
\label{ssec:complexity}

The semi-exhaustive search optimization still requires significant computation due to the large potential number of combinations of $\{{a_{k,k'}}\}$ and $\{{b_{k,i}}\}$.
The efficient alternate optimization is much more computationally efficient, and as we will see in Section \ref{ssec:optconv},
its observed solutions have comparable performance
to that of the semi-exhaustive search optimization.

Considering the computational complexities of both algorithms with respect to the integer variables, we have the following lemma:
\begin{lemma}
    \label{lem:complexity}
    With respect to the task assignment and subchannel allocation variables, the semi-exhaustive search optimization (Algorithm \ref{al:Exh})  has $\mathcal{O}((KS-S+1)^K)$ and the efficient alternate optimization (Algorithm \ref{al:alt}) has $\mathcal{O}({K^3}S)$, where $K$ and $S$ are the number of nodes and number of subchannels, respectively.
\end{lemma}
The proof is relegated to Appendix B.
The computational complexity of the semi-exhaustive search optimization is worse than exponential in the number of nodes, while that of the efficient alternate optimization is  polynomial.
%The efficient alternate optimization is much more computationally efficient than the semi-exhaustive search optimization. 
For example, if we consider $K=10$ and $S=2$, the semi-exhaustive search optimization already has up to $ 19^{10}$ combinations of binary variables
to consider (depending on condition \eqref{eq:con:b1}), and
the optimization for non-integer variables will be performed for each combination. 
In contrast, the efficient alternate optimization limits the number of combinations to at most  $2000$, depending on how many combinations provide a positive offloading benefit.
Further, the full optimization over non-integer variables is performed once the best combination is determined, i.e., it is not performed for every binary combination.

\vspace{-1mm}
\section{Performance Evaluation and Discussion}
\label{sec:eval}

\begin{figure*}[h!]
\vspace{-2mm}
\minipage{0.32\textwidth}
  \includegraphics[width=\linewidth]{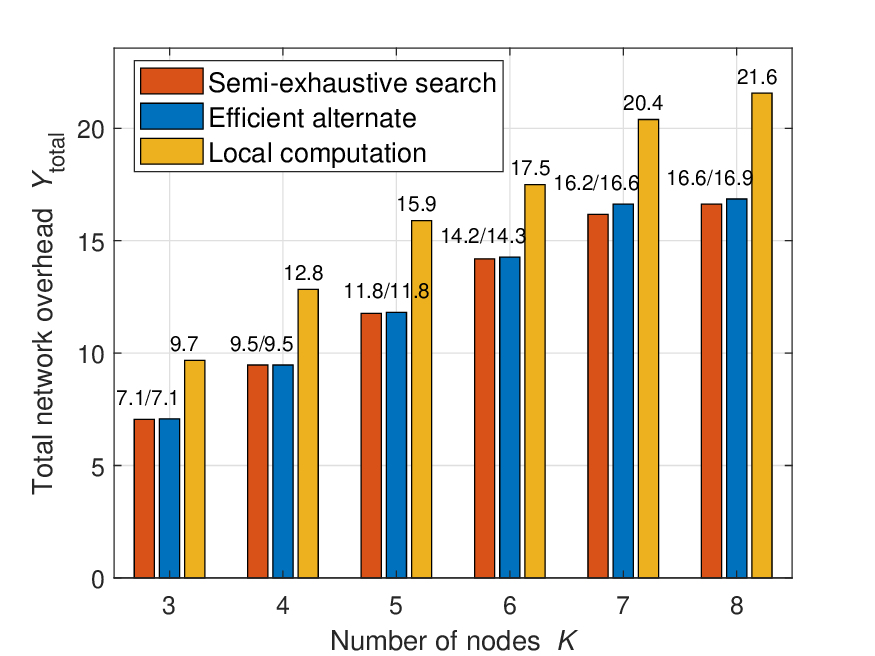}
  \vspace{-5.8mm}
  \caption{The total network overhead obtained by the semi-exhaustive search optimization, the efficient alternate optimization and local computation where $S=2$ and $N=5$.}
\label{fig:sim:optimality}
\endminipage\hfill
\minipage{0.32\textwidth}
  \includegraphics[width=\linewidth]{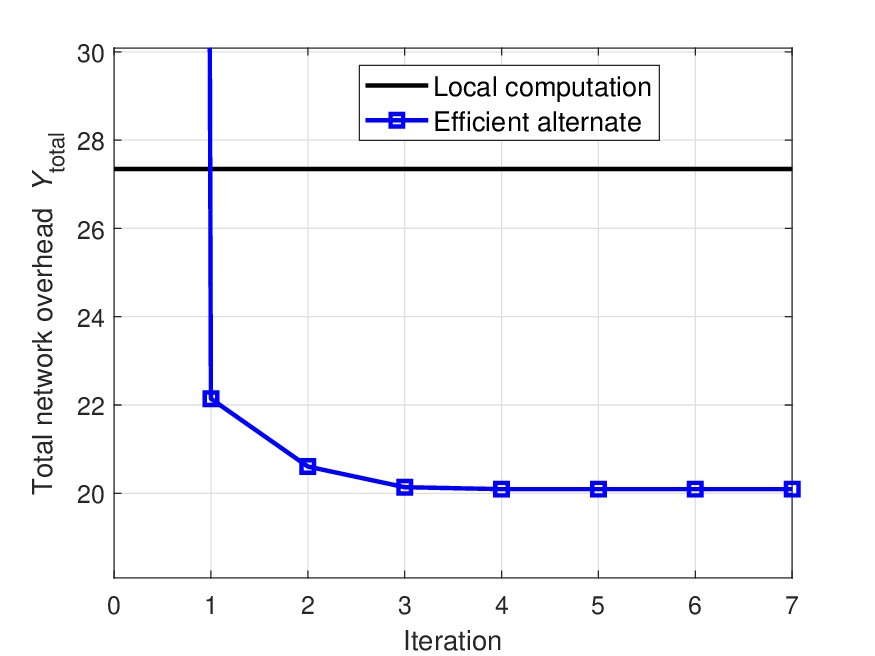}
  \vspace{-5.8mm}
  \caption{Convergence behavior of the efficient alternate optimization algorithm when $K=10$, $S=2$, and $N=5$. 
The total network overhead  converges within a few iterations. 
%reaching a  {23\%} improvement over local computation.
}
\label{fig:sim:convergence}
\endminipage\hfill
\minipage{0.32\textwidth}%
  \includegraphics[width=\linewidth]{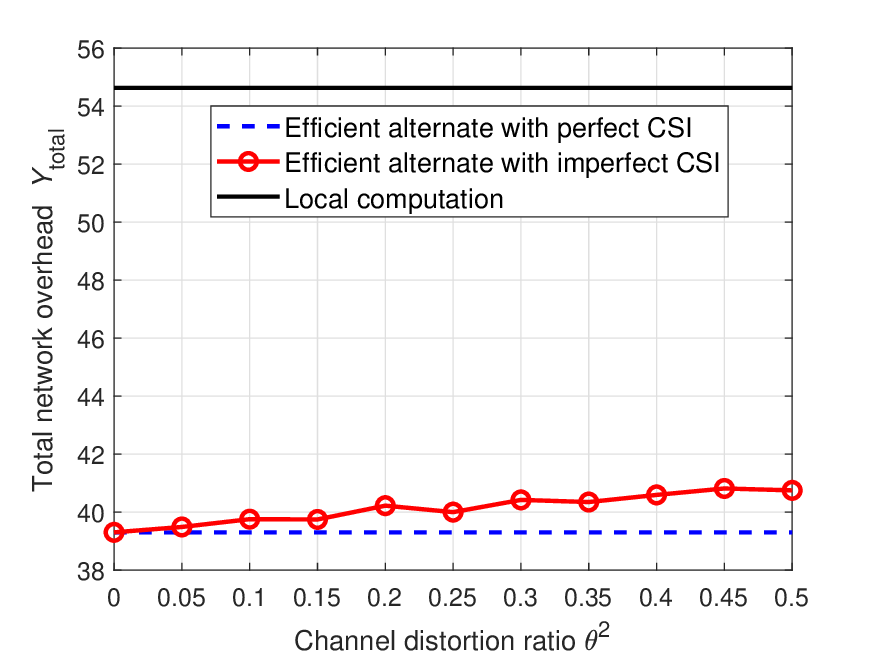}
  \vspace{-5.8mm}
  \caption{
Effect of imperfect CSI on performance of the efficient alternate optimization
% Total network overhead under perfect and imperfect CSI 
with $K=20$, $S=2$, and $N=5$. 
Significant improvements are still obtained as $\theta^2$ increases.}
% It still obtains a $22 \%$ improvement even with imperfect CSI.
% At $\theta^2 = 0.5$, still 
% Even with imperfect CSI, a $22 \%$ improvement is obtained.
%in network overhead 
%compared to local computing, while  $24 \%$ with perfect CSI.
\label{fig:sim:CSI}
\endminipage
\vspace{-5mm}
\end{figure*}

In this section, we conduct experiments to validate our methods for minimizing the total network overhead in D2D networks. 
After discussing our setup (Section \ref{ssec:sim:setup}), in
Section \ref{ssec:optconv}, we will 
quantify improvements relative to local processing and compare the efficient alternate optimization to the semi-exhaustive search optimization.
Then, in Sections \ref{ssec:off} to \ref{ssec:sca}, we will evaluate the performance of the efficient alternate optimization under variation of different network parameters.
% in different network settings.

\vspace{-2mm}
\subsection{Experimental Setup}
\label{ssec:sim:setup}

\subsubsection{Parameter values} 
For all of our experiments, 
we select values that are common for mobile computing environments \cite{liu2019joint,arnold2010power}. 
%The data size, CPU, processing density, transmission power limit, and energy coefficient of devices are chosen similar to .
%the data size, CPU, processing density, transmission power limit, and energy coefficient of devices are chosen reasonably considering the MEC environment such as in \cite{liu2019joint}.
Each channel ${\bf{H}}_{k,k'}^{(i)}$ is modeled as 
%a realization of a spatially uncorrelated 
a Rayleigh fading channel where the entries are i.i.d. following $\mathcal{CN}(0,\beta_{k,k'})$~\cite{Tse05}.
Here, $\beta_{k,k'} \negmedspace = \negmedspace \beta_0 -10\alpha \log_{10}(d_{k,k'}/d_0)$ (dB) denotes the large-scale fading factor between nodes $k$ and $k'$,
where $\beta_0 = -30$ dB is the path loss at the distance $d_0 = 1$ m, and $\alpha=3.5$ is the path loss exponent in  cellular networks~\cite{seidel1992914}.
The distance between nodes $k$ and $k'$, $d_{k,k'}$, is randomly generated as $d_{k,k'} \negmedspace \sim \negmedspace \mathcal{U}(10,30)$ (m),
where $\mathcal{U}(a,b)$ denotes  the uniform distribution on the interval $[a,b]$.
We assume that the individual transmit power limit is $P_k \negmedspace = \negmedspace P \negmedspace = \negmedspace 3$ dBW \cite{liu2019joint} for $k \in \mathcal{K}$,  the noise power is $\sigma^2 \negmedspace = \negmedspace -90$ dBW~\cite{LTEstandard}, the circuit power is $P_{\rm c} \negmedspace = \negmedspace -20$ dBW \cite{arnold2010power}, and the subchannel bandwidth is $W\negmedspace = \negmedspace 1$ MHz.
The beamformers $ {\bf {f}}_k $ and combiners $ {\bf {z}}_{k',i} $ are initially generated to be uniformly distributed on the complex sphere~\cite{au2007performance}  with radius $ \sqrt{P}$ and 1, respectively, for $k,k' \in \mathcal{K}$ and $i \in \mathcal{S}$.

% As a computation task, we consider facial recognition, which is a popular application found on mobile devices \cite{chen2015efficient,lyu2016multiuser}.
As a computation task, we consider image recognition \cite{chen2015efficient,lyu2016multiuser}, which is a popular application on mobile devices (e.g., for user authentication via facial recognition).
To emulate heterogeneous devices, we consider different task sizes and CPUs across the nodes.
For the task size,
we sample $I_k \sim \mathcal{U}(1 , 8)$ in Mbits, i.e.,  $\mathcal{U}(0.128, 1)$ in Mbytes, for each node $k$.
%, where $\mathcal{U}(a,b)$ denotes  the uniform distribution on the interval $[a,b]$.
These represent common image sizes found on mobile devices~\cite{chen2015efficient,lyu2016multiuser}.
For CPU, we consider a bimodal distribution for each node $k'$:
$F_{k'} \sim \frac{3}{4} \mathcal{U}(0.1, 0.2) + \frac{1}{4} \mathcal{U}(0.9, 1)$,
%With the probability 3/4, the distribution follows ${F}_{k'} \sim \mathcal{U}(0.1 , 0.3)$ and with the probability 1/4, ${F}_{k'} \sim \mathcal{U}(1 , 3)$
with units of GHz. 
This selection generates a composition of resource-hungry and resource-rich devices, reflecting common processor clock speeds found in mobile devices~\cite{chen2015efficient,lyu2016multiuser,liu2019joint}.
We assume constant processing density $\mu_k = 200$ cycles/bit, and energy coefficients
$\kappa_{k'} = 3.5 \times 10^{-27}$ across all nodes, as in \cite{liu2019joint}.
%The parameters $I_k$, ${F}_{k'}$, $\mu_k$, and $\kappa_{k'}$ are chosen similar to \cite{liu2019joint}.
The overhead factor $\beta_k$ is assumed to be the same for all nodes, i.e., $\beta_k=\beta$ for all $k$. Unless otherwise stated, $\beta=0.5$.
All nodes are considered to have $N$ transmit and receive antennas, i.e., $N_{k}=N $ for all $k$.
 {
Each experiment is averaged over 20 different samplings of task sizes, CPUs, and channel realizations.
For the efficient alternate optimization, we consider 10 different initializations of $\{a_{k,k'\!}\}$ and $\{b_{k,i}\}$, and choose the best solution.
The threshold for Algorithms \ref{al:MIMO}\&\ref{al:alt} is $\varepsilon = 10^{-4}$.}
%Each experiment is averaged over 100 random channel realizations.
%For the efficient alternate optimization, we consider 20 different initializations of $\{a_{k,k'}\}$ and $\{b_{k,i}\}$, and choose the best solution.
%The threshold for Algorithm \ref{al:MIMO} and \ref{al:alt} is $\varepsilon = 10^{-4}$.

\subsubsection{Baselines} 
\label{sec:eval:set:base}
We compare the proposed algorithms with three different baselines.
The first baseline is local computation, where all the nodes locally process their own tasks  without offloading. The total network overhead for local processing is
\vspace{-1.5mm}
\begin{equation}
    Y_{\rm total} = \sum\limits_{k=1}^K Y_{\rm comp} (k,k).
    \vspace{-1.5mm}
\end{equation}
This baseline will be used to assess the improvements obtained via our offloading optimization methodology.

The second baseline is the efficient alternate optimization with the weighted minimum mean square error (WMMSE) approach \cite{Qingjiangshi} used in place of Algorithm 2.
%This baseline consists of WMMSE and the greedy algorithm (Algorithm \ref{al:gre}).
%The second baseline is the algorithm that consists of the WMMSE method and greedy algorithm.
WMMSE is an existing method for beamformer design 
with a sum-utility maximization objective, proposed in \cite{Qingjiangshi}.
%to maximize the sum utility problem, which is proposed in \cite{Qingjiangshi}.
Specifically, in place of \eqref{eq:bf:obj}-\eqref{eq:bf:con}, with WMMSE, we minimize the total communication time as
%With WMMSE, total communication time can be minimized as one of the conventional sum-utility maximization problem,
\vspace{-1.5mm}
\begin{equation}
    \mathop {\text{minimize}}_{\{ {{\bf{f}}_k}\}, \{ {{\bf{z}}_{k',i}}\} } 
    \quad \sum\limits_{k=1}^K \sum\limits_{k' \ne k}^K a_{k,k'} T_{\rm comm}(k,k').
    \vspace{-1.5mm}
\end{equation}
This baseline will allow us to assess the importance of balancing time and energy as competing objectives in overhead minimization.

The third baseline is the efficient alternate optimization but with equal CPU allocation. 
For a given task assignment, the CPU is equally allocated across the requested tasks.
Specifically, in Algorithm \ref{al:gre},
%The algorithm of the third baseline is composed of MCOB (Algorithm \ref{al:MIMO}) and the greedy algorithm without CPU optimization.
%In the greedy algorithm, 
we do not consider the minimization problem with respect to $\{F_{k,k'}\}$ in \eqref{eq:Cloc} and \eqref{eq:Coff}.
This baseline, together with the second baseline, will assess the importance of our formulation as a joint optimization over communication and computation resources.

%The second and third baselines can be considered as the partially optimized solution.
%%
%The comparison with the first baseline will show the novelty of the proposed framework and optimization algorithm.
%Further, the comparisons with the second and third baselines will show the performance of the proposed inner algorithm intensively and stress the importance of joint optimization for overall problem.

\begin{figure*}[!t]
\vspace{-1mm}
\minipage{0.32\textwidth}
\includegraphics[width=\linewidth]{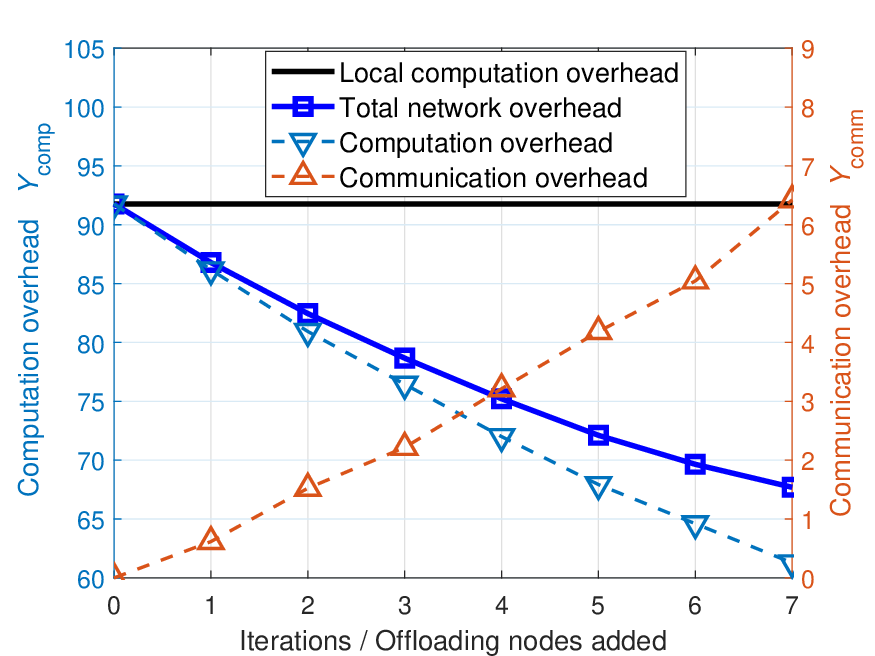}
\vspace{-5.8mm}
\caption{
Evolution of the computation (left axis), communication (right), and total (left) network overheads after each iteration of the greedy algorithm, for $K = 30$, $S = 2$, and $N = 5$. 
%Each iteration adds an offloading node as long as 
%the decrease in $Y_{\rm comp}$ outweighs the increase in $Y_{\rm comm}$.
}
\label{fig:sim:greedy} 
\endminipage\hfill
\minipage{0.32\textwidth}
\includegraphics[width=\linewidth]{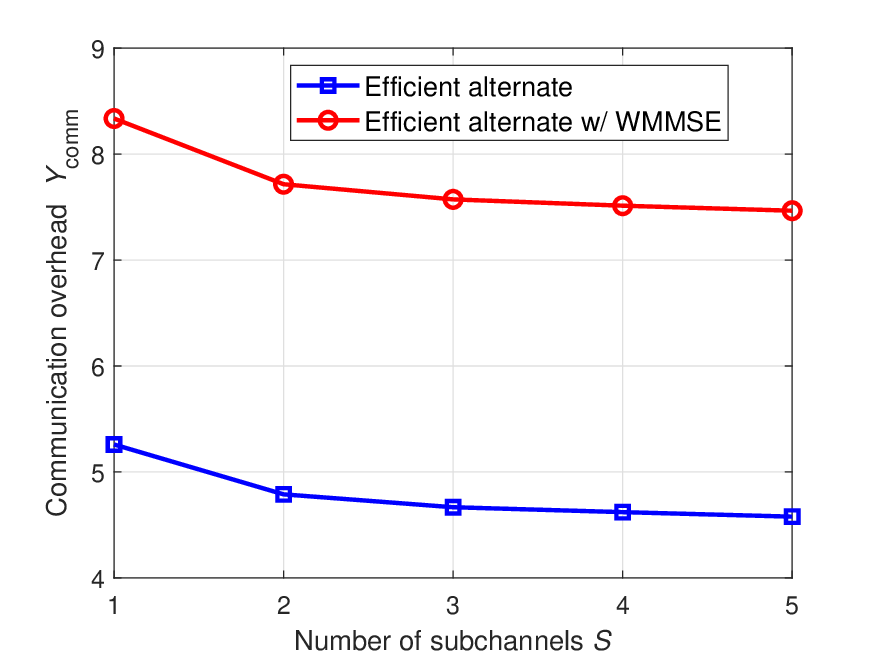}
\vspace{-5.8mm}
\caption{Communication overhead with varying subchannels $S$ for $K=30$ and $N=5$. 
Our method leverages additional subchannels for overhead reduction via interference mitigation. 
%We obtain an improvement of roughly  {30\%} 
%over the case where WMMSE is used for beamforming design.
}
\label{fig:sim:subchannel}
\endminipage\hfill
\minipage{0.32\textwidth}%
\includegraphics[width=\linewidth]{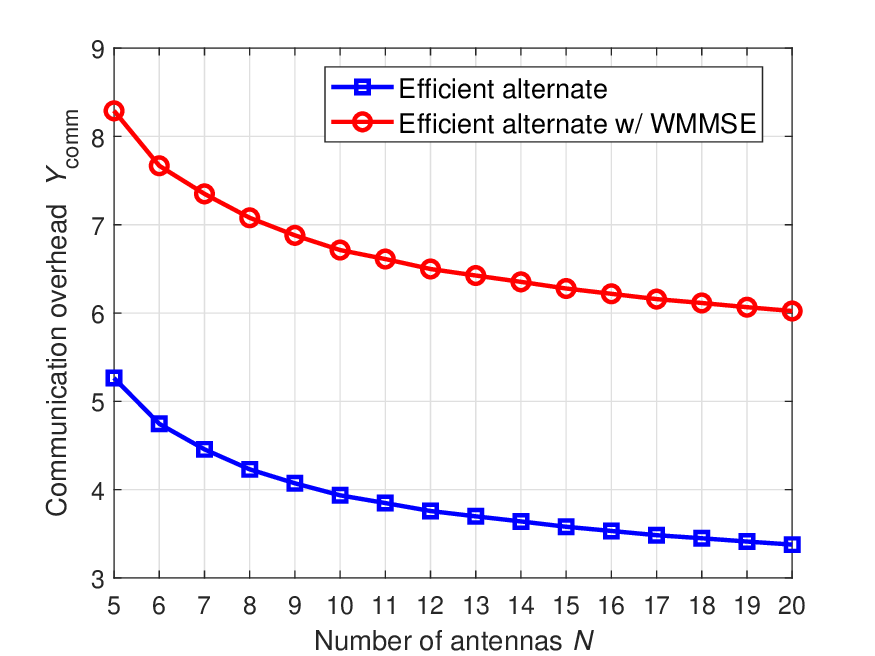}
\vspace{-5.8mm}
\caption{Communication overhead with varying antennas $N$ for $K=30$ and $S=1$.
Interferences can be suppressed further with a larger number of antennas.
%due to the directionality introduced by transmit beamforming and receive combining.
}
\label{fig:sim:antenna}
\endminipage
\vspace{-5mm}
\end{figure*}

%\vspace{-2mm}
%%%%%%%%%%%%%%%%%%%%%%%%%%%%%%%%%%%%%%%%%%%%

\vspace{-3mm}
 {\subsection{Optimality, Convergence, and Imperfect CSI}
\label{ssec:optconv}}

%\begin{figure}[t]
%%
%\centering
%\includegraphics[width=.8\linewidth]{figures/sim/optimality.eps}
%\caption{The total network overhead obtained by the semi-exhaustive search optimization, the efficient alternate optimization and local computation where $S=2$ and $N=5$. Even for small networks, offloading enables a high reduction in total network overhead compared to local computation.}
%\label{fig:sim:optimality}
%\end{figure}

Our first experiment compares the total network overhead incurred by semi-exhaustive search, efficient alternate optimization, and local computation for different numbers of nodes $K$.
Fig. \ref{fig:sim:optimality} 
shows the results as $K$ varies from $3$ to $8$ in a small network
with $S=2$ and $N=5$.
Compared to the local computation, 
the offloading through
our methodology results in a significant decrease between 19\% and 27\% in the total network overhead  even for small D2D networks.
{The semi-exhaustive search optimization  provides a lower bound on the minimum achievable overhead required by the efficient alternate optimization, as discussed in Section \ref{ssec:analysis}.
% If we can afford the implementation cost of the semi-exhaustive search optimization, then employing it is desirable due to the chance of obtaining a smaller overhead than the efficient alternate.
However, we find that the implementation of the semi-exhaustive search is computationally infeasible for more than $K = 8$ nodes, consistent with its computational complexity given in Lemma \ref{lem:complexity}.
% \footnote{ {
% If we can afford the implementation cost of the semi-exhaustive search optimization, we will always obtain better performance compared to the efficient alternate.}}
Furthermore, the efficient alternate optimization gives almost the same overhead performance as the semi-exhaustive search optimization.
%impossible for the case with a large number of nodes.
Henceforth, we will present results based on the efficient alternate optimization.
An experiment on the runtime growth rate of the efficient alternate optimization is provided in Appendix D-A, verifying its polynomial complexity.}

%Therefore, the semi-exhaustive search optimization is expected to yield better performances than the efficient alternate optimization.

%\begin{figure}[t]
%%
%\centering
%\includegraphics[width=.8\linewidth]{figures/sim/convergence.eps}
%\caption{Convergence behavior of the efficient alternate optimization algorithm when $K=10$, $S=2$, and $N=5$. The total network overhead  converges within a few iterations, reaching a  {23\%} improvement over local computation.
%}
%\label{fig:sim:convergence}
%\end{figure}

Fig. \ref{fig:sim:convergence} shows the convergence behavior of the efficient alternate optimization, plotting the total network overhead obtained after each iteration of Algorithm \ref{al:alt}, for the same settings in Fig. \ref{fig:sim:optimality} and $K = 10$ nodes.
%which is the alternate method between the beamforming design and greedy algorithm.
%Along the iteration number,
%the total network overhead is decreasing and eventually below the total network overhead of local computation. 
After the first iteration, the total network overhead decreases dramatically due to the high reduction in communication overhead obtained from the beamformer design.
We observe in our experiments that, the objective function generally converges within a few iterations.

%\subsection{Practical Implementation Imperfect Channels}
%\label{ssec:CSI}

%\begin{figure}[t]
%\centering
%\includegraphics[width=.8\linewidth]{figures/sim/CSI_total.eps}
%\caption{
% {
%Total network overhead under perfect CSI and imperfect CSI with $K=20$, $S=2$, and $N=5$. 
%With imperfect CSI, the proposed efficient alternate algorithm still obtains  substantial improvement of about $22 \%$ in network overhead as compared to local computing, while with perfect CSI, approximately $24 \%$ improvement is obtained. }
%}
%\label{fig:sim:CSI}
%\end{figure}

Fig. \ref{fig:sim:CSI} demonstrates 
the effect of imperfect channels on our proposed framework and algorithms.
%
% To emulate  imperfect CSI,
% channel distortion value is added to each component of the perfect channels. 
Imperfect CSI is modeled by adding a channel distortion value to the actual CSI.
Each distortion value follows a Gaussian distribution with zero mean and variance of $\theta^2 \|\mathbf{H}_{k,k'}^{(i)}\|_2^2$ , i.e.,  $\mathcal{CN}(0,\theta^2 \|\mathbf{H}_{k,k'}^{(i)}\|_2^2)$, where $\theta^2$ denotes the channel distortion ratio.
% With imperfect CSI, the proposed efficient alternate optimization still obtains a substantial improvement of about $22 \%$ in terms of the total network overhead as compared to local computing. Note that with perfect CSI, an improvement of approximately $24 \%$ is obtained.
As the distortion ratio increases to $\theta^2 = 0.5$,
% which is half of the variance on the channel matrices $\mathbf{H}_{k,k'}^{(i)}$, 
the overhead improvement relative to local computing only decreases from 28\% to 25\%.
This shows that our methodology is still applicable for
% the proposed framework and algorithm could be potentially used for imperfect CSI scenarios for 
minimizing  total D2D network overhead in the presence of imperfect CSI.
% Therefore, in the rest of our simulations, we consider perfect CSI  to clearly observe the effects of other factors.}

%%%%%%%%%%%%%%%%%%%%%%%%%%%%%%%%%%%%%%%%%%%%
\vspace{2mm}
\subsection{Communication-Computation Overhead Tradeoff}
\label{ssec:off}
% \vspace{-2mm}

Our next experiment assesses the benefit provided by each offloading node that the greedy algorithm adds in the efficient alternate optimization. Specifically, Fig. \ref{fig:sim:greedy}  shows the change in overhead as more data streams $(k, k', i)$ are added for offloading in Algorithm \ref{al:gre}, for $K  =  30$, $S  = \negmedspace 2$, and $N  =  5$. We show the evolution of the communication overhead $Y_{\rm comm} \negmedspace = \negmedspace \sum\nolimits_{k=1}^K \sum\nolimits_{k' \ne k}^K a_{k,k'} Y_{\rm comm}(k,k')$, the computation overhead $Y_{\rm comp} \negmedspace = \negmedspace \sum\nolimits_{k=1}^K \sum\nolimits_{k'=1}^K a_{k,k'} Y_{\rm comp}(k,k')$, and the total overhead $Y_{\rm total} = Y_{\rm comm} + Y_{\rm comp}$.
Overall, we see that the total network overhead is decreasing at each iteration, which is consistent with the operation of the greedy algorithm. This is obtained by trading an increase in communication overhead for a more substantial decrease in computation overhead. The algorithm successively exploits low-cost opportunities for offloading from resource constrained to resource-rich nodes, until such opportunities are no longer cost-effective. In this case,  {23\% of the nodes (7 out of 30)} become offloading nodes by the time the algorithm terminates.

%\begin{figure}[t]
%%
%\centering
%\includegraphics[width=.8\linewidth]{figures/sim/Offloading.eps}
%\caption{
%Evolution of the computation (left axis), communication (right axis), and total (left axis) network overheads after each iteration of the greedy algorithm, for $K = 30$, $S = 2$, and $N = 5$. Each iteration adds an offloading node as long as the increase in $Y_{\rm comm}$ is outweighed by the decrease in $Y_{\rm comp}$.
%%Total network overhead along the iteration number of the greedy algorithm where $K=20$, $S=2$, and $N=5$. In the greedy algorithm, the iteration implies the number of offloading nodes. The total network overhead is decreased by allowing more offloading nodes.
%}
%\label{fig:sim:greedy} 
%\end{figure}

%%%%%%%%%%%%%%%%%%%%%%%%%%%%%%%%%%%%%%%%%%%%
\vspace{-2mm}
\subsection{Varying Interference Management Resources}
\label{ssec:int}
% \vspace{-1mm}

Our next experiments assess the communication overhead reduction obtained by our methodology from leveraging interference management resources.
%Recall that inter-channel interference is inevitably incurred by restricted communication resources, such as the number of subchannels and  number of antennas.
When the number of subchannels $S$ and number of antennas $N$ are limited, we expect that communication overhead will be higher due to decreasing transmission data rates from inter-channel interferences.
Fig. \ref{fig:sim:subchannel} shows 
the effect of $S$ on $Y_{\rm comm}$ for both the efficient alternate optimization and the baseline using WMMSE, when $N = 5$ and there are $K = 30$ devices. We see that the total communication overhead decreases as the number of subchannels increases because more subchannels enable avoiding  interferences by allocating non-overlapping subchannels to different data streams.
Moreover, the efficient alternate optimization with MCOB gives better performance than that with WMMSE -- with improvements of roughly  37\% for each choice of $S$ --
because MCOB is designed to minimize the total communication overhead, while WMMSE  minimizes only the total communication time.

Fig. \ref{fig:sim:antenna} shows
the effect on communication overhead as more antennas are employed for $K=30$ and $S=1$. With a limited number of subchannels available ($S=1$), the beamforming strategy plays a significant role in communication overhead reduction.
As $N$ increases, our methodology suppresses the interference further due to the increased spatial degrees of freedom.
%In terms of the communication overhead reduction, the efficient alternate optimization with MCOB is better than that with WMMSE, similar in Fig. \ref{fig:sim:subchannel}.
The gap in communication overhead between the efficient alternate optimization with MCOB vs. WMMSE increases with more antennas, 
reaching roughly a 43\% improvement.

%with the same reason in the Fig. \ref{fig:sim:subchannel}.

%\begin{figure}[t]
%\centering
%\includegraphics[width=.8\linewidth]{figures/sim/subchannel.eps}
%\caption{Communication overhead with varying subchannels $S$ for $K=30$ and $N=5$. 
%Our method leverages additional subchannels for overhead reduction via interference mitigation. We obtain an improvement of roughly  {30\%} over the case where WMMSE is used for beamforming design.
%}
%\label{fig:sim:subchannel}
%\end{figure}

%\begin{figure}[t]
%\centering
%\includegraphics[width=.8\linewidth]{figures/sim/antenna.eps}
%\caption{Communication overhead with varying antennas $N$ for $K=30$ and $S=1$.
%Interferences can be suppressed further with a larger number of antennas due to the directionality introduced by transmit beamforming and receive combining.}
%\label{fig:sim:antenna}
%\end{figure}

\begin{figure*}[!t]
%
% \vspace{-2mm}
\minipage{0.32\textwidth}
   \includegraphics[width=\linewidth]{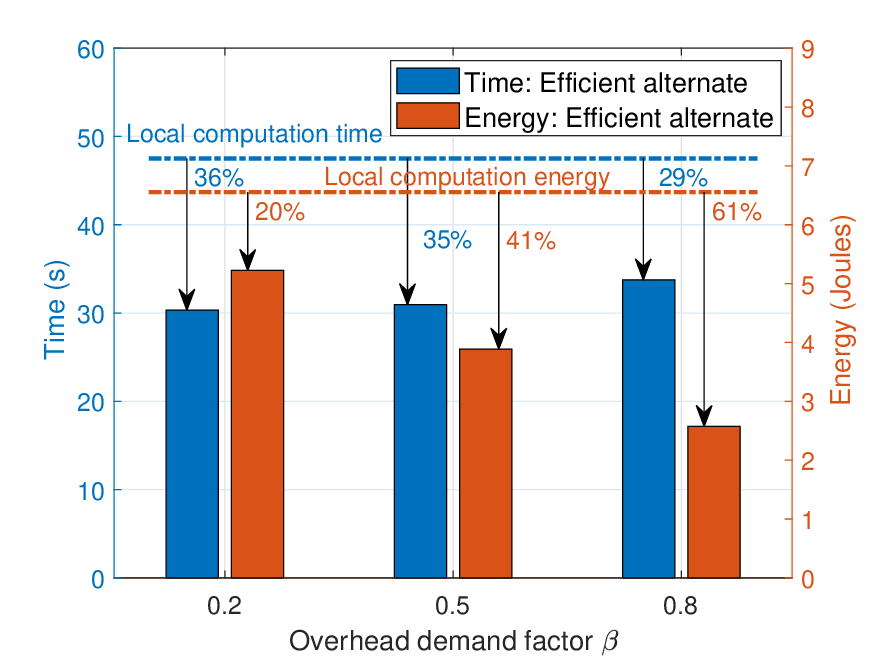}
   \vspace{-5.8mm}
 \caption{Total time delay and energy consumption with different $\beta$ for $K=10$, $S=2$, and $N=5$. 
 The tradeoff in the optimization objective is  adjusted according to $\beta$.}
% The total time is further reduced when $\beta=0.2$, while more  energy is reduced when $\beta=0.8$.
\label{fig:sim:beta_total}
\endminipage\hfill
\minipage{0.32\textwidth}
 \includegraphics[width=\linewidth]{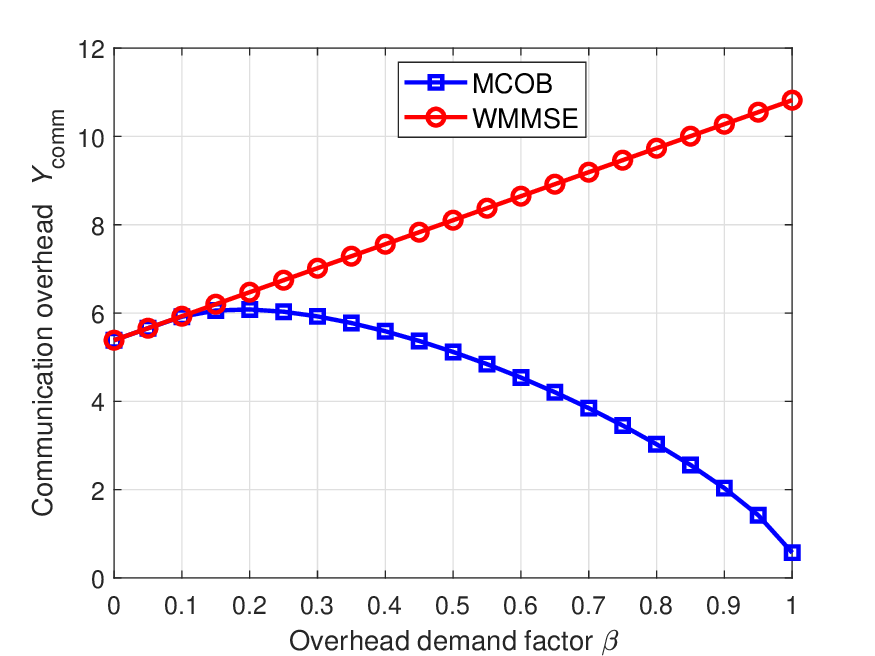}
 \vspace{-5.8mm}
\caption{Communication overhead varying the overhead factor $\beta$ for $K=30$, $S=2$, and $N=5$.
The optimization with MCOB outperforms WMMSE as $\beta$ increases.
%As more weight is placed on energy consumption ($\beta$ increases), the optimization with MCOB outperforms WMMSE.
}
\label{fig:sim:beta}
\endminipage\hfill
\minipage{0.32\textwidth}%
 \includegraphics[width=\linewidth]{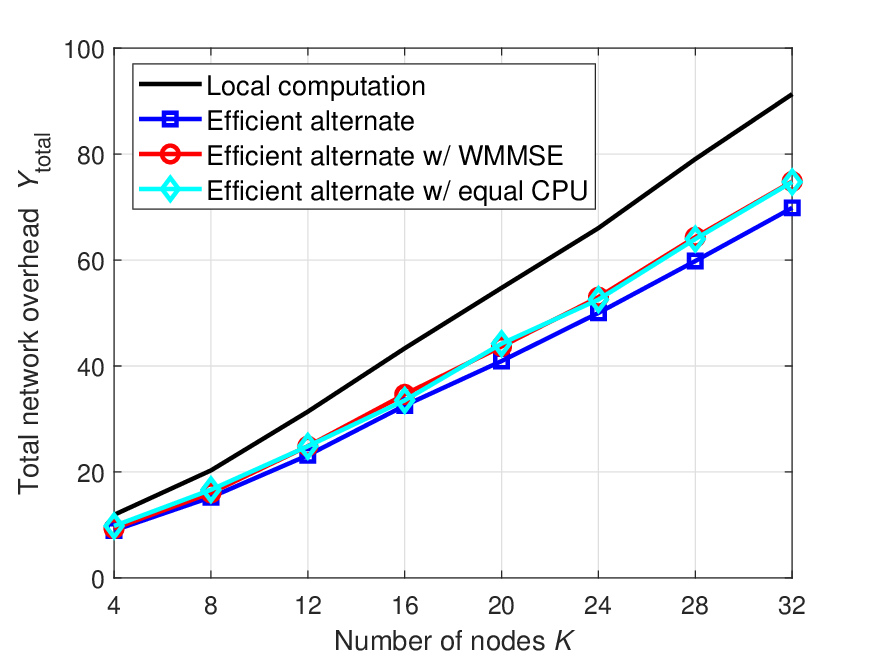}
 \vspace{-5.8mm}
\caption{
Impact of the number of nodes $K$ on the total network overhead for $S=2$ and $N=5$.
The efficient alternate emphasizes the benefit of our holistic optimization approach.}
\label{fig:sim:scalability}
\endminipage
\vspace{-5mm}
\end{figure*}

%%%%%%%%%%%%%%%%%%%%%%%%%%%%%%%%%%%%%%%%%%%%
\vspace{-3mm}
\subsection{Varying Time/Energy Optimization Importance}
\label{ssec:beta}

% \vspace{-.5mm}

%\begin{figure}[t!]
% \centering
% \includegraphics[width=.8\linewidth]{figures/sim/detail_total.eps}
% \caption{ {The total time and energy consumption with different values of $\beta$ for $K=10$, $S=2$, and $N=5$. The total time is further reduced when $\beta=0.2$, while more  energy is reduced when $\beta=0.8$.}}
%\label{fig:sim:beta_total}
%\end{figure}

 {We are also interested in the impact of the importance placed on time vs. energy in the total network overhead optimization.
Fig. \ref{fig:sim:beta_total} demonstrates the effect of the overhead demand factor $\beta$ on the total time delay and energy consumption incurred, aggregating all tasks and over both communication and computation overhead components,
for $K=10$, $S=2$, and $N=5$.
The individual time delay and energy consumption incurred by each task for different $\beta$
is provided in Appendix D-B.
The results in Fig. \ref{fig:sim:beta_total} are consistent with design of the optimization objective: compared to the case of all local computation, total time is reduced the most (36\%) when $\beta= 0.2$, while more energy reduction (61\%) is achieved when $\beta= 0.8$. This experiment confirms that the objective can be  adjusted according to the tradeoff importance requirement of a specific use case.}
% With small $\beta$, the optimization is geared toward to minimizing total time, while with large $\beta$, total energy minimization is more focused.
% The optimization results are consistent with the objective: total time is further reduced when $\beta=0.2$, while more energy reduction is achieved when $\beta=0.8$.
% This implies that 
% the optimization objective can be flexibly adjusted by setting an appropriate $\beta$, according to the requirement of network demand.

%\begin{figure}[t]
%\centering
%\includegraphics[width=.8\linewidth]{figures/sim/betaTotal.eps}
%\caption{Communication overhead varying the overhead factor $\beta$ for $K=30$, $S=2$, and $N=5$.
%%The proposed MCOB (Algorithm \ref{al:MIMO}) outperforms WMMSE when $\beta>0.3$. 
%As more weight is placed on energy consumption ($\beta$ increases), the optimization with MCOB outperforms WMMSE, since
%MCOB is designed to incorporate both factors. 
%%\textcolor{red}{MCOB yields the highest value at $\beta = 0.5$ where both competing objectives are given equal importance.
%}
%\label{fig:sim:beta}
%\end{figure}

%Beamforming design algorithm is assessed in this experiment with varying overhead factor $\beta$. Note that $\beta$ denotes the weight parameter between time and energy consumption for objective minimization.
 {Now, we consider the communication overhead optimization to evaluate the performance of MCOB, according to different values of $\beta$.
%We are also interested in the impact of the importance placed on time vs. energy in our communication overhead optimization.
%Recall that the importance on time vs. energy is controlled by the overhead factor $\beta$
%: i
}
%If $\beta=0$, the overall problem aims to minimize time consumption, and if $\beta=1$, the problem shifts to minimizing energy consumption.
Fig. \ref{fig:sim:beta} shows total communication overhead as $\beta$ varies from 0 to 1 with MCOB and WMMSE, for $K=10$, $S=2$, and $N=5$.
%\
 {
When $\beta$ is small, the performances of MCOB and WMMSE are almost identical, as the emphasis is on completion time minimization; in other words, WMMSE is a special case of MCOB for $\beta = 0$.
Since WMMSE does not factor in energy consumption minimization, at
any value of $\beta > 0.15$, MCOB 
shows a substantial improvement 
% in terms of communication overhead 
compared to WMMSE.
This emphasizes the importance of considering both time delay and energy consumption as components of the overhead model.}
% This is because the energy consumption, as well as the time consumption, is included in the communication overhead as the optimization objective.
%
 {
% We can also observe that the communication overhead of WMMSE increases linearly as $\beta$ increases:
% since WMMSE only solves for total time minimization and outputs the same beamformers regardless of the value of $\beta$, the values of $T_{\rm comm}$ and $E_{\rm comm}$ in  \eqref{eq:Ccomm} are constant. 
Note that the communication overhead of WMMSE increases linearly as $\beta$ increases.
This is because  the values of $T_{\rm comm}$ and $E_{\rm comm}$ in \eqref{eq:Ccomm} are constant as a result of WMMSE, and $E_{\rm comm}$ is larger than $T_{\rm comm}$ in this specific D2D scenario. 
% Therefore, as $\beta$ increases,  $Y_{\rm comm}$ increases.
Additionally, the achievable communication overhead by MCOB drops as $\beta$ increases beyond $0.15$. This is also due to the specific D2D scenario being considered: in \eqref{eq:Ccomm}, as a result of MCOB, the value of $E_{\rm comm}$ (in Joules) is further reduced than $T_{\rm comm}$ (in seconds) as $\beta$ increases.}
%
%
% The two key takeaways from the figure are that (i) it is important to include both time delay and energy consumption as components of the overhead model, and (ii) WMMSE is a special case of MCOB for $\beta=0$, i.e., MCOB can be considered as a generalized beamforming scheme.}
% This experiment shows that MCOB can be considered as a generalized beamformer design scheme with respect to communication overhead reduction.

%%%%%%%%%%%%%%%%%%%%%%%%%%%%%%%%%%%%%%%%%%%%
\vspace{-2mm}
\subsection{Varying the Number of Nodes}
\label{ssec:sca}

%\begin{figure}[t]
%\centering
%\includegraphics[width=.8\linewidth]{figures/sim/scalability.eps}
%\caption{
%Impact of the number of nodes $K$ on the total network overhead for $S=2$ and $N=5$. 
%The offloading-enabled methods scale better than local computation with respect to the network size. The improvement of the efficient alternate optimization relative to the two partially-optimized baselines emphasizes the importance of our holistic optimization approach.}
%\label{fig:sim:scalability}
%\end{figure}

In this experiment, we compare the total network overhead obtained by efficient alternate optimization and the three baselines as the size of the D2D network changes.
%stated in Section \ref{sec:eval:set:base} for different numbers of nodes $K$.
Fig. \ref{fig:sim:scalability} plots  
$Y_{\rm total}$ as $K$ increases for
 $S=2$ and $N=5$. 
Compared to local computation, the other three schemes each yield significant reduction in the total network overhead due to the benefit of offloading.
Furthermore, the offloading-enabled methods scale better as the size of the network increase: the performance gap widens and the improvement of the efficient alternate optimization stays around 20\% to 30\% consistently. With more nodes, there are offloading opportunities, leading to more overhead reduction.

In comparing the offloading-enabled methods, we note that the efficient alternate optimization consistently outperforms the equal CPU allocation and WMMSE baselines  (by 7\% and 6\%, respectively), which are partially optimized solutions. This emphasizes the importance of considering a joint optimization of communication and computation resources to obtain the lowest overhead in an environment of heterogeneous wireless devices. The equal CPU allocation baseline is a lower complexity algorithm, however, given it does not solve the CPU optimization problem. This could be a necessary tradeoff if optimization speed is critical, which depends on the timescale at which the solver is employed in practice.
We also provide a comparison between the theoretical and observed computational complexity of our method in Appendix D-A.

\vspace{-2mm}
\section{Conclusion and Future Work}
\label{sec:conc}

In this paper, we proposed a novel optimization methodology that  minimizes the total network overhead required to process a set of tasks in wireless D2D edge networks. Our optimization model consists of several computation and communication resources including topology configuration, CPU allocations, subchannel allocations, and beamforming design for MIMO transmitters and receivers.
Given that the problem is a non-convex MIP, we proposed two methods to solve it: semi-exhaustive search optimization and efficient alternate optimization.
%Semi-exhaustive search optimization yields
%a practical solution that we can realistically obtain but still requires tremendous computations for large networks.
%As an alternative for computational efficiency, efficient alternate optimization was proposed, which scales well to the large networks and is shown to have comparable performance to the semi-exhaustive search optimization in the simulation.
In analyzing the optimality and computational complexity of the proposed methods, we showed that the semi-exhaustive search can be regarded as a best effort 
for optimality, while the efficient alternate optimization has much smaller computational complexity.
%For beamforming design, we propose minimum communication overhead beamforming (MCOB) that aims to minimize both time and energy consumption as communication cost.
%
Through our numerical experiments, we showed the total network overhead can be reduced
significantly by leveraging offloading opportunities to resource-rich nodes in D2D networks.
% due to the benefit by offloading.
Further, in comparison with solutions that only optimize a subset of the variables, our results showed
%for efficient offloading, we show 
that joint communication and computation resource optimization is critical to obtaining the highest reductions in network overhead.

There are several potential avenues of future work.
First,
while we have focused on tasks generated in a single time period, dynamic task generation at nodes can also be considered. This can be accomplished by augmenting our methodology with
% to extend to a dynamic task generation scenario,
% a key direction is to develop 
a stochastic optimization that models task generation over a long-term time horizon.
% develop the proposed optimization by incorporating efficient congestion control and advanced resource management schemes for the long-term overhead minimization.
Second, 
% to develop more scalable frameworks, e.g., with linear time complexity for runtime, a potential approach is to develop
while we have focused on centralized control for optimizing distributed data processing, it will be desirable to decentralize this optimization in certain settings. One possibility is to investigate
a fully distributed implementation of Algorithm 3, where each node solves its local beamforming and resource design sub-problems based on information provided by adjacent neighbors.
Lastly, 
% we could extend our overhead model 
our overhead model can be extended
to include components outside of data transmission and processing, e.g., waiting latency, handover latency, and latency of receiving the processed data result back at the origin.

\bibliographystyle{IEEEtran}
\vspace{-2mm}
\bibliography{ref}

\vspace{-7mm}
\begin{IEEEbiography}
    {Junghoon Kim} received the M.S. degree in electrical engineering from Korea Advanced Institute of Science and Technology (KAIST) in 2014. Since 2019, he has worked towards the Ph.D. degree at Purdue University.
%     He is currently pursuing the Ph.D. degree in
% electrical and computer engineering with Purdue University.
\end{IEEEbiography}

\vspace{-7mm}
\begin{IEEEbiography}
{Taejoon Kim} (Senior Member, IEEE) received the Ph.D. degree in electrical and computer engineering from Purdue University in 2011. He is currently an Assistant Professor of electrical engineering and computer science at the University of Kansas (KU). His research interest includes 5G-and-beyond wireless systems and millimeter-wave and terahertz wireless. He was the recipient of the Miller Faculty Award from the KU School of Engineering and The IEEE Communications Society Stephen O. Rice Prize.
\end{IEEEbiography}

\vspace{-7mm}
\begin{IEEEbiography}
    {Morteza Hashemi} is an Assistant Professor with the Department of Electrical Engineering and Computer Science at the University of Kansas, Lawrence, Kansas. He received his MSc and PhD degrees in Electrical Engineering from Boston University in 2013 and 2015, respectively. Before joining KU in 2019, he was a postdoctoral researcher and senior lecturer at the Ohio State University. His research interests span the areas of wireless communications, information systems, real-time data networking, and networked cyber-physical~systems.
\end{IEEEbiography}

\vspace{-7mm}
\begin{IEEEbiography}
    {Christopher G. Brinton} (Senior Member, IEEE) received the Ph.D. degree in electrical engineering from Princeton University in 2016. He is currently an Assistant Professor of electrical and computer engineering with Purdue University. Since joining Purdue University in 2019, he has won several awards including the Seed for Success Award and the Ruth and Joel Spira Outstanding Teacher Award.
\end{IEEEbiography}

\vspace{-7mm}
\begin{IEEEbiography}
    {David J. Love} (S’98 - M’05 - SM'09 - F'15)  is the Nick Trbovich Professor of Electrical and Computer Engineering at Purdue University.  His research interests are in the design and analysis of broadband wireless communication systems, beyond-5G wireless systems, multiple-input multiple-output (MIMO) communications, millimeter wave wireless, software defined radios and wireless networks, coding theory, and MIMO array processing.  
\end{IEEEbiography}

\clearpage
\appendices

\section{Proof of Theorem 1}
\label{sec:appThm}

We first rewrite the problem \eqref{eq:opt:MIMO2}-\eqref{eq:opt:MIMO2:const} to an equivalent form by introducing an auxiliary variable $\gamma_k \in \mathbb{R}^{++}$ for $k \in {K_{{\rm{Tx}}}}$ as
\begin{align}
    & \text{minimize} & & 
    \label{eq:opt:MIMO3}
    \sum\limits_{k \in {K_{{\rm{Tx}}}}}^{} {I_k{\gamma _k}} 
    \\
    \label{eq:opt:MIMO3:const1}
    & \text{subject to} & &  \frac{{{g_k}({{\bf{f}}_k})}}{{{{u_k}(\{ {{\bf{f}}_k}\} ,{{\bf{z}}_{k',i}},{w_k})}}} \le {\gamma _k}{\rm{  }},
    \\
    \label{eq:opt:MIMO3:const2}
    & & & ||{\bf{f}}_k||_2^2 \le P_k \;\; \forall k \in \mathcal{K}_{\rm Tx}
    \\
    & \text{variables} & &  \{{\bf{f}}_{k}\}, \; \{{\bf{z}}_{k',i}\}, \; \{w_k\}, \; \{\gamma_k\}.  \nonumber
\end{align}
%where $h_k$ and $f_k$ are given in \eqref{eq:h} and \eqref{eq:f}, respectively. 
%Note that the problem \eqref{eq:opt:MIMO3}-\eqref{eq:opt:MIMO3:const2} is not convex with respect to $\{{{\bf{f}}_k}\}$, either.
%The first constraint denotes the communication demand constraint, and the second constraint denotes the transmit power constraint for transmit node $k$.
Introducing the Lagrange multipliers $\{\lambda_k\}$ and $\{\mu_k\}$ for the  two inequality constraints in \eqref{eq:opt:MIMO3:const1}-\eqref{eq:opt:MIMO3:const2}, we obtain the Lagrange function $\mathcal{L}(\cdot)$ of the problem \eqref{eq:opt:MIMO3}-\eqref{eq:opt:MIMO3:const2} as
\begin{align}
    & \mathcal{L}(\{ {{\bf{f}}_k}\} ,\{ {{\bf{z}}_{k',{i}}}\}, \{{w_k}\}, \{{\gamma _k}\}, \{{\lambda _k}\}, \{{\mu _k}\}) =
    \\
    & \sum\limits_{k \in \mathcal{K}_{\rm Tx} }^{} {I_k{\gamma _k}}  + \sum\limits_{k \in \mathcal{K}_{\rm Tx}} {{\lambda _k}({g_k} - {\gamma _k}{u_k}})  
    + \sum\limits_{k \in \mathcal{K}_{\rm Tx}} {{\mu _k}(||{\bf{f}}_k||_2^2 - P_k)} 
    \nonumber
\end{align}
where $u_k(\{ {{\bf{f}}_k}\} ,{{\bf{z}}_{k',i}} , {w_k} )$ and $g_k({{\bf{f}}_k})$ are denoted as $u_k$ and $g_k$ for simplicity. 

Assuming that $\{ {{\tilde {\bf{f} }}_k}\}$, $\{ {{\tilde{\bf z}}_{k',{i}}}\}$, $\{ {\tilde w_k}\}$, and $\{ {\tilde \gamma_k}\}$ are the solutions of the problem \eqref{eq:opt:MIMO3}-\eqref{eq:opt:MIMO3:const2},
they must satisfy the KKT conditions
\begin{gather}
    \frac{\partial }{{\partial {\bf{f}}_k}} \mathcal{L}( \cdot ) = {\bf{0}},
    \;\;
    \frac{\partial }{\partial {\bf{ z}}_{k',{i}} } \mathcal{L}( \cdot ) = {\bf{0}},
    \;\;
    \frac{\partial }{{\partial w_k}} \mathcal{L}( \cdot ) = 0,
    \nonumber
    \\
    \frac{\partial }{{\partial \gamma_k}} \mathcal{L}( \cdot ) = I_k -\lambda_k u_k = 0,
    \label{eq:prf:st}
    \\
    {\lambda _k}({g_k} - {\gamma _k}{u_k}) = 0,
    \;\;
    {\mu _k}(||{\bf{f}}_k||_2^2 - P_k) = 0,
    \label{eq:prf:cs}
    \\
    g_k \le \gamma_k u_k,
    \;\;
    ||{\bf{f}}_k||_2^2 \le P_k,
    \label{eq:prf:pf}
    \\
    \lambda_k \ge 0,
    \;\;
    \mu_k \ge 0
    \;\;\; \forall k \in \mathcal{K}_{\rm Tx},
    \label{eq:prf:df}
\end{gather}
where \eqref{eq:prf:st}-\eqref{eq:prf:df} represent the conditions of stationarity, complementary slackness, primal feasibility, and dual feasibility.
% with the length $N_k$ or $N_{k'}$.
%Note that the stationarity \eqref{eq:prf:st}, complementary slackness \eqref{eq:prf:cs}, primal feasibility \eqref{eq:prf:pf}, and dual feasibility condition \eqref{eq:prf:df}.

From \eqref{eq:h} and \eqref{eq:f}, ${u_k} \ge 0$ and ${g_k} > 0$. Furthermore, with the optimal solutions $\{ {{\tilde{\bf f}}_k}\}$, $\{ {{\tilde{\bf z}}_{k',{i}}}\}$, and $\{ {\tilde w_k}\}$, 
the inequality $u_k>0$ is guaranteed. 
Otherwise, it will drive the value of the objective function in \eqref{eq:opt:MIMO2} to infinity.
Since $u_k > 0$, the last condition in \eqref{eq:prf:st} and first condition in \eqref{eq:prf:cs} yield
% $\lambda_k$ and $\gamma_k$ as
%Therefore, for the optimal solution  $\{ {{\tilde{\bf f}}_k}\}$, $\{ {{\tilde{\bf z}}_{k',{i}}}\}$, and $\{ {\tilde w_k}\}$, the followings should be satisfied
\begin{align}
    \lambda_k = \frac{I_k}{u_k},
    \;\;\;
    \gamma_k = \frac{g_k}{u_k}.
    \label{eq:prf:sys}
\end{align}

Then, the remaining conditions, i.e., the first three conditions in \eqref{eq:prf:st}, the second condition in \eqref{eq:prf:cs}, the second condition in \eqref{eq:prf:pf}, and the second condition in \eqref{eq:prf:df}, are exactly the KKT conditions of the problem below:
\begin{align}
    & \text{minimize} & & 
    \sum\limits_{k \in {K_{{\rm{Tx}}}}}^{} {{\lambda _k}({g_k}- {\gamma _k}{u_k})} 
    \label{eq:prf:prob2}
    \\
    & \text{subject to} & & ||{\bf{f}}_k||_2^2 \le P_k \;\; \forall k \in \mathcal{K}_{\rm Tx}
    \label{eq:prf:prob2:const}
    \\
    & \text{variables} & &  \{{\bf{f}}_{k}\}, \; \{{\bf{z}}_{k',i}\}, \; \{w_k\}. \nonumber
\end{align}

In summary, if $\{ {{\tilde{\bf f}}_k}\}$, $\{ {{\tilde{\bf z}}_{k',{i}}}\}$, $\{ {\tilde w_k}\}$, and $\{ {\tilde \gamma_k}\}$ are solutions of the  problem \eqref{eq:opt:MIMO3}-\eqref{eq:opt:MIMO3:const2}, then $\{ {{\tilde{\bf f}}_k}\}$, $\{ {{\tilde{\bf z}}_{k',{i}}}\}$, and $\{ {\tilde w_k}\}$ are solutions of the problem \eqref{eq:prf:prob2}-\eqref{eq:prf:prob2:const} while simultaneously satisfying \eqref{eq:prf:sys}.
The contrary conclusion can be obtained in the opposite direction, which leads to the proof of Theorem 1.
If $\{ {{\tilde{\bf f}}_k}\}$, $\{ {{\tilde{\bf z}}_{k',{i}}}\}$, and $\{ {\tilde w_k}\}$ are 
solutions of the problem \eqref{eq:prf:prob2}-\eqref{eq:prf:prob2:const} and also simultaneously satisfy the system equations with $\tilde \lambda_k$ and $\tilde \gamma_k$ in \eqref{eq:prf:sys}, then
$\{ {{\tilde{\bf f}}_k}\}$, $\{ {{\tilde{\bf z}}_{k',{i}}}\}$, $\{ {\tilde w_k}\}$, $\{\tilde \lambda_k \}$, and  $\{\tilde \gamma_k \}$ satisfy  all of the KKT conditions \eqref{eq:prf:st}-\eqref{eq:prf:df}.
%of the problem \eqref{eq:opt:MIMO3}-\eqref{eq:opt:MIMO3:const2}.
This means that $\{ {{\tilde{\bf f}}_k}\}$, $\{ {{\tilde{\bf z}}_{k',{i}}}\}$, $\{ {\tilde w_k}\}$,  and $\{\tilde \gamma_k \}$ are the solutions of \eqref{eq:opt:MIMO3}-\eqref{eq:opt:MIMO3:const2}.
It follows that $\{ {{\tilde{\bf f}}_k}\}$, $\{ {{\tilde{\bf z}}_{k',{i}}}\}$, $\{ {\tilde w_k}\}$ are optimal solutions of \eqref{eq:opt:MIMO2}-\eqref{eq:opt:MIMO2:const}. % filename in curly brackets
\section{Proof of Lemma 2}
\label{sec:appLem2}

For computational complexity of the two proposed methods, 
we only need to compare
how many combinations of the binary variables $\{{ a_{k,k'}}\}$ and $\{{ b_{k,i}} \}$ are addressed for optimization. Note that non-integer variables are optimized when the binary variables are given.
First, we deal with the computational complexity of the semi-exhaustive search optimization.
From the condition \eqref{eq:con:a1}, each $k$ must choose one $k'$ where $k' \in \mathcal{K}$. If $k' \ne k$, we also must choose one $i$ from the condition \eqref{eq:con:b1} where $i \in \mathcal{S}$.
Then, we have $(K-1)S+1$ cases for each $k$. This is performed for every $k \in \mathcal{K}$, and we get $(KS-S+1)^K$ cases. Therefore, we have $\mathcal{O}((KS-S+1)^K)$.
Although condition \eqref{eq:con:b1} can reduce the total number of cases, we consider the worst case scenario for computational complexity. 
%This is also applied for the efficient alternate optimization.

For the computational complexity of the efficient alternate optimization,
% is given as $\mathcal{O}({K^3}S)$.
%To obtain $\mathcal{O}({K^3}S)$, 
a few steps need to be described.
As the first iteration of the greedy search,
the number of pairs among all $K$ nodes is $K(K-1)$. For each pair, we consider $S$ cases from the subchannel allocation condition  \eqref{eq:con:b1}. Therefore, we obtain $K(K-1)S$ cases at the first iteration. 
%For computational convenience, we set $K^2S$.
At the second iteration, the transmit candidate set $\mathcal{K}_{\rm Tx}$ is updated with $|\mathcal{K}_{\rm Tx}| = K-2$. Note that $|\mathcal{K}_{\rm Rx}| = K$.
Then, we have $(K-2)(K-1)S$ cases.
At the third iteration, we have $|\mathcal{K}_{\rm Tx}| = K-3$ or $K-4$.
If the larger case $|\mathcal{K}_{\rm Tx}| = K-3$ is considered as worst case scenario, total cases will be $(K-3)(K-1)S$.
This would continue to $1 \cdot (K-1)S$. 
%The total combinations will be $(K-1)S \sum\nolimits_{k = 1}^K {k} = (K-1)K(K+1)S/2$. 
We can apply the upper bound and calculate the total combinations approximately as $(K-1)S \sum\nolimits_{k = 1}^K {k} = (K-1)K(K+1)S/2$. 
%the total combinations considered for optimization are approximated to
%$(K-1)S \sum\nolimits_{k = 1}^K {k} = (K-1)K(K+1)S/2$. 
%= S\sum\nolimits_{p = 2}^K {\frac{{p(p - 1)}}{2}}  \approx {S{K^3}}/{6}
Therefore, we have $\mathcal{O}({K^3}S)$. % filename in curly brackets
 {\section{Derivation of the Closed-Form Solution for ${\bf f}_k$}
\label{sec:appbeam}}

 {
To solve problem \eqref{eq:opt:MIMOeach}-\eqref{eq:opt:MIMOeach:const}, for each ${\bf{f}}_{k}$ where $k \in \mathcal{K}_{\rm Tx}$, we introduce the Lagrange multiplier $\nu_k$ for the inequality constraint \eqref{eq:opt:MIMOeach:const} and obtain the Lagrangian function as
\begin{multline}
    L({{\bf{f}}_k},{\nu_k}{\rm{)}} = {\lambda _k}{\beta _k}{\bf{f}}_k^H{{\bf{f}}_k} - 2{\lambda _k}{\gamma _k}w_k^{ - 1}{\mathop{\rm Re}\nolimits} [{\bf{z}}_{k',i}^H{\bf{H}}_{k,k'}^{({i})}{{\bf{f}}_k}] 
    \\
    + {\bf{f}}_k^H{{\boldsymbol \Sigma} _k}{{\bf{f}}_k} + {\nu _k}({\bf{f}}_k^H{{\bf{f}}_k} - P_k).
\end{multline}
The first-order optimality condition yields 
\begin{equation}
    {\lambda _k}{\beta _k}{{\bf{f}}_k} - {\lambda _k}{\gamma _k}w_k^{ - 1}{\bf{H}}{_{k,k'}^{({i})H}} {{\bf{z}}_{k',i}} + {{\boldsymbol \Sigma} _k}{{\bf{f}}_k} + {\nu _k}{{\bf{f}}_k} = {\bf 0}.
\end{equation}
We can obtain the solution ${{\bf{f}}_k} $ as a function of $\nu_k$ to be
\begin{equation}
    \label{eq:sol:g}
    {{\bf{f}}_k}({\nu _k}) = {\lambda _k}{\gamma _k}w_k^{ - 1}{({{\boldsymbol \Sigma} _k} + {\nu _k}{\bf{I}} + {\lambda _k}{\beta _k}{\bf{I}})^{ - 1}}{\bf{H}}{_{k,k'}^{({i})H}} {{\bf{z}}_{k',i}}.
\end{equation}}

 From  complementary slackness,  dual feasibility, and  primal feasibility, the following conditions should be satisfied:
 ${\nu _k} ({\bf{f}}_k^H({\nu _k}){{\bf{f}}_k}({\nu _k}) - P_k) = 0$, $\nu_k \ge 0$, and ${\bf{f}}_k^H({\nu _k}){{\bf{f}}_k}({\nu _k}) \le P_k$. 
In other words, if ${\bf{f}}_k^H(0){{\bf{f}}_k}(0) \le P_k$, then ${{\bf{f}}_k}(0) $ is the optimal solution.
If ${\bf{f}}_k^H(0){{\bf{f}}_k}(0) > P_k$, then ${{\bf{f}}_k}({\nu^\star _k}) $ is the optimal solution  where ${\bf{f}}_k^H({\nu^\star _k}){{\bf{f}}_k}({\nu^\star _k}) = P_k$ should be satisfied for some $\nu^\star_k$.
In this case, the remaining part is to determine $\nu^\star_k$.
Using the eigendecomposition $\Psi_k{\Lambda _k}\Psi _k^H = {{\boldsymbol \Sigma}_k} + {\lambda _k}{\beta _k}{\bf{I}}$ and denoting $ \Phi _k = \Psi _k^H{\bf{H}}_{k,k'}^{(i)H}{{\bf{z}}_{k',i}}$, we can represent
${\bf{f}}_k^H({\nu _k}){{\bf{f}}_k}({\nu _k})$  as
\begin{equation}
    \label{eq:gHg}
    {\bf{f}}_k^H({\nu _k}){{\bf{f}}_k}({\nu _k}) = {\left| {{\lambda _k}{\gamma _k}w_k^{ - 1}} \right|^2} \sum\limits_{m = 1}^{{N_k}} {\frac{{{{\left| {{{[{\Phi _k}]}_m}} \right|}^2}}}{{{{({{[{\Lambda _k}]}_{m,m}} + {\nu _k})}^2}}}},
\end{equation}
where ${[{\Phi _k}]}_m$ denotes the $m$-th element of ${\Phi _k}$ and ${[{\Lambda _k}]}_{m,m}$ denotes the $m$-th diagonal element of $\Lambda_k$. The formula in \eqref{eq:gHg} is a non-increasing function of $\nu_k$. Therefore, a finite-interval one-dimensional search method such as the bisection method~\cite{bisection}
can be used to determine $\nu^\star_k$ to satisfy ${\bf{f}}_k^H({\nu^\star _k}){{\bf{f}}_k}({\nu^\star _k}) = P_k$.
$\\$
\section{Additional Experimental Results} 
\label{sec:appfigures}

\subsection{Runtime Growth Rate of the Efficient Alternate Optimization}
\label{ssec:runtime}

\begin{figure}[t]
\centering
  \includegraphics[width=.7\linewidth]{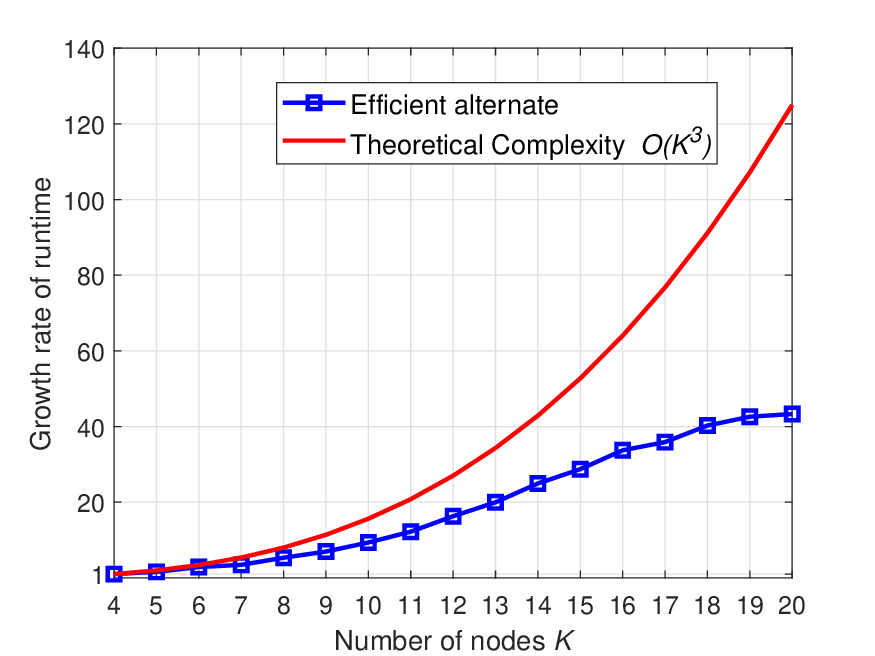}
\caption{
 {
The runtime growth rate of the efficient alternate optimization along the number of nodes $K$, for our experimental setting with $S=1$ and $N=5$. While the worst-case theoretical complexity is on the order of $K^3$, we observe a better growth rate in practice (approximately $K^{2.4}$ according to curve fitting).}}
\label{fig:sim:runtime}
%\vspace{-2mm}
\end{figure}

 {In Fig. \ref{fig:sim:runtime}, we consider the runtime growth rate of the efficient alternate optimization over the number of nodes compared to the theoretical complexity given in Section \ref{ssec:complexity}.
We consider $S=1$ subchannels and $N=5$ antennas. The runtime at $K=4$ is normalized to 1 as a basis, and the growth rate along the number of nodes is plotted. While the growth rate of the worse-case complexity derived in Lemma 2 is $K^3$, we observe a growth rate is lower than $K^3$ (approximately $K^{2.4}$ according to polynomial curve fitting).}

 {Overall, we see that the efficient alternate optimization has polynomial time complexity, which is considered a desirable upper bound in algorithm design. We leave additional improvements in runtime, e.g., to linear time complexity, to future work. In this regard, it may be desirable to have a fully distributed implementation of 
the efficient alternate optimization.
% Algorithm 3 in certain scenarios, e.g., if no network operator is available to centralize the execution.
To do so, future work
can consider distributed algorithms for the (i) beamforming and (ii) resource design sub-problems. For (i), one can investigate a version of MCOB where each node solves for its own beamformer and combiner, leveraging channel reciprocity, with network information provided by neighbors. For (ii), one can investigate a distributed consensus version of the greedy algorithm, where each node chooses its local candidate for link addition and the nodes collaboratively discover a final decision. 
The incurred communication overhead from such approaches can also be factored into the overall network overhead optimization.}
% We believe that this will be an interesting future work.

\subsection{Example of Task Offloading and Processing}
\label{ssec:specific}

\begin{figure*}[t!]
\centering
\begin{subfigure}{.32\linewidth}
  \centering
  \includegraphics[width=\linewidth]{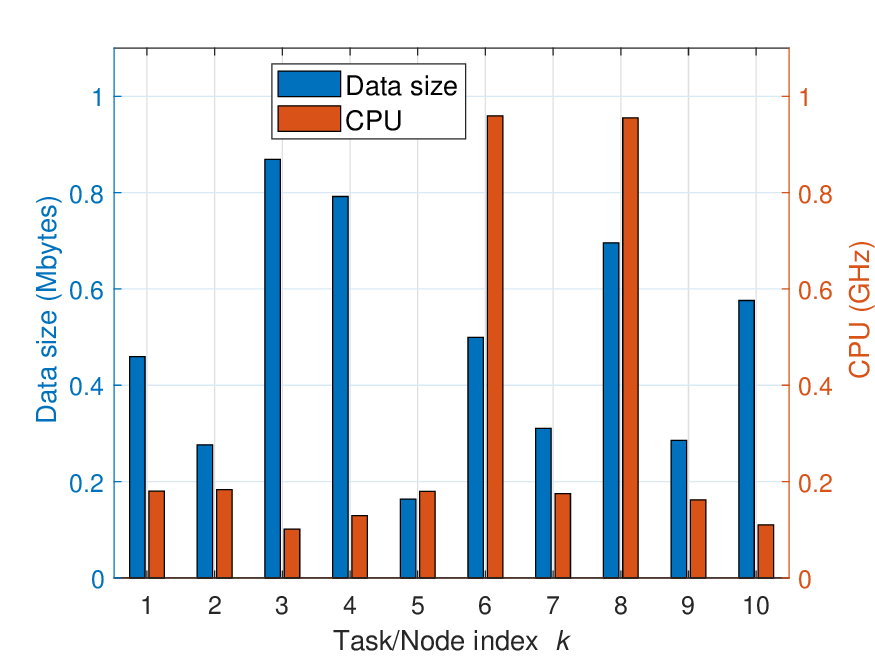}
  \caption{{The size of task $k$ and available CPU at node $k$}.}
\end{subfigure}
\begin{subfigure}{.32\linewidth}
  \centering
  \includegraphics[width=\linewidth]{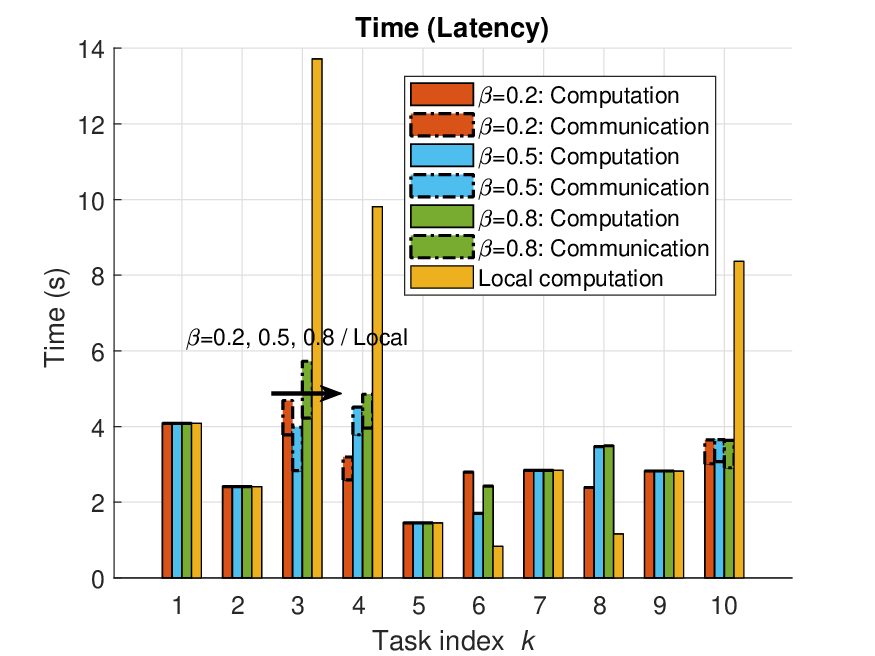}
  \caption{{Time delay (latency) for processing each task}.}
\end{subfigure}
\begin{subfigure}{.32\linewidth}
  \centering
  \includegraphics[width=\linewidth]{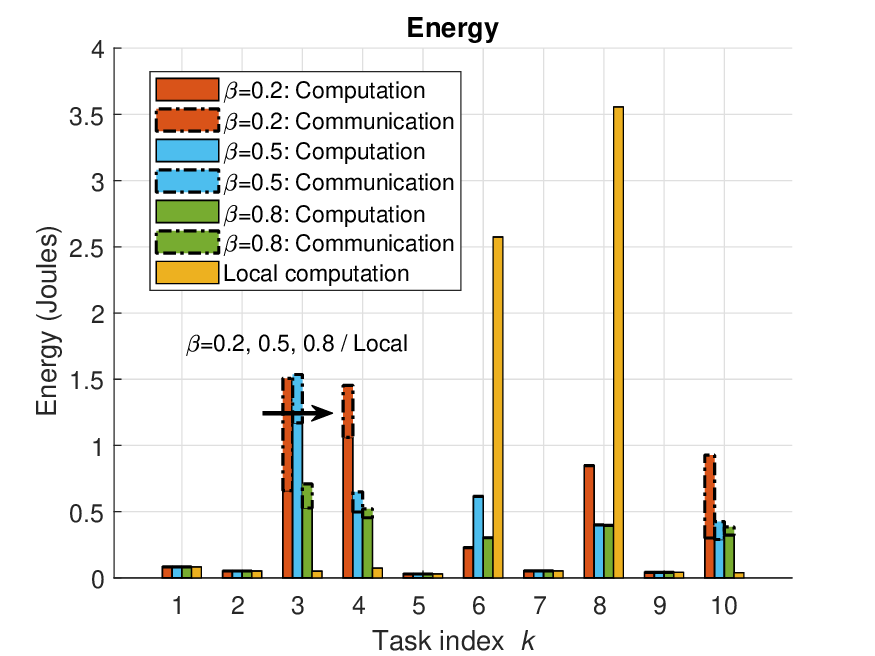}
  \caption{{Energy consumption for processing each task.}}
\end{subfigure}
%
%\begin{subfigure}{.48\linewidth}
% \centering
% \includegraphics[width=.8\linewidth]{figures/sim/detail_total.eps}
% \caption{Total time and energy consumption with different $\beta$ }
%\end{subfigure}
%
\caption{
Specific breakdown of task offloading and processing for the experiment in Fig. \ref{fig:sim:beta_total}.
% An example of task offloading and processing for each task in D2D network with $K=10$, $N=5$, and $S=2$. 
From (a), intuitively, the nodes ($k=3,4,10$) with low CPU and large data size are more likely to offload their tasks to other nodes ($k=6,8$) with high CPU. 
After the efficient alternate optimization (Algorithm 3) is executed for each $\beta = 0.2, 0.5, 0.8$, the individual time delay and energy consumption values incurred by each task are depicted in (b) and (c).
}
\label{fig:specific}
%\vspace{-2mm}
\end{figure*}

In Fig. \ref{fig:specific}, we show a more detailed version of the result in Fig. \ref{fig:sim:beta_total}, i.e., individual  time and energy consumption incurred by each task for different values of $\beta$ through our optimization.
% in Fig. \ref{fig:specific}.
% The experiment setting is the same as for Fig. \ref{fig:sim:beta_total} where $K=10$, $S=2$, and $N=5$.
The size of each task and available CPU at each node are demonstrated in Fig. \ref{fig:specific}(a). 
Intuitively, the nodes ($k=3,4,10$) with low CPU and large data size are more likely to offload tasks to other nodes ($k=6,8$) with high CPU as a result of our optimization.
Given the computation and communication resources including subchannels, task size, and available CPU, we apply our efficient alternate optimization methodology for each
% solve the optimization with
% the proposed efficient alternate optimization for each 
$\beta = 0.2, 0.5, 0.8$, 
with the results summarized in Fig. \ref{fig:specific}(b)\&(c).
Fig. \ref{fig:specific}(b)\&(c) break down the time and energy overhead incurred as a result of communication and computation for each task.
%
% and obtain the results, such as task allocation and corresponding time/energy consumption.
%
For $\beta = 0.2$, tasks
$k=3,4,10$ are offloaded to nodes $k=6,8,6$, respectively.
For $\beta=0.5,0.8$, which increases the emphasis on minimizing energy consumption, tasks $k=3,4,10$ are offloaded to nodes $k=6,8,8$, respectively.
The rest of the tasks are processed locally.

In order to explore how the time (latency) and energy consumption for processing the data are changed through  task offloading, we take a look the specific behavior of task $4$ offloaded to node $8$ as an example.
In Fig. \ref{fig:specific}(b), for any $\beta$, we see that the time overhead for task 4, including computation and communication overhead, is greatly reduced as compared to local task processing, because task 4 is offloaded to node 8 which has high CPU.
This also leads to increasing the computation time overhead for  task 8, because less CPU is allocated to it due to CPU sharing with the offloaded task 4.
In Fig. \ref{fig:specific}(c), the energy consumption for processing task 4 increases as compared to local task processing, because a larger amount of CPU is used to process the offloaded task (at node 8).
At the same time, the energy consumption for task 8 decreases because less CPU (at node 8) is now allocated to this task due to CPU sharing.
Overall, task offloading is conducted to minimize the optimization objective, which is the weighted sum of  total time and energy consumption for processing the data throughout the network.
The overall performance in terms of the total time and energy consumption is shown in Fig. \ref{fig:sim:beta_total}.

\subsection{Dynamic Task Generation}
\label{ssec:queue}

\begin{figure}[t!]
\centering
  \includegraphics[width=.7\linewidth]{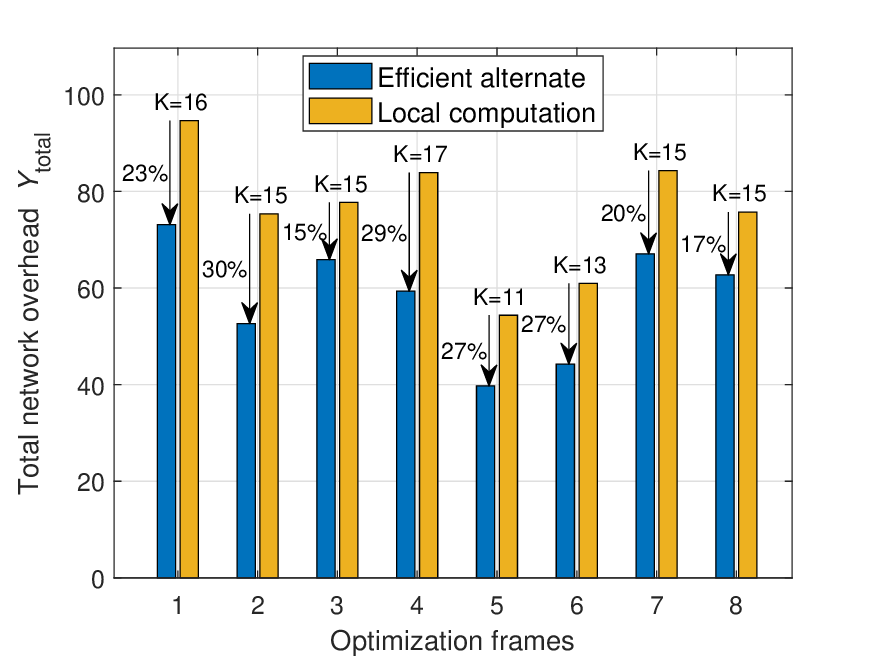}
\caption{
Total network overhead over multiple optimization frames under a dynamic task generation scenario. The total number of nodes in the D2D network is $K_{\max} = 30$, where some of nodes will participate in each optimization frame if they have a task in their queue. Task generation is assumed to follow a Poisson distribution at each node, with task arrival rate $\phi=1/10$ (tasks/s).}
\label{fig:sim:queue}
\end{figure}

In this experiment, we assess our proposed framework and algorithms under a dynamic task scenario. 
% If a new task is generated at node, it can be queued as a new task at each node.
When a new task is generated at a node, we assume that it is  queued until there is a time frame available for it to be processed.
Then, the optimization proceeds as a series of frames, where each frame will consider one task from each node in a first in first out (FIFO) manner.
Since nodes will not always have tasks in their queue, let $\mathcal{K}(t)$ denote the set of participating nodes in optimization frame $t$.
In Fig. \ref{fig:sim:queue}, we show a simulation which considers
the total network overhead  obtained by our efficient alternate optimization algorithm 
over $t = 1,...,8$ optimization frames, where the total nodes is $K_{max} = 30$ and $\mathcal{K}(t)$ varies from  tasks being generated according to a standard $M/M/1$ queue model \cite{ross2014introduction}.
% In this experiment, the total number of nodes is $K_{\max} = 30$, where some of the nodes will participate in each optimization frame when they have a queued task to be processed.
For the $M/M/1$ queue model, 
Poisson distributed traffic is considered with task arrival rate $\phi=1/10$ (tasks/sec), and the traffic is assumed to be independent across nodes. 
We consider the period of each optimization frame to be $T_s = 5$ sec.
% , which implies that the probability that a node generates a task is  50\% in each optimization frame.
The task size is assumed to be $I_k = 1$ Mbytes for each node.
In every optimization frame $t$, the number of participating nodes, $K(t)$, is 
% varying according to dynamic nature of task generation, which is
indicated above the bars in Fig. \ref{fig:sim:queue}. 
Overall, the total network overhead 
is reduced substantially in each time period
% can be highly reduced by 16 \% to 29 \%, 
compared to local computation: between 15\% and 30\%. This is consistent with our experiments for the non-dynamic case.

The above experiment shows that our framework and algorithms can be applied in the dynamic task setting.
Moreover, extensions of the approach we have presented here are noteworthy for future work. For example, optimizing the frame rate $T_s$ over time may provide additional gains in terms of task processing delay and energy efficiency.
Doing so would require stochastic optimization to capture dynamic task generation over a long-term time horizon.

\end{document}